\documentclass[11pt]{article}

\usepackage{setspace,graphicx,epstopdf,amsmath,amsfonts,amssymb,amsthm}
\usepackage{marginnote,enumitem,fancyvrb}
\usepackage{hyperref,float}
\usepackage[longnamesfirst]{natbib}
\setcitestyle{authoryear,round,semicolon,aysep={, },yysep={, }}
\usepackage[dvipsnames]{xcolor}
\usepackage{microtype}

\usepackage[margin=1in]{geometry}
\usepackage{booktabs}
\usepackage{pdflscape}
\usepackage{chngcntr}
\usepackage{multirow}
\usepackage{threeparttable}
\usepackage{dcolumn}
\usepackage{needspace}
\usepackage{caption}
\usepackage{array}
\usepackage{tabularx}
\usepackage{adjustbox}
\usepackage{longtable}
\usepackage{makecell}
\usepackage{tikz}
\usetikzlibrary{arrows.meta,positioning}
\usepackage[T1]{fontenc}
\usepackage[utf8]{inputenc}

\definecolor{ijfgreen}{RGB}{0,150,105}

\hypersetup{
  colorlinks=true,
  citecolor=ijfgreen,
  linkcolor=ijfgreen,
  urlcolor=ijfgreen,
  pdftitle={Do Carbon Price Forecasts Improve Compliance Procurement? Evidence from European Union Allowances},
  pdfauthor={Muzi Chen, Difang Huang, Shouyang Wang, and Xinghan Xia},
  pdfsubject={Forecast-origin European Union Allowance forecasting and compliance procurement},
  pdfkeywords={EU Allowances, carbon price forecasting, forecast-origin evaluation, multi-horizon forecasting, compliance procurement, optimal execution}
}
\providecommand*{\theHpage}{}
\renewcommand*{\theHpage}{main.\arabic{page}}

\newcolumntype{d}[1]{D{.}{.}{#1}}
\newcolumntype{Y}{>{\raggedright\arraybackslash}l}
\newcolumntype{Z}{>{\centering\arraybackslash}c}


\newcommand{\tablenote}[1]{}
\newcommand{\figurenote}[1]{}
\newcommand{\inserttable}[1]{}
\newcommand{\insertfigure}[1]{}
\newcommand{\standardtablelayout}{%
  \footnotesize
  \setstretch{1.0}%
  \setlength{\tabcolsep}{4.8pt}%
  \renewcommand{\arraystretch}{1.08}%
}
\newcommand{\landscapetablelayout}{%
  \small
  \setstretch{1.0}%
  \setlength{\tabcolsep}{5.2pt}%
  \renewcommand{\arraystretch}{1.10}%
}

\newenvironment{landscapetablepage}
  {\clearpage
   \begin{landscape}
   \thispagestyle{empty}
   \begin{center}
   \vspace*{\fill}
   \begin{minipage}{0.99\linewidth}
   \centering
   \captionsetup{width=\linewidth,hypcap=false}
   \landscapetablelayout}
  {\end{minipage}
   \vspace*{\fill}
   \vspace*{0.30in}
   {\normalfont\normalsize\thepage\par}
   \end{center}
   \end{landscape}
   \clearpage}

\newcommand{\directionalwidepanel}[2]{%
  {\footnotesize\textbf{#2}}\par\vspace{0.10em}
  \includegraphics[width=0.97\textwidth]{#1}\par
}

\newcommand{\papertitlelineone}{Do Carbon Price Forecasts Improve Compliance Procurement?}
\newcommand{\papertitlelinetwo}{Evidence from European Union Allowances}
\newcommand{\titlepagenote}{We thank participants at various seminars for helpful comments and suggestions. This work was supported by the National Natural Science Foundation of China (Grant Nos.\ 72471257, 72403045, 72503232, and 72574227). Authors are listed alphabetically. All remaining errors are our own.}

\begin{document}

\setlist{noitemsep}
\onehalfspacing

\hypersetup{pageanchor=false}
\title{{\Large\bfseries \papertitlelineone\\\papertitlelinetwo}\thanks{\titlepagenote}}
\author{
  {\bfseries Muzi Chen} \\
  Central University of Finance and Economics \\
  {\bfseries Difang Huang} \\
  Chinese Academy of Sciences \\
  {\bfseries Shouyang Wang} \\
  Chinese Academy of Sciences \\
  {\bfseries Xinghan Xia} \\
  Central University of Finance and Economics
}
\date{\today}

\maketitle
\thispagestyle{empty}

\clearpage
\hypersetup{pageanchor=true}
\pagenumbering{arabic}

\doublespacing

\begin{center}{\Large\bfseries \papertitlelineone\\\papertitlelinetwo}\end{center}

\vspace*{0.35in}

\begin{center}
\begin{minipage}{0.94\textwidth}
\setstretch{1.5}
\begin{center}\textbf{Abstract}\end{center}

\noindent Firms covered by emissions trading systems need forecasts not only to value allowances, but also to decide when to buy them. This paper asks whether European Union Allowance (EUA) prices contain short-horizon predictability that survives a forecast-origin information design and improves simulated compliance procurement. Using daily data from 2019 to 2025, we produce direct forecasts for one to five trading days ahead. All predictors are observable at the forecast origin, and calibration and model-selection rules are fixed before the final holdout. The released forecast has the lowest point-estimate RMSE at every horizon among fourteen benchmarks, with the strongest loss-difference evidence at horizons three and four. Relative to a random walk, out-of-sample $R^2$ rises from 1.2\% at one day to 15.5\% at five days. We then use the forecast path in a constrained procurement problem with execution costs, market impact, capacity limits, and tail risk; sensitivity exercises add demand uncertainty. For a fixed 100,000-EUA order, optimized schedules lower average realized costs by 8.5 to 38.5 basis points relative to uniform execution across horizons $h=2$ to $h=5$. The gains come from reallocating purchases within a fixed window, not from reliable next-day directional timing.


\end{minipage}
\end{center}

\newpage


\section{Introduction}

The European Union Emissions Trading System (EU ETS) turns carbon emissions into a traded compliance liability. European Union Allowance (EUA) prices affect firms' compliance costs, power producers' dispatch and hedging decisions, and valuations of carbon-intensive assets \citep{ellerman2016,martin2016}. For a covered firm with an allowance shortfall, however, the relevant short-run problem is not simply whether the price rises tomorrow. The firm must allocate purchases across a limited execution window while prices, liquidity, and the eventual compliance requirement remain uncertain. In this setting, forecast usefulness depends on ranking the coming price path and improving a feasible procurement decision.

EUA prices may contain short-horizon information because the market absorbs several economically distinct signals. Fuel prices and electricity-market conditions affect fuel-switching incentives and emissions demand, while industrial activity and weather affect generation demand \citep{alberola2008drivers,chevallier2009,hintermann2010,lutz2013nonlinearity,zheng2019regional,mi2019cities,mi2020household}. Financial conditions affect the valuation of traded allowances, and compliance deadlines and investor attention can concentrate trading pressure \citep{tan2017dependence,tan2020connectedness,zhang2022attentioncarbon}. These relationships are unlikely to be stable. EUA prices are set in a policy-constrained market whose scarcity rules, participant composition, and links to energy and financial markets evolve over time \citep{su2013competitive,su2014multiregion,sun2024embodied,zhang2017burden,karydas2019green}. Apparent predictability may therefore depend on the horizon, the market state, and the information that was genuinely available when the forecast was issued.

The carbon-price forecasting literature has developed increasingly flexible methods for extracting these signals. Earlier studies relied on time-series and fundamentals-based specifications \citep{koop2013forecasting,zhu2013arimalssvm,zhu2018multiscale}. Recent work uses nonlinear models, signal decomposition, ensemble learning, deep neural networks, and interval-valued forecasts \citep{yu2008emd,zhao2017deep,huang2022euaohcl,chen2024multiscale,yang2024multiscaleinterval,wang2025dynamicinterval}. This literature reports forecast gains from energy, financial, and attention variables. Its primary emphasis, however, remains the construction and comparison of candidate models. Three decision-relevant conditions are less settled: whether gains survive when every input and transformation respects forecast-origin information, whether calibration and forecast selection are fixed before the evaluation period, and whether lower statistical losses generate value for a compliance buyer facing execution costs and constraints.

The crude-oil forecasting literature is the closest benchmark because, like EUAs, oil is a traded commodity whose short-run price combines fundamentals, scarcity, financial conditions, and shifting macro states. \citet{alquist2013} show that in-sample predictability need not translate into out-of-sample accuracy and emphasize the no-change forecast as a demanding benchmark. \citet{baumeister2012} show that commodity-price forecasts must be constructed from data vintages and information available in real time, while \citet{baumeister2015} show that forecast performance changes with the economic environment. More broadly, real-time econometrics requires a clear separation between model development, information arrival, and genuine out-of-sample evaluation \citep{tashman2000,pesaran2005,giacomini2006}. Because this paper forecasts a one- to five-day path, the comparison must also be coherent across horizons rather than tied to a single endpoint \citep{quaedvlieg2021}.

Statistical accuracy is only an intermediate criterion. \citet{granger2000} argue that forecasts should be assessed by the decisions they support and the losses faced by the forecast user. This distinction matters in the EU ETS. A speculative trading rule and a compliance-procurement problem have different objectives, constraints, and benchmarks. A compliance buyer must complete an allowance order within a finite window, trading off expected purchase costs, bid--ask spreads, fees, market impact, price risk, and demand uncertainty subject to execution-capacity limits. This setting is related to optimal execution, where a large order is distributed over time to balance price exposure against transaction and market-impact costs \citep{bertsimas1998,almgren2001}. Evaluating EUA forecasts in this setting tests whether statistical improvements are economically usable.

This paper asks whether EUA prices exhibit forecast-origin predictability across short horizons and whether that predictability improves simulated compliance procurement relative to prespecified benchmarks. We use daily data from 2019 to 2025 and issue forecasts at the day-$t$ market close for each of the next five trading days. The design has two identifying restrictions. First, every predictor, transformation, and rolling state is observable at the forecast origin. Second, calibration and horizon-specific source selection are fixed on a validation sample before evaluation in a reserved final holdout. Test-period outcomes update only information that would mechanically become available to a forecaster. They do not reopen calibration or model selection.

The empirical analysis yields three main findings. First, the released forecast has lower point-estimate RMSE than all fourteen benchmark models at every horizon from one to five trading days in the final holdout. Predictive gains are modest at one day but increase over the multi-day path; loss-difference evidence is strongest at horizons three and four. The random-walk-relative out-of-sample $R^2$ reaches 15.5\% at five days. The gains are concentrated in price-level forecasting and path ranking, not in consistently predicting the sign of the next daily price change.

Second, the validation-fixed release rule changes the interpretation of the forecasting gain. It improves accuracy relative to the uncalibrated candidate path at every horizon and reduces level and path-slope bias. The reported series is therefore the output of a pre-specified operational rule, not an ex post selection from competing candidate paths.

Third, the forecast path has economic value for compliance procurement. We embed the released forecast in a constrained multi-day execution problem with transaction costs, market impact, capacity limits, tail risk, and allowance-demand uncertainty. For a fixed order of 100,000 EUAs, the optimized schedule reduces realized procurement costs relative to uniform execution for horizons $h=2,\ldots,5$. Savings rise from 8.5 basis points at $h=2$ to 38.5 basis points at $h=5$. These horizons correspond to three and six execution days because each window includes day 0 through day $h$. The gains are strongest when the forecast reallocates purchases within a predetermined execution window.

The paper makes three contributions. First, it applies a forecast-origin evaluation design to the daily EU carbon market by enforcing information availability at each forecast date and fixing calibration and selection before the final holdout. Second, it provides cross-horizon evidence on short-run EUA price predictability, showing that economically meaningful gains arise primarily along the multi-day price path. Third, it links statistical forecast accuracy to a compliance-specific procurement decision with realistic execution frictions and demand risk.

Additional diagnostics identify when the forecast is most informative. Predictive accuracy varies with the composition of forecast-origin information, proximity to the compliance deadline, and market volatility. Economically useful EUA predictability is strongest at multi-day horizons, under a strictly chronological forecast process, and when evaluated through the procurement decision the forecast is meant to support.

The remainder of the paper is organized as follows. Section 2 defines the real-time information set, variables, and chronological sample roles. Section 3 presents the direct multi-horizon forecasting system, validation-fixed release protocol, benchmark models, and statistical tests. Section 4 reports the final-holdout forecast results. Section 5 evaluates compliance-procurement value. Section 6 examines predictive information and state dependence. Section 7 concludes.
\section{Data and Real-Time Information Set}

\subsection{EU ETS and forecast target}

The forecast target is the daily closing price of the rolling nearest-December ICE EUA futures quotation. The price history is downloaded from the Investing.com Carbon Emissions Futures record, and the quoted contract corresponds to the physically delivered ICE Endex EUA Futures product (screen symbol C; 1,000 EUAs per lot). The series rolls from one December delivery contract to the next and is therefore a continuous EUA futures series, not an EUA spot price or a single fixed-expiry contract.

The close provides a stable daily decision timestamp and avoids arbitrary intraday sampling. It also corresponds to the price observable when a procurement schedule is set after the current trading session. Forecasts for $h=1,\ldots,5$ map into the next one to five EUA trading days. A procurement problem at horizon $h$ uses the current close on day 0 and the next $h$ forecast trading days. The execution window therefore contains $h+1$ days indexed by $j=0,\ldots,h$. The one-day forecast supports a two-day execution choice, while horizons two through five support allocation over three through six execution days.

\subsection{Real-time information boundary}

Forecasts are issued after the day-$t$ EUA close. Every input must be observable at that origin, rather than aligned retrospectively by calendar date. The admissible information set is

\begin{equation}
\mathcal{I}_t
=
\left\{
P_{\tau},\,
Z^{\mathrm{dir}}_{\tau},\,
M_{\tau},\,
S_{\tau}
:
\tau\le t,\ 
\text{and each component is observable by the close of day } t
\right\}.
\label{eq:day_t_information_set}
\end{equation}

The market cutoff is 17:00 London time. The EUA close and European energy, power, currency, and equity observations finalized before that cutoff may enter at origin $t$. A same-calendar-day observation finalized later, including a U.S.-session close, first enters at $t+1$. On holidays and non-overlapping trading days, the most recently released value is carried forward. Later observations are never backfilled. Industrial production enters only after its statistical release. Google Trends enters only after the relevant monthly observation is complete and observable. Compliance distance is known at the origin, and all rolling transformations use histories ending at $t$.

\subsection{Market-information blocks and rolling state representation}

The day-$t$ representation contains raw observables, five information indices, and three rolling state components:

\begin{equation}
x_t=
\left(
x_t^{\mathrm{raw}},
x_t^{\mathrm{info}},
x_t^{\mathrm{state}}
\right),
\qquad
x_t^{\mathrm{raw}}\in\mathbb{R}^{8},\ 
x_t^{\mathrm{info}}\in\mathbb{R}^{5},\ 
x_t^{\mathrm{state}}\in\mathbb{R}^{3}.
\label{eq:feature_block_partition}
\end{equation}

The raw layer retains EUA history and timely energy, power, financial, compliance, and attention measurements. The indices compress related variables into fuel-substitution, power-load, financial-pricing, compliance-timing, and attention blocks. Fuel and power conditions proxy thermal substitution and emissions demand. Utilities, exchange rates, yields, and equity prices capture valuation and risk-pricing conditions. Compliance distance and attention represent institutional timing and information arrival \citep{tan2017dependence,gronwald2011copula,wang2018spillover,tan2020connectedness}. Financial variables also capture uncertainty, correlation, and cross-market timing \citep{yu2023valuation,yu2023uncertainty,chen2022dynamiccorr,wu2023timezone}.

The rolling state is not an additional external information block. It summarizes endogenous residual price dynamics after the observable blocks have been filtered through a causal rolling fit \citep{chen2021dynamicanalyses}. This lets the price state evolve without duplicating the raw inputs. Coal, gas, electricity, EU utilities, EUR/USD, deadline days, and attention enter directly. Oil, industrial production, bond yields, and temperature enter through index construction or diagnostics. Table~\ref{tab:main_forecast_variables} reports the maintained inputs, and Appendix~A gives the full variable map and descriptive statistics.
\inserttable{\ref{tab:main_forecast_variables}}

\subsection{Chronological sample roles}

The raw price history begins on 2 January 2019. After an 80-observation causal burn-in, the usable sample contains 1,721 trading days from 25 April 2019 through 31 December 2025. The sample spans the end of Phase III and the first five years of Phase IV. Fixed 80\%--10\%--10\% chronological proportions yield training through 26 August 2024, validation from 27 August 2024 to 30 April 2025, and testing from 1 May to 31 December 2025. The cutoffs were fixed before test outcomes were examined and do not follow market events. The 173 test observations yield 169 complete five-day origins.

Training estimates the Transformer and GRU correction branch; validation alone determines hyperparameters, early stopping, calibration, and horizon-specific source selection. Testing only generates and scores forecasts under frozen model weights and release rules.

\begin{equation}
x_{j,t}\in\mathcal{I}_t
\quad\Longleftrightarrow\quad
r_j(t)\le t,
\label{eq:release_timing_constraint}
\end{equation}

Here $r_j(t)$ is the earliest public-release time for observation $j$. A realized price or forecast error enters a later origin only after its target date. The error of $\widehat P_{t+h|t}$ becomes available on $t+h$ and cannot affect an earlier decision. Realized outcomes expand admissible histories but never change model weights, calibration, hyperparameters, or source-selection rules. Table~\ref{tab:main_sample_split} summarizes the chronological roles.
\inserttable{\ref{tab:main_sample_split}}
\section{Forecast Construction and Validation-Fixed Release}

\subsection{Direct multi-horizon task}

At each day-$t$ origin, the realized target path and its forecast counterpart are
\begin{equation}
\begin{aligned}
\mathbf{P}_{t+1:t+5}
&=
\left(P_{t+1},\ldots,P_{t+5}\right),\\
\widehat{\mathbf{P}}_{t|t}
&=
\left(\widehat{P}_{t+1|t},\ldots,\widehat{P}_{t+5|t}\right).
\end{aligned}
\label{eq:path_target_definition}
\end{equation}
The target is the five-dimensional path of future price levels, not returns or directional indicators. The model predicts all five horizons jointly from a rolling tensor of day-$t$ variables,
\begin{equation}
X_t
=
\left[
x_{t-29},x_{t-28},\ldots,x_t
\right]^\top
\in\mathbb{R}^{30\times16}.
\label{eq:rolling_input_tensor}
\end{equation}
The 30-day lookback is rolled forward causally at each origin. Forecasting is direct rather than recursive. No predicted value is fed back to generate a later horizon. Each horizon has its own squared-error loss and release decision, so short- and medium-horizon price-level accuracy need not follow the same ranking.

\subsection{Candidate forecasts}

The primary candidate path is produced by a compact direct multi-step sequence forecaster trained on the rolling tensor,
\begin{equation}
\widehat{\mathbf P}^{\,\mathrm{main}}_{t|t}
=
f_{\theta}(X_t),
\qquad
\widehat{\mathbf P}^{\,\mathrm{main}}_{t|t}\in\mathbb{R}^{5}.
\label{eq:main_candidate_path}
\end{equation}
The main Transformer path maps the 30-day input directly into five outputs. Appendix~B reports the architecture, optimization, and training settings.

A residual-correction branch generates a second candidate path by modeling residual dynamics left by the main path,
\begin{equation}
\begin{aligned}
e_{\tau}
&=
P_{\tau}
-
\widehat{P}^{\,\mathrm{main}}_{\tau|\tau-1},\\
\widehat{\mathbf P}^{\,\mathrm{corr}}_{t|t}
&=
\widehat{\mathbf P}^{\,\mathrm{main}}_{t|t}
+
g_{\phi}(R_t),
\qquad
R_t=(e_{t-L+1},\ldots,e_t).
\end{aligned}
\label{eq:corrected_candidate_path}
\end{equation}
The GRU path adds a five-output forecast of the residual sequence to the main path \citep{cho2014gru}.

Persistence sets all five forecasts equal to $P_t$. Drift extrapolates the training-sample mean daily change. The release rule indexes the four candidate sources by
\begin{equation}
\mathcal{C}
=
\left\{
\mathrm{main},
\mathrm{corr},
\mathrm{pers},
\mathrm{drift}
\right\}.
\label{eq:candidate_set}
\end{equation}
Each label $c\in\mathcal C$ maps to a candidate path $\widehat{\mathbf P}^{\,c}_{t|t}$ observed at origin $t$.

\subsection{Validation-fixed release}

The released forecast is determined by rules fixed on the validation window before any final-holdout scoring. Calibration first maps each candidate path into the reporting scale,
\begin{equation}
\widetilde{P}^{\,c}_{t+h|t}
=
a_{c,h}
+
b_{c,h}\widehat{P}^{\,c}_{t+h|t},
\qquad
c\in\mathcal{C},\ h=1,\ldots,5.
\label{eq:calibration_map}
\end{equation}
The calibrated candidate set therefore contains calibrated main-path forecasts, calibrated corrected-path forecasts, persistence, and drift.

The release sequence is
\begin{center}
\begin{tikzpicture}[
  node distance=0.30cm,
  release step/.style={
    draw,
    rounded corners=1.5pt,
    align=center,
    text width=2.20cm,
    minimum height=1.85cm,
    inner xsep=2pt,
    inner ysep=0pt,
    font=\small
  },
  release arrow/.style={-{Latex[length=2mm,width=1.3mm]},semithick}
]
\node[release step] (data) {Forecast\\origin\\inputs};
\node[release step,right=of data] (candidate) {Candidate\\paths};
\node[release step,right=of candidate] (calibration) {Validation\\calibration};
\node[release step,right=of calibration] (selection) {\mbox{Horizon-wise}\\source selection};
\node[release step,right=of selection] (freeze) {Frozen\\release rule};
\node[release step,right=of freeze] (test) {Final\\holdout};
\draw[release arrow] (data) -- (candidate);
\draw[release arrow] (candidate) -- (calibration);
\draw[release arrow] (calibration) -- (selection);
\draw[release arrow] (selection) -- (freeze);
\draw[release arrow] (freeze) -- (test);
\end{tikzpicture}
\end{center}

For each horizon, the validation sample fixes both the calibration map and the candidate source,
\begin{equation}
\begin{aligned}
c_h^{\ast}
&=
\arg\min_{c\in\mathcal{C}}
\frac{1}{|\mathcal V|}
\sum_{t\in\mathcal V}
\left(
P_{t+h}
-
\widetilde{P}^{\,c}_{t+h|t}
\right)^2,\\
\widehat{P}^{\,\mathrm{rep}}_{t+h|t}
&=
\widetilde{P}^{\,c_h^{\ast}}_{t+h|t},
\qquad
t\in\mathcal T^{\mathrm{test}},\ h=1,\ldots,5.
\end{aligned}
\label{eq:reported_forecast_rule}
\end{equation}
Equation~\eqref{eq:reported_forecast_rule} minimizes validation MSE separately for each horizon. Because the square root is monotone, it produces the same source ranking as horizon-specific RMSE. The released forecast uses the calibrated main Transformer path at $h=1$ through $h=4$ and the calibrated GRU-corrected path at $h=5$.

During the final holdout, the Transformer and GRU parameters, calibration coefficients, horizon-specific source choices, and release rules remain fixed. At each origin, the loop updates state variables and residual inputs only from prices already observed by that date.

\subsection{Benchmarks and statistical tests}

The benchmark panel contains four groups. The first comprises Random Walk and drift. The second contains AR/ARX-style dynamic regression, LASSO, Elastic Net, and Spline-Ridge GAM. The third contains Random Forest, Gradient Boosting, SVR, and KNN. The fourth contains LSTM, Temporal Fusion Transformer, and Informer sequence models. Historical Mean is retained as an additional naive reference. All models use the same forecast origins and day-$t$ information boundary, with model-appropriate representations. Hyperparameters are chosen on validation data by average RMSE across the five horizons and then frozen. Appendix~C reports the representations, grids, and selected settings \citep{hyndman2020history,makridakis2020m4,wang2023forecastcombinations,sun2021tvma}.

Let $\mathcal T^{\mathrm{test}}$ denote the 169 final-holdout forecast origins with realized outcomes for all five horizons, and let $N_{\mathrm{test}}=|\mathcal T^{\mathrm{test}}|$. RMSE is the primary ranking metric because the released forecast is a level path rather than a directional trading rule,
\begin{equation}
\mathrm{RMSE}_h
=
\left[
\frac{1}{N_{\mathrm{test}}}
\sum_{t\in\mathcal T^{\mathrm{test}}}
\left(
P_{t+h}-\widehat P_{t+h|t}
\right)^2
\right]^{1/2},
\qquad
h=1,\ldots,5.
\label{eq:rmse_definition}
\end{equation}
MAE and mean signed error (bias) describe absolute and systematic level errors. Relative performance versus the Random Walk is summarized by Campbell--Thompson out-of-sample $R^2$ \citep{campbell2008},
\begin{equation}
R^2_{OOS,h}
=
1-
\frac{
\sum_{t\in\mathcal T^{\mathrm{test}}}
\left(
P_{t+h}-\widehat P_{t+h|t}
\right)^2
}{
\sum_{t\in\mathcal T^{\mathrm{test}}}
\left(
P_{t+h}-\widehat P^{\,RW}_{t+h|t}
\right)^2
}.
\label{eq:roos_definition}
\end{equation}
The main comparison reports the best benchmark at every horizon together with RMSE, MAE, bias, and $R^2_{OOS}$. Diebold--Mariano tests compare squared forecast loss with the horizon-specific best benchmark. These pairwise $p$-values condition on the displayed comparator and are not adjusted for selecting that comparator by final-holdout RMSE. Clark--West adjusted-loss statistics provide the corresponding one-sided comparison, with positive values favoring the released forecast. Appendix~D reports the full benchmark panel and pairwise matrices \citep{samuels2017mcs}.

\section{Cross-Horizon Forecast Results}

\subsection{Main benchmark comparison}

Table~\ref{tab:main_holdout_accuracy} reports the main cross-horizon benchmark comparison. The released forecast records the lowest final-holdout RMSE at all five horizons. The horizon-specific best competitors are KNN at $h=1$, Spline-Ridge GAM at $h=2$, Gradient Boosting at $h=3$, Random Forest at $h=4$, and Informer at $h=5$.
\inserttable{\ref{tab:main_holdout_accuracy}}

The one-day advantage is small. RMSE is 0.976 for the released forecast and 0.979 for KNN, a difference of 0.003. The gap becomes larger from three to five days. Released-forecast RMSE is lower than the best competing RMSE by 0.043 at $h=3$, 0.062 at $h=4$, and 0.050 at $h=5$. Relative to the Random Walk, $R^2_{OOS}$ increases from 1.2\% at one day to 15.5\% at five days.

These results concern forecasts of the future EUA price level. They show that the model becomes more useful for representing and ranking the multi-day price path. They do not imply consistent predictability of the sign of the next daily price change.

\subsection{Statistical significance and loss sensitivity}

Table~\ref{tab:main_loss_sensitivity} reports horizon-specific inference and loss sensitivity. Squared-loss gains relative to the best competing model are 0.007, 0.047, 0.126, 0.205, and 0.173 from $h=1$ to $h=5$. Newey--West Diebold--Mariano tests use lag $h-1$ and reject equal squared-error loss at $h=3$ and $h=4$, with $p$-values of 0.032 and 0.019. The inference is pairwise against the displayed best competitor rather than selection-adjusted across the benchmark family. A moving-block bootstrap with block length five gives the same inference. The 95\% intervals exclude zero at $h=3$ and $h=4$.
\inserttable{\ref{tab:main_loss_sensitivity}}

The loss ranking is also stable under absolute error beyond the first horizon. Released-forecast MAE is lower than that of the best competing model by 0.009, 0.033, 0.054, and 0.031 from $h=2$ to $h=5$. At $h=1$, released MAE is 0.001 higher despite its lower RMSE. A five-horizon Newey--West Wald test of the joint equal-squared-loss null yields $\chi^2(5)=9.09$ and $p=0.106$. The cross-horizon ordering is therefore consistent in point estimates, but joint statistical precision remains limited in the 169-anchor final holdout.

Directional accuracy provides a boundary result. It is 0.491 at $h=1$ and ranges from 0.580 to 0.633 over $h=2$ through $h=5$. The one-day value is indistinguishable from chance. The directional results are not the basis for either the RMSE ranking or the procurement application.

\subsection{Candidate versus released forecast}

Table~\ref{tab:main_holdout_reporting_comparison} evaluates whether the validation-fixed release rule improves the uncalibrated candidate path without using final-holdout outcomes. The candidate forecasts, realized prices, and 169 forecast origins are held fixed. Only the calibration and horizon-specific source choices determined on the validation sample distinguish the released series.
\inserttable{\ref{tab:main_holdout_reporting_comparison}}

RMSE declines at every horizon, from 1.064 to 0.976 at $h=1$, from 1.344 to 1.316 at $h=2$, from 1.586 to 1.424 at $h=3$, from 1.640 to 1.615 at $h=4$, and from 1.750 to 1.700 at $h=5$. Statistical strength is not uniform. The candidate-versus-released Diebold--Mariano test rejects equal squared-error loss at $h=1$ ($p=0.022$), is marginal at $h=3$ ($p=0.079$), and does not reject at the remaining horizons.

The changes are concentrated in level and path-slope bias. Candidate bias is positive at $h=1$, $h=2$, $h=3$, and $h=5$, reaching 0.541 at $h=3$. The released path shifts these biases toward a lower trajectory, with signed bias between $-0.107$ and $-0.379$ across horizons. The release comparison therefore measures the holdout performance of a validation-fixed operational rule, not a separate general forecasting method.

\section{Compliance Procurement Value}

\subsection{Decision setting and scenario construction}

At forecast origin $t$, a compliance buyer must procure a required quantity $Q_t$. For horizon $h$, execution may occur at the current close on day 0 or on any of the next $h$ trading days. The procurement window is therefore $\{0,1,\ldots,h\}$ and contains exactly $h+1$ execution days. The optimized schedule is compared with uniform execution over the same window and immediate purchase of the full order on day 0. The objective minimizes expected procurement cost plus tail risk relative to the stronger of these two feasible benchmarks.

\Needspace{14\baselineskip}
Let $\mathbf q_t=(q_{t,0},\ldots,q_{t,h})$ denote the purchase schedule, and let $V_t$ be rolling 20-day median market volume in EUA. The fixed-demand problem requires
\begin{equation}
\begin{aligned}
\sum_{j=0}^{h}q_{t,j}&=Q_t,
&
q_{t,0}&\geq 0,\\
0\leq q_{t,j}&\leq \bar q_t\quad (j=1,\ldots,h),
&
\bar q_t&=\min\{0.10V_t,0.50Q_t\}.
\end{aligned}
\label{eq:procurement_quantity}
\end{equation}
For each decision, 250 future price paths are formed by adding centered, jointly resampled historical forecast-error paths to the released forecast. Only error paths fully realized before origin $t$ enter the scenario library. The current price is fixed at $P_t$ in every scenario.

\Needspace{16\baselineskip}
\subsection{Execution costs and optimization}

Let $P_{t+j}^{(k)}$ denote the scenario price on execution day $j$, $f$ the exchange fee per EUA, and $s$ the half-spread in basis points. Scenario cost for demand $D$ is
\begin{align}
C_t^{(k)}(\mathbf q_t,D)
={}&
\sum_{j=0}^{h}q_{t,j}
\left(P_{t+j}^{(k)}+f+\frac{s}{10^4}P_{t+j}^{(k)}\right)
\;+\;
\sum_{j=0}^{h}
\frac{\eta P_{t+j}^{(k)}}{V_t}q_{t,j}^{2}
\;+\;
\Pi_t^{(k)}(\mathbf q_t,D), \label{eq:procurement_cost}\\
\Pi_t^{(k)}(\mathbf q_t,D)
={}&
\left(P_{t+h}^{(k)}+\pi\right)
\left(D-\sum_{j=0}^{h}q_{t,j}\right)^{+}
-
\left(P_{t+h}^{(k)}-v\right)
\left(\sum_{j=0}^{h}q_{t,j}-D\right)^{+}.
\label{eq:procurement_imbalance}
\end{align}

The main fixed-demand evaluation uses a 100,000-EUA order. Market impact is calibrated to five basis points at 10\% of rolling median daily volume. Future-day purchases are capped at 10\% of rolling median daily volume and 50\% of the total order. Demand risk is represented by
\[
D\in\{0.75Q_t,Q_t,1.25Q_t\},
\qquad
\Pr(D)=(0.25,0.50,0.25).
\]
The objective combines expected unit cost with relative tail risk:
\begin{align}
\min_{\mathbf q_t}\quad
&\frac{\mathbb E_{k,D}\!\left[C_t^{(k)}(\mathbf q_t,D)\right]}{Q_t}
\;+\;
\lambda\,\operatorname{CVaR}_{0.95}
\left(
\frac{\mathbb E_D[C_t^{(k)}(\mathbf q_t,D)]-B_t^{(k)}}{Q_t}
\right),
\label{eq:procurement_objective}\\
B_t^{(k)}
={}&
\min\!\left\{
\mathbb E_D[C_t^{(k)}(\mathbf q_t^{\mathrm{TWAP}},D)],
\mathbb E_D[C_t^{(k)}(\mathbf q_t^{\mathrm{Imm}},D)]
\right\},
\label{eq:procurement_benchmark}
\end{align}
where $\lambda=0.02$. The TWAP schedule divides $Q_t$ uniformly across day 0 through day $h$, while immediate purchase assigns the full order to day 0. Appendix~F reports exchange fees, the spread, lot size, imbalance valuation, scenario construction, and the complete parameter calibration.

\subsection{Fixed-demand results and benchmark comparison}

Table~\ref{tab:main_procurement_savings} reports fixed-demand results for a 100,000-EUA order. The optimized schedule lowers average realized cost relative to TWAP at every horizon. Mean savings increase from 8.5 basis points at $h=2$ to 17.7, 28.7, and 38.5 basis points at $h=3$, $h=4$, and $h=5$.
\inserttable{\ref{tab:main_procurement_savings}}

The corresponding monetary gains rise from EUR 6.7 thousand per procurement decision at $h=2$ to EUR 14.1 thousand at $h=3$, EUR 22.0 thousand at $h=4$, and EUR 29.3 thousand at $h=5$.

Newey--West inference uses lag $h$ to account for overlap between windows spanning day 0 through day $h$. The two-sided $p$-values are 0.160, 0.032, 0.005, and 0.001 from $h=2$ through $h=5$. Over the same horizons, the share of origins at which the optimized schedule costs less than TWAP rises from 52.9\% to 56.2\%, 58.8\%, and 65.9\%. Both statistical precision and the win rate strengthen as the execution window expands.

Against immediate purchase, which locks the full order at the current close, mean savings are $-2.1$, 1.9, 5.8, and 7.9 basis points from $h=2$ through $h=5$. For a 100,000-EUA order, these differences equal EUR $-1.6$ thousand, EUR 1.4 thousand, EUR 3.9 thousand, and EUR 5.3 thousand per decision.

The Newey--West $p$-values are 0.511, 0.692, 0.521, and 0.521, so this benchmark does not provide precise evidence of average savings. Forecast value therefore arises mainly from improving allocation within a predetermined execution window rather than from delaying the full purchase.

\subsection{Demand and implementation sensitivity}

Table~\ref{tab:main_procurement_sensitivity} reports the principal sensitivity results at every horizon from $h=2$ through $h=5$. Under demand uncertainty and central realized demand, savings relative to TWAP move from $-2.1$ basis points at $h=2$ to 38.8 basis points at $h=5$; relative to immediate purchase, they move from $-12.5$ to 8.5 basis points. On this displayed comparison, demand risk delays the economic advantage until the longer execution windows.
\inserttable{\ref{tab:main_procurement_sensitivity}}

Order size changes the share of the gross forecasting gain that survives execution frictions. At $h=5$, savings relative to TWAP decline from 39.2 basis points for 10,000 EUA to 29.6 basis points for one million EUA, while the corresponding monetary gain rises from EUR 3.0 thousand to EUR 228.2 thousand.

Non-overlapping windows retain every $(h+1)$th procurement origin. Savings relative to TWAP are 15.6, 10.0, 28.9, and 35.8 basis points from $h=2$ through $h=5$; the $h=2$, $h=4$, and $h=5$ estimates remain significant at the 5\% level. The immediate-purchase comparison remains less precise.

Fees and the bid--ask spread affect every feasible schedule. Market impact and daily capacity determine how aggressively purchases can be shifted toward forecast-low-price days. Larger impact coefficients and tighter capacity limits reduce the scope for concentrating execution.

\Needspace{4\baselineskip}
Risk aversion enters through $\lambda$. Lower values place more weight on expected cost, whereas higher values place more weight on adverse relative-cost scenarios. Appendix~F reports the full demand grid, four order sizes, non-overlapping estimates, and the fee, spread, market-impact, capacity, and risk-aversion specifications.

\section{Predictive Information and State Dependence}

\subsection{Information-block removal}

Table~\ref{tab:main_block_removal} reports re-estimated structural comparisons. Each reduced specification retains the chronological split and the fixed architecture and optimization settings. It then re-estimates the forecasting and correction models and refits the validation-stage calibration and horizon-specific release rule. The exercise changes the available information without carrying a release rule fitted to a different input set into the final holdout.
\inserttable{\ref{tab:main_block_removal}}

Removing all information indices raises horizon-five RMSE from 1.700 to 1.740, while retaining only the key raw predictors raises it to 1.759. Removing the constructed residual state while retaining observed price history raises horizon-five RMSE to 1.811. The deterioration is concentrated at the end of the forecast path and remains moderate in magnitude.

\subsection{Alternative state representations}

Table~\ref{tab:main_state_representations} compares VMD with EMD, autoregressive residual lags, an EWMA residual level and volatility state, a local-level Kalman state, and a specification without a constructed residual state. All alternatives are causal and use only information available by the forecast origin. They are evaluated under the same re-estimation and validation-fixed release procedure.
\inserttable{\ref{tab:main_state_representations}}

VMD records the lowest horizon-five RMSE at 1.700, but the local-level Kalman state is nearly identical at 1.705. Autoregressive residual lags produce 1.718, followed by EWMA at 1.789, EMD at 1.791, and no constructed state at 1.811. The ordering is not uniform at shorter horizons. The Kalman and autoregressive-lag specifications slightly improve on VMD at $h=1$, while the no-state specification is marginally lower at $h=2$ and $h=3$. The evidence supports a maintained rolling state at the path tail, but it does not identify VMD as uniquely necessary.

\subsection{Attribution and conditional sensitivity}

Grouped SHAP is retained as an auxiliary diagnostic of activity inside the fitted rule \citep{lundberg2017unified}. Table~\ref{tab:main_shap_top_blocks} ranks the financial index first at horizons one, two, and five, attention first at $h=3$, and power-market information first at $h=4$. At $h=5$, the grouped scores are 1.549 for financial information, 1.224 for attention, and 1.099 for power-market information. These local attributions summarize how the maintained prediction is formed.
\inserttable{\ref{tab:main_shap_top_blocks}}

Fuel sensitivity is interpreted through fuel-switching costs and relative generation economics rather than through a unilateral coal-price perturbation. The relevant objects are the clean dark spread, the clean spark spread, and the coal-to-gas switching signal. Together, they determine the relative cost of coal- and gas-fired generation after fuel, carbon, and power prices are combined. The reported fuel information index captures common fuel-cost pressure and the coal-to-gas price differential. A coal-only shock has no invariant sign once gas prices, power prices, plant efficiencies, and emissions intensities are held apart. It is therefore not used as evidence of a directional predictive response.

The regime-partitioned SHAP results in Table~\ref{tab:main_shap_partition_scores} provide a further conditional diagnostic. Within the low-volatility and far-from-compliance rows, financial information and attention rank ahead of the other non-price information groups. Within the high-volatility row, the state representation has the largest grouped score. These patterns describe conditional attribution within each fitted partition, not causal effects across regimes.
\inserttable{\ref{tab:main_shap_partition_scores}}

\subsection{Forecastability across states}

Table~\ref{tab:main_regime_accuracy} reports group sizes and forecast accuracy for the compliance-distance and volatility partitions. Differences in RMSE are evaluated with a circular moving-block bootstrap of five consecutive forecast origins. This preserves short-range dependence induced by overlapping forecast windows.
\inserttable{\ref{tab:main_regime_accuracy}}

At $h=1$, RMSE is 0.236 lower in the transition group than far from compliance ($p=0.050$) and 0.248 lower near compliance ($p=0.007$). The near-compliance group contains only 22 observations, so its point estimate is not treated as a stable ranking. At $h=5$, the transition and near-compliance differences are $-0.371$ and $-0.450$, with $p$-values of 0.087 and 0.076. The longer-horizon compliance differences are therefore suggestive rather than conventionally significant.

Volatility separates one-day precision more clearly. Relative to the 54 low-volatility observations, RMSE is 0.211 higher in the mid-volatility group ($p=0.027$) and 0.414 higher in the high-volatility group ($p=0.001$). At $h=5$, the high-minus-low difference is 0.416 but is not statistically distinguishable from zero ($p=0.221$); the mid-minus-low difference is $-0.110$ ($p=0.651$). Directional accuracy does not move monotonically across either partition.

These partitions describe conditional forecast performance, not causal effects of compliance proximity or volatility. Their role is to locate where the released forecast is more or less precise within the holdout. The small near-compliance cell and the uncertainty in the five-day group differences preclude a stronger interpretation.

\section{Conclusion}

This paper examines whether EUA prices contain economically useful short-horizon predictability. We construct direct one- to five-day price forecasts under a forecast-origin design in which all predictors and transformations respect the information available when the forecast is issued. Calibration and horizon-specific forecast selection are completed on a validation sample and remain fixed throughout the final holdout. The evaluation therefore concerns the forecast that would have been released in practice.

The released forecast has lower RMSE than all fourteen benchmark models at every horizon in the final holdout. Predictive gains are limited at one day but increase over the multi-day price path, with random-walk-relative out-of-sample $R^2$ reaching 15.5\% at five days. Validation-fixed release also improves the candidate forecast at each horizon by correcting level and path errors. These results identify a clear horizon structure in EUA predictability. Economically meaningful information is stronger for forecasting and ranking the coming price path than for predicting the direction of the next daily price change.

The multi-day forecasts also improve compliance procurement. When the released price path is embedded in an execution problem with transaction costs, market impact, capacity limits, tail risk, and uncertain allowance demand, the optimized schedule reduces realized costs relative to uniform execution across all procurement horizons. For a fixed order of 100,000 EUAs, savings increase from 8.5 basis points at $h=2$ to 38.5 basis points at $h=5$, corresponding to execution windows of three and six days. The economic value of the forecast rises with the scope for reallocating purchases across the execution window and is strongest at horizons four and five.

The diagnostic results show that forecast performance depends on observable market information and on the representation of the evolving price state. Financial conditions, market attention, fuel and power-market information, and compliance timing contribute differently across horizons and market states. Forecast accuracy is generally higher near the compliance deadline and under lower volatility, indicating that the value of short-horizon prediction varies with the economic environment.

Overall, the findings connect three dimensions of carbon-price forecasting that are often evaluated separately: real-time information discipline, cross-horizon statistical accuracy, and decision value. EUA predictability becomes economically relevant when the forecast process is fixed before evaluation, the full multi-day price path is considered, and forecast performance is measured against the procurement decision faced by compliance buyers. The central implication is that short-horizon carbon-price forecasts are most useful for improving allowance purchases within a defined execution window, not for unconditional market timing.


\clearpage
\singlespacing
\bibliographystyle{apalike}

\clearpage
\doublespacing

\clearpage
\begin{table}[p]
\centering
\standardtablelayout
\caption[Forecast Variables and Roles]{\textbf{Forecast Variables and Roles.} The forecasting inputs are organized as raw variables, information indices, state components, and auxiliary series. Direct inputs enter the maintained 30-day sequence, while auxiliary series enter only through index construction or supplementary diagnostics. The forecast target is the rolling nearest-December ICE EUA futures close.}
\label{tab:main_forecast_variables}
\begin{tabularx}{0.94\textwidth}{>{\raggedright\arraybackslash}p{0.17\textwidth}>{\raggedright\arraybackslash}p{0.19\textwidth}>{\raggedright\arraybackslash}X>{\centering\arraybackslash}p{0.10\textwidth}}
\toprule
Block & Series or index & Use in the forecasting system & Direct input \\
\midrule
Target and raw input & Rolling ICE EUA futures close & Current and lagged EUA futures prices enter the 30-day input window; future closes are targets. & Yes \\
Raw input & Coal price & Observed fuel-cost series in the sequence input. & Yes \\
Raw input & Gas price & Observed fuel-cost series in the sequence input. & Yes \\
Raw input & EU power price & Observed power-market series in the sequence input. & Yes \\
Raw input & EU utilities & Observed equity-market series in the sequence input. & Yes \\
Raw input & EUR/USD & Observed financial-pricing series in the sequence input. & Yes \\
Raw input & Deadline days & Observed compliance-timing series in the sequence input. & Yes \\
Raw input & GT attention & Observed attention series in the sequence input. & Yes \\
Information block & Fuel idx. & Rolling fuel-substitution index from fuel prices and the coal-gas spread. & Yes \\
Information block & Power-load idx. & Rolling power-load index from power price, industrial production, and temperature. & Yes \\
Information block & Financial idx. & Rolling financial-pricing index from utilities, Stoxx 50, EUR/USD, and bond yields. & Yes \\
Information block & Compliance idx. & Rolling compliance-calendar index from the negative standardization of deadline days. & Yes \\
Information block & Attention idx. & Rolling market-attention index from standardized Google Trends information. & Yes \\
State block & Three VMD modes & Rolling VMD modes from the endogenous residual; terminal values enter as state inputs. & Yes \\
Auxiliary only & Oil, Stoxx 50, bond yield, industrial production, temperature & Used only for index construction or supplementary diagnostics. & No \\
Comparison only & Three EMD modes & Used only in the decomposition comparison and the raw-layer ablation. & No \\
\bottomrule
\end{tabularx}
\end{table}
 
\clearpage
\begin{table}[t]
\centering
\standardtablelayout
\caption[Chronological Sample Roles]{\textbf{Chronological Sample Roles.} Dates and observations refer to the usable sample after the 80-observation causal burn-in. Training estimates model parameters, validation fixes hyperparameters and release rules, and the final holdout evaluates the frozen forecast. The test block contains 173 observations and 169 origins with complete five-horizon outcomes.}
\label{tab:main_sample_split}
\begin{tabular*}{0.92\textwidth}{@{\extracolsep{\fill}}YZZZZ}
\toprule
Subset & Start date & End date & Observations & Share \\
\midrule
Training set & 2019-04-25 & 2024-08-26 & 1,376 & 79.95\% \\
Validation set & 2024-08-27 & 2025-04-30 & 172 & 9.99\% \\
Test set & 2025-05-01 & 2025-12-31 & 173 & 10.05\% \\
Full sample & 2019-04-25 & 2025-12-31 & 1,721 & 100.00\% \\
\bottomrule
\end{tabular*}
\end{table}
 
\clearpage
\begin{landscapetablepage}
\captionof{table}[Cross-Horizon Benchmark Accuracy]{\textbf{Cross-Horizon Benchmark Accuracy.} The released forecast and the lowest-RMSE competing model at each horizon are evaluated on the same 169 final-holdout origins. Bias is the mean signed forecast error. $R^2_{OOS}$ is measured relative to the Random Walk, with positive values favoring the released forecast. Best-benchmark comparisons are pairwise and not selection-adjusted.}
\label{tab:main_holdout_accuracy}
\begin{adjustbox}{width=\linewidth}
\begin{tabular}{lrrrlrrrr}
\toprule
Horizon & Released RMSE & Released MAE & Released bias & Best benchmark & Bench. RMSE & Bench. MAE & Bench. bias & $R^2_{OOS}$ (\%) \\
\midrule
$h=1$ & 0.976 & 0.781 & -0.107 & KNN & 0.979 & 0.779 & -0.138 & 1.2 \\
$h=2$ & 1.316 & 1.065 & -0.153 & Spline-Ridge GAM & 1.333 & 1.074 & -0.285 & 4.6 \\
$h=3$ & 1.424 & 1.179 & -0.173 & Gradient Boosting & 1.467 & 1.213 & -0.297 & 9.7 \\
$h=4$ & 1.615 & 1.313 & -0.379 & Random Forest & 1.677 & 1.367 & -0.471 & 9.3 \\
$h=5$ & 1.700 & 1.360 & -0.275 & Informer Sequence & 1.750 & 1.391 & -0.315 & 15.5 \\
\bottomrule
\end{tabular}
\end{adjustbox}
\end{landscapetablepage}
 
\clearpage
\begin{landscapetablepage}
\captionof{table}[Forecast-Loss Inference and Sensitivity]{\textbf{Forecast-Loss Inference and Sensitivity.} Each horizon compares the released forecast with its lowest-RMSE competing model over the same 169 final-holdout origins. Squared-loss gain and MAE gain equal benchmark loss minus released-forecast loss, so positive values favor the released forecast. The 95\% moving-block-bootstrap intervals use 2,000 draws with block length five; Diebold--Mariano $p$-values are two-sided with Newey--West lag $h-1$, and Clark--West $p$-values are one-sided. Reported pairwise $p$-values are not selection-adjusted for choosing the displayed comparator by final-holdout RMSE. DA denotes released-forecast directional accuracy; the joint Newey--West Wald test gives $\chi^2(5)=9.09$ with $p=0.106$.}
\label{tab:main_loss_sensitivity}
\begin{adjustbox}{width=\linewidth}
\begin{tabular}{llrrrrrr}
\toprule
Horizon & Best benchmark & Squared-loss gain & MAE gain & MBB 95\% interval & DM $p$-value & CW $p$-value & DA \\
\midrule
$h=1$ & KNN & 0.007 & -0.001 & $[-0.002,\ 0.016]$ & 0.207 & 0.071 & 0.491 \\
$h=2$ & Spline-Ridge GAM & 0.047 & 0.009 & $[-0.017,\ 0.119]$ & 0.224 & 0.014 & 0.580 \\
$h=3$ & Gradient Boosting & 0.126 & 0.033 & $[0.011,\ 0.237]$ & 0.032 & $<0.001$ & 0.627 \\
$h=4$ & Random Forest & 0.205 & 0.054 & $[0.042,\ 0.375]$ & 0.019 & $<0.001$ & 0.604 \\
$h=5$ & Informer Sequence & 0.173 & 0.031 & $[-0.141,\ 0.482]$ & 0.289 & 0.002 & 0.633 \\
\bottomrule
\end{tabular}
\end{adjustbox}
\end{landscapetablepage}

\clearpage
\begin{landscapetablepage}
\captionof{table}[Released and Candidate Forecast Accuracy]{\textbf{Released and Candidate Forecast Accuracy.} The validation-fixed released forecast and uncalibrated candidate path are evaluated on the same 169 final-holdout origins. RMSE reduction is $100\times(\text{Candidate RMSE}-\text{Released RMSE})/\text{Candidate RMSE}$, and bias is the mean signed forecast error. Diebold--Mariano statistics use squared-error loss and Newey--West lag $h-1$; negative statistics favor the released forecast, and reported $p$-values are two-sided.}
\label{tab:main_holdout_reporting_comparison}
\begin{tabular*}{\linewidth}{@{\extracolsep{\fill}}lccccccc}
\toprule
Horizon & Released RMSE & Candidate RMSE & RMSE reduction (\%) & Released bias & Candidate bias & DM stat. & $p$-value \\
\midrule
$h=1$ & 0.976 & 1.064 & 8.3 & -0.107 & 0.353 & -2.312 & 0.022 \\
$h=2$ & 1.316 & 1.344 & 2.1 & -0.153 & 0.077 & -0.593 & 0.554 \\
$h=3$ & 1.424 & 1.586 & 10.2 & -0.173 & 0.541 & -1.767 & 0.079 \\
$h=4$ & 1.615 & 1.640 & 1.6 & -0.379 & -0.192 & -0.425 & 0.671 \\
$h=5$ & 1.700 & 1.750 & 2.8 & -0.275 & 0.049 & -0.793 & 0.429 \\
\bottomrule
\end{tabular*}
\end{landscapetablepage}

\clearpage
\begin{table}[p]
\centering
\standardtablelayout
\caption[Fixed-Demand Procurement Value]{\textbf{Fixed-Demand Procurement Value.} Results use a 100,000-EUA order and an execution window from day 0 through day $h$, containing $h+1$ trading days. The optimized schedule includes exchange fees, bid--ask spread, quadratic market impact, future-day capacity, and 95\% CVaR. Savings equal benchmark cost minus optimized cost, so positive values indicate lower procurement cost; EUR columns report average savings per decision, and win rate is the share of origins with positive savings. HAC $p$-values are two-sided and use Newey--West lag $h$.}
\label{tab:main_procurement_savings}
\begin{adjustbox}{max width=\textwidth}
\begin{tabular}{lrrrrrrr}
\toprule
& \multicolumn{4}{c}{Relative to TWAP} & \multicolumn{3}{c}{Relative to immediate purchase} \\
\cmidrule(lr){2-5}\cmidrule(lr){6-8}
Forecast horizon & Savings (bps) & Savings (EUR) & Win rate & HAC $p$ & Savings (bps) & Savings (EUR) & HAC $p$ \\
\midrule
$h=2$ & 8.5 & 6,708 & 0.529 & 0.160 & $-2.1$ & $-1,649$ & 0.511 \\
$h=3$ & 17.7 & 14,053 & 0.562 & 0.032 & 1.9 & 1,420 & 0.692 \\
$h=4$ & 28.7 & 21,972 & 0.588 & 0.005 & 5.8 & 3,932 & 0.521 \\
$h=5$ & 38.5 & 29,290 & 0.659 & 0.001 & 7.9 & 5,324 & 0.521 \\
\bottomrule
\end{tabular}
\end{adjustbox}
\end{table}

\clearpage
\begin{landscapetablepage}
\captionof{table}[Procurement Sensitivity]{\textbf{Procurement Sensitivity.} Forecast horizon $h$ defines $h+1$ execution days from day 0 through day $h$, and columns report horizons two through five. Entries are mean realized savings in basis points for central realized demand, with positive values indicating lower optimized cost than the stated benchmark. Full-sample results use all eligible origins, while non-overlap retains every $(h+1)$th origin. The uncertain-demand schedule is optimized over $0.75Q_t$, $Q_t$, and $1.25Q_t$.}
\label{tab:main_procurement_sensitivity}
\begin{adjustbox}{width=\linewidth}
\begin{tabular}{llrrrrrrrr}
\toprule
Sensitivity dimension & Specification & \multicolumn{4}{c}{Relative to TWAP} & \multicolumn{4}{c}{Relative to immediate purchase} \\
\cmidrule(lr){3-6}\cmidrule(lr){7-10}
& & $h=2$ & $h=3$ & $h=4$ & $h=5$ & $h=2$ & $h=3$ & $h=4$ & $h=5$ \\
\midrule
Central fixed demand & 100,000 EUA, full sample & 8.5 & 17.7 & 28.7 & 38.5 & $-2.1$ & 1.9 & 5.8 & 7.9 \\
Demand uncertainty & 100,000-EUA reference, full sample & $-2.1$ & 6.5 & 21.9 & 38.8 & $-12.5$ & $-9.1$ & $-0.7$ & 8.5 \\
Order size & 10,000 EUA, fixed demand & 8.8 & 17.6 & 29.0 & 39.2 & $-2.1$ & 1.3 & 5.8 & 8.2 \\
Order size & 1,000,000 EUA, fixed demand & 5.5 & 12.8 & 17.7 & 29.6 & $-1.4$ & 0.8 & $-1.2$ & 2.9 \\
Window overlap & 100,000 EUA, fixed demand, non-overlap & 15.6 & 10.0 & 28.9 & 35.8 & $-4.5$ & $-3.9$ & 2.3 & 7.2 \\
\bottomrule
\end{tabular}
\end{adjustbox}
\end{landscapetablepage}
 
\clearpage
\begin{landscapetablepage}
\captionof{table}[Information and State Removal]{\textbf{Information and State Removal.} Each reduced specification uses the original chronological split and fixed architecture and optimization settings. The forecasting model, residual correction, validation calibration, and horizon-specific release rule are re-estimated for the available inputs. $\Delta$RMSE$_{h=5}$ is measured against the full released forecast, with positive values indicating deterioration.}
\label{tab:main_block_removal}
\begin{tabular*}{\linewidth}{@{\extracolsep{\fill}}YYZZZZZZ}
\toprule
Specification & Available representation & RMSE$_1$ & RMSE$_2$ & RMSE$_3$ & RMSE$_4$ & RMSE$_5$ & $\Delta \mathrm{RMSE}_5$ \\
\midrule
Full released forecast & Information indices, raw predictors, VMD state & 0.976 & 1.316 & 1.424 & 1.615 & 1.700 & 0.000 \\
Drop information indices & Raw predictors and VMD state & 0.983 & 1.339 & 1.492 & 1.623 & 1.740 & +0.040 \\
Key raw predictors only & Key raw predictors and VMD state & 0.971 & 1.345 & 1.461 & 1.687 & 1.759 & +0.059 \\
No constructed state & Information indices and raw predictors & 0.970 & 1.305 & 1.422 & 1.661 & 1.811 & +0.111 \\
\bottomrule
\end{tabular*}
\end{landscapetablepage}
 
\clearpage
\begin{landscapetablepage}
\captionof{table}[Alternative State Representations]{\textbf{Alternative State Representations.} State inputs are constructed from residual histories available at each forecast origin. Every specification uses the original chronological split and fixed architecture and optimization settings. The forecasting model, correction path, validation calibration, and horizon-specific release rule are re-estimated, with lower RMSE indicating greater forecast accuracy.}
\label{tab:main_state_representations}
\begin{tabular*}{\linewidth}{@{\extracolsep{\fill}}YZZZZZ}
\toprule
State representation & RMSE$_1$ & RMSE$_2$ & RMSE$_3$ & RMSE$_4$ & RMSE$_5$ \\
\midrule
VMD & 0.976 & 1.316 & 1.424 & 1.615 & 1.700 \\
EMD & 0.986 & 1.335 & 1.490 & 1.741 & 1.791 \\
AR residual lags & 0.970 & 1.314 & 1.517 & 1.595 & 1.718 \\
EWMA level and volatility & 0.980 & 1.345 & 1.500 & 1.648 & 1.789 \\
Local-level Kalman state & 0.968 & 1.311 & 1.645 & 1.591 & 1.705 \\
No constructed state & 0.970 & 1.305 & 1.422 & 1.661 & 1.811 \\
\bottomrule
\end{tabular*}
\end{landscapetablepage}
 
\clearpage
\begin{table}[p]
\centering
\standardtablelayout
\caption[Leading SHAP Blocks by Horizon]{\textbf{Leading SHAP Blocks by Horizon.} Each row ranks the three largest grouped SHAP scores for the released forecast at one horizon. Price history and the constructed residual state are treated as separate groups. Larger scores indicate stronger local attribution within the row.}
\label{tab:main_shap_top_blocks}
\begin{tabular*}{0.92\textwidth}{@{\extracolsep{\fill}}YYZYZYZ}
\toprule
Horizon & Rank 1 block & Score & Rank 2 block & Score & Rank 3 block & Score \\
\midrule
$h=1$ & Financial idx. & 1.134 & Attention idx. & 1.132 & Price history & 0.887 \\
$h=2$ & Financial idx. & 1.158 & Attention idx. & 1.053 & Price history & 1.043 \\
$h=3$ & Attention idx. & 1.225 & Power-load idx. & 1.128 & Financial idx. & 1.091 \\
$h=4$ & Power-load idx. & 1.238 & Attention idx. & 1.177 & Financial idx. & 1.138 \\
$h=5$ & Financial idx. & 1.549 & Attention idx. & 1.224 & Power-load idx. & 1.099 \\
\bottomrule
\end{tabular*}
\end{table}

\clearpage
\begin{landscapetablepage}
\captionof{table}[SHAP Attribution across Horizons and States]{\textbf{SHAP Attribution across Horizons and States.} Grouped SHAP scores are computed separately for each forecast horizon, compliance state, and volatility state. Price history and the constructed residual state are treated as separate groups. Scores are comparable within each row, with larger values indicating stronger local attribution.}
\label{tab:main_shap_partition_scores}
\begin{tabular*}{\linewidth}{@{\extracolsep{\fill}}YYZZZZZZZ}
\toprule
Dimension family & Dimension & Fuel & Power-load & Fin. & Comp. & Attn. & Price history & State \\
\midrule
Horizon & $h=1$ & 0.108 & 0.819 & 1.134 & 0.183 & 1.132 & 0.887 & 0.712 \\
Horizon & $h=2$ & 0.085 & 0.821 & 1.158 & 0.208 & 1.053 & 1.043 & 0.637 \\
Horizon & $h=3$ & 0.094 & 1.128 & 1.091 & 0.204 & 1.225 & 0.916 & 0.506 \\
Horizon & $h=4$ & 0.080 & 1.238 & 1.138 & 0.230 & 1.177 & 0.733 & 0.567 \\
Horizon & $h=5$ & 0.074 & 1.099 & 1.549 & 0.202 & 1.224 & 0.498 & 0.514 \\
Compliance & Far from compliance ($>90$ days) & 0.105 & 1.302 & 1.641 & 0.226 & 1.545 & 1.147 & 0.742 \\
Compliance & Transition (31--90 days) & 0.056 & 0.243 & 0.160 & 0.154 & 0.102 & 0.062 & 0.336 \\
Compliance & Near compliance ($\leq 30$ days) & 0.075 & 1.213 & 1.257 & 0.207 & 1.427 & 0.721 & 0.348 \\
Volatility & Low volatility & 0.085 & 1.700 & 2.047 & 0.252 & 1.982 & 1.401 & 0.671 \\
Volatility & Mid volatility & 0.078 & 1.216 & 1.431 & 0.153 & 1.370 & 0.989 & 0.598 \\
Volatility & High volatility & 0.103 & 0.138 & 0.157 & 0.212 & 0.130 & 0.062 & 0.495 \\
\bottomrule
\end{tabular*}
\end{landscapetablepage}

\clearpage
\begin{landscapetablepage}
\captionof{table}[Forecast Accuracy across Market States]{\textbf{Forecast Accuracy across Market States.} Compliance and volatility states are formed within the 169 final-holdout origins using days to the compliance deadline and thirds of rolling volatility. RMSE differences are measured relative to the far-from-compliance or low-volatility group, and DA denotes directional accuracy. Parentheses report two-sided $p$-values from a circular moving-block bootstrap with block length five.}
\label{tab:main_regime_accuracy}
\begin{tabular*}{\linewidth}{@{\extracolsep{\fill}}YYZZZZZZZ}
\toprule
Regime family & Regime & Samples & RMSE$_1$ & DA$_1$ & $\Delta$RMSE$_1$ ($p$) & RMSE$_5$ & DA$_5$ & $\Delta$RMSE$_5$ ($p$) \\
\midrule
Compliance & Far from compliance ($>90$ days) & 104 & 1.061 & 0.510 & --- & 1.842 & 0.635 & --- \\
Compliance & Transition (31--90 days) & 43 & 0.825 & 0.395 & $-0.236$ (0.050) & 1.471 & 0.581 & $-0.371$ (0.087) \\
Compliance & Near compliance ($\leq 30$ days) & 22 & 0.813 & 0.591 & $-0.248$ (0.007) & 1.391 & 0.727 & $-0.450$ (0.076) \\
Volatility & Low volatility & 54 & 0.748 & 0.519 & --- & 1.579 & 0.704 & --- \\
Volatility & Mid volatility & 57 & 0.959 & 0.474 & $+0.211$ (0.027) & 1.469 & 0.632 & $-0.110$ (0.651) \\
Volatility & High volatility & 58 & 1.162 & 0.483 & $+0.414$ (0.001) & 1.995 & 0.569 & $+0.416$ (0.221) \\
\bottomrule
\end{tabular*}
\end{landscapetablepage}
 
\clearpage
\appendix
\begin{center}
	{\bf \Large Internet Appendix}
\end{center}

\doublespacing
\pagenumbering{arabic}
\renewcommand*{\thepage}{Appendix-\arabic{page}}
\renewcommand*{\theHpage}{appendix.\arabic{page}}

\counterwithin*{table}{section}
\counterwithin*{figure}{section}
\setcounter{table}{0}
\setcounter{figure}{0}
\renewcommand{\thetable}{\Alph{section}.\arabic{table}}
\renewcommand{\thefigure}{\Alph{section}.\arabic{figure}}


This Internet Appendix reports the construction details and robustness evidence that support the main text. It documents the data-timing rules, model settings, benchmark grids, supplementary forecast tests, information and state diagnostics, procurement sensitivity exercises, and directional trading diagnostics. The appendix is organized to make the empirical design reproducible and to show where the main conclusions are strong, where they are horizon-specific, and where they are sensitive to auxiliary choices.

\section{Data Timing and Variable Construction}
\label{app:data-timing}

This appendix documents the variables, transformations, and aligned sample used in forecast construction. Availability at each forecast origin is determined by the release timing and transformation sequence applied to every series. The central design choice is to align each predictor to what was observable by the EUA market close, rather than to what is known ex post on the same calendar date.

\subsection{Raw Variable List}

Table~\ref{tab:app_raw_variables} reports the source frequency, forecast role, and construction route of each variable. Direct inputs comprise carbon-price history, seven additional raw predictors, five information indices, and three rolling VMD state components. Auxiliary variables enter through index construction or conditional diagnostics rather than as standalone columns in the 30-day sequence. This separation keeps the forecasting tensor compact while retaining the economic channels emphasized in the main text: fuel substitution, power-market load, financial pricing, compliance timing, and market attention.
\inserttable{\ref{tab:app_raw_variables}}

\clearpage
\begin{landscapetablepage}
\captionof{table}[Raw Variables and Their Uses]{\textbf{Raw Variables and Their Uses.} Series are aligned to the trading calendar. Direct inputs enter the 30-day sequence; auxiliary variables enter only through index construction or supplementary diagnostics.}
\label{tab:app_raw_variables}
\begin{tabularx}{\linewidth}{>{\raggedright\arraybackslash}p{0.20\linewidth}>{\raggedright\arraybackslash}p{0.26\linewidth}>{\raggedright\arraybackslash}p{0.15\linewidth}>{\raggedright\arraybackslash}X}
\toprule
Variable or block & Source and frequency & Forecast role & Construction note \\
\midrule
Carbon allowance price & EUA settlement / close price; trading day & Target and direct input & Lagged price history enters the 30-day input window and anchors the cumulative-delta output path. \\
Coal price & Coal benchmark series; daily market data & Direct input & Used as a raw predictor and in the fuel-substitution block. \\
Natural gas price & Gas benchmark series; daily market data & Direct input & Used as a raw predictor and in the fuel-substitution block. \\
EU electricity price & Germany-France electricity average; daily / day-ahead data & Direct input & Used as a raw predictor and in the power-load block. \\
EU utilities equity index & EU utilities equity index; daily market data & Direct input & Used as a raw predictor and in the financial-pricing block. \\
EUR/USD exchange rate & EUR/USD series; daily market data & Direct input & Used as a raw predictor and in the financial-pricing block. \\
Days to compliance deadline & Trading-day distance to the compliance deadline; calendar series & Direct input & Used as a raw timing variable and in the compliance-calendar block. \\
Google Trends attention & Carbon-related Google Trends attention; monthly data & Direct input & Monthly values are backward-merged to the trading calendar; the rolling standardized version forms the attention index. \\
Oil price & Oil benchmark series; daily market data & Auxiliary only & Used in the fuel-substitution index, but not retained as a separate direct input. \\
Euro Stoxx 50 index & Euro Stoxx 50 series; daily market data & Auxiliary only & Used in the financial-pricing index, but not retained as a separate direct input. \\
German bond yield & German bond-yield series; daily market data & Auxiliary only & Used in the financial-pricing index, but not retained as a separate direct input. \\
EU industrial production & EU industrial-production release; monthly data & Auxiliary only & Carried forward after release and used in the power-load index. \\
Temperature aggregate & Trading-calendar temperature aggregate; daily data & Auxiliary only & Aligned to the trading calendar and used in the power-load index. \\
\bottomrule
\end{tabularx}
\end{landscapetablepage}

\subsection{Descriptive Statistics}

Table~\ref{tab:app_descriptive_statistics} reports distributional summaries for the target, direct predictors, information indices, and rolling state variables. All statistics use the common 1,721-day trading-calendar sample, with constructed variables expressed in their transformed-signal units. The sample is economically heterogeneous: the EUA price averages 60.399 with a range from 16.120 to 98.010, electricity prices have a high right tail, and the compliance-calendar and attention indices vary substantially over the trading calendar. These features motivate the paper's emphasis on short-horizon forecasting under changing market states rather than on a single stationary linear relation.
\inserttable{\ref{tab:app_descriptive_statistics}}

\clearpage
\begin{table}[p]
\centering
\standardtablelayout
\caption[Descriptive Statistics]{\textbf{Descriptive Statistics.} The sample contains 1,721 trading days from April 25, 2019 to December 31, 2025, with all series aligned to the EUA trading calendar. Information indices and VMD components are transformed variables and are reported in constructed-signal units.}
\label{tab:app_descriptive_statistics}
\begin{tabular*}{0.92\textwidth}{@{\extracolsep{\fill}}YZZZZZZ}
\toprule
Variable & Obs. & Mean & Std. dev. & Min & Median & Max \\
\midrule
EUA futures close & 1,721 & 60.399 & 23.275 & 16.120 & 67.960 & 98.010 \\
Coal price & 1,721 & 152.482 & 102.387 & 48.500 & 126.800 & 457.800 \\
Natural gas price & 1,721 & 3.345 & 1.820 & 1.210 & 2.720 & 23.860 \\
EU electricity price & 1,721 & 119.469 & 112.139 & 35.525 & 79.700 & 1052.500 \\
EU utilities index & 1,721 & 16.240 & 5.864 & 7.771 & 14.440 & 35.235 \\
EUR/USD exchange rate & 1,721 & 1.112 & 0.055 & 0.960 & 1.105 & 1.234 \\
Days to deadline & 1,721 & 203.983 & 121.450 & 0.000 & 203.000 & 518.000 \\
Google Trends attention & 1,721 & 12.467 & 18.913 & 4.000 & 8.000 & 95.000 \\
Fuel substitution index & 1,721 & 0.001 & 1.408 & -4.888 & -0.230 & 5.600 \\
Power-load index & 1,721 & 0.029 & 0.895 & -2.797 & 0.007 & 3.239 \\
Financial pricing index & 1,721 & 0.180 & 0.662 & -2.247 & 0.210 & 2.120 \\
Compliance calendar index & 1,721 & 0.376 & 1.679 & -7.980 & 1.202 & 2.299 \\
Market attention index & 1,721 & 0.432 & 1.098 & -2.062 & 0.553 & 2.638 \\
VMD IMF 1 & 1,721 & 1.513 & 5.480 & -12.148 & 0.615 & 39.501 \\
VMD IMF 2 & 1,721 & -0.298 & 3.514 & -31.667 & 0.001 & 37.279 \\
VMD IMF 3 & 1,721 & 0.027 & 1.446 & -15.868 & -0.005 & 24.291 \\
\bottomrule
\end{tabular*}
\end{table}
  
\clearpage
\section{Forecast Construction and Hyperparameters}
\label{app:forecast-construction}

This appendix documents the maintained architecture, training procedure, rolling-state construction, and evaluation settings used in the main chronological split. The details are reported separately from the main text because they are implementation constraints rather than additional modeling choices made after seeing the holdout.

\subsection{Hyperparameter Summary}

Table~\ref{tab:app_model_architecture} reports the input-output structure and model dimensions, Table~\ref{tab:app_training_settings} reports optimization and regularization settings, and Table~\ref{tab:app_state_evaluation_settings} reports rolling-state and final-holdout settings. These settings are determined from the training and validation samples and remain fixed during final-holdout evaluation. The architecture is intentionally modest: a two-layer Transformer with 128 hidden units is used for the main path, and a single-layer GRU with 64 hidden units is used only for residual correction. The forecasting task therefore relies on a disciplined information design and validation-fixed release rule, not on an unusually large neural network.
\inserttable{\ref{tab:app_model_architecture}}
\inserttable{\ref{tab:app_training_settings}}
\inserttable{\ref{tab:app_state_evaluation_settings}}

\clearpage
\begin{table}[p]
\centering
\standardtablelayout
\caption[Forecasting Model Architectures]{\textbf{Forecasting Model Architectures.} The table reports the architecture and input-output dimensions of the Transformer and GRU paths. Both use a 30-day lookback and produce five direct forecast outputs.}
\label{tab:app_model_architecture}
\begin{tabularx}{0.88\textwidth}{p{0.18\textwidth}p{0.18\textwidth}X>{\raggedleft\arraybackslash}p{0.16\textwidth}}
\toprule
Model & Component & Parameter & Value \\
\midrule
Transformer & Architecture & Hidden dimension & 128 \\
Transformer & Architecture & Attention heads & 8 \\
Transformer & Architecture & Encoder layers & 2 \\
Transformer & Architecture & Dropout rate & 0.1 \\
Transformer & Input & Lookback window & 30 days \\
Transformer & Output & Forecast horizon & 5 days \\
GRU & Architecture & Hidden dimension & 64 \\
GRU & Architecture & GRU layers & 1 \\
GRU & Architecture & Dropout rate & 0.1 \\
GRU & Input & Lookback window & 30 days \\
GRU & Output & Residual forecast horizon & 5 days \\
\bottomrule
\end{tabularx}
\end{table}

\clearpage
\begin{table}[p]
\centering
\standardtablelayout
\caption[Model Training and Optimization Settings]{\textbf{Model Training and Optimization Settings.} The table reports the maintained optimization, regularization, and stopping settings for the Transformer and GRU paths. Training settings are fixed before final-holdout evaluation.}
\label{tab:app_training_settings}
\begin{tabularx}{0.82\textwidth}{p{0.20\textwidth}X>{\raggedleft\arraybackslash}p{0.20\textwidth}}
\toprule
Model & Parameter & Value \\
\midrule
Transformer & Optimizer & AdamW \\
Transformer & Learning rate & $1 \times 10^{-3}$ \\
Transformer & Weight decay & $1 \times 10^{-4}$ \\
Transformer & Batch size & 64 \\
Transformer & Maximum epochs & 120 \\
Transformer & Early stopping patience & 20 epochs \\
Transformer & Direction-loss weight & 0.25 \\
Transformer & Direction temperature & 0.06 \\
Transformer & Gradient clipping norm & 1.0 \\
Transformer & Random seed & 42 \\
GRU & Optimizer & AdamW \\
GRU & Learning rate & $1 \times 10^{-3}$ \\
GRU & Weight decay & $1 \times 10^{-4}$ \\
GRU & Batch size & 64 \\
GRU & Maximum epochs & 120 \\
GRU & Early stopping patience & 20 epochs \\
\bottomrule
\end{tabularx}
\end{table}

\clearpage
\begin{table}[p]
\centering
\standardtablelayout
\caption[Rolling-State and Evaluation Settings]{\textbf{Rolling-State and Evaluation Settings.} The table reports VMD, rolling-feature, residual-state, and final-holdout settings. Every rolling construction uses information available at the forecast origin.}
\label{tab:app_state_evaluation_settings}
\begin{tabularx}{0.88\textwidth}{p{0.24\textwidth}X>{\raggedleft\arraybackslash}p{0.22\textwidth}}
\toprule
Component & Parameter & Value \\
\midrule
VMD & Number of modes ($K$) & 3 \\
VMD & Quadratic penalty ($\alpha$) & 2,000 \\
VMD & Noise tolerance ($\tau$) & 0.0 \\
VMD & DC component & 0 (off) \\
VMD & Initialization & 1 (uniform) \\
VMD & Convergence tolerance & $1 \times 10^{-7}$ \\
Information indices & Z-score window / minimum & 90 / 40 obs. \\
Attention index & Z-score window / minimum & 8 / 4 monthly obs. \\
Residual state & Fit window / ridge penalty & 180 / $1 \times 10^{-3}$ \\
Rolling decomposition & Window / minimum & 120 / 80 obs. \\
Final holdout & Train / validation / test split & 80\% / 10\% / 10\% \\
Final holdout & Test origins & 169 \\
\bottomrule
\end{tabularx}
\end{table}
  
\clearpage
\section{Benchmark Specifications}
\label{app:benchmark-specifications}

This appendix documents the construction, validation tuning, and diagnostic measures of the fourteen competing models. Every benchmark uses the same chronological split and day-$t$ information boundary as the released forecast, with model-specific input representations. This common evaluation frame is essential because benchmark underperformance would be uninformative if competing models were allowed weaker input timing, different forecast origins, or ex post tuning.

\subsection{Benchmark Design and Tuning}

Table~\ref{tab:app_benchmark_specifications} reports the input representation, validation grid, and retained hyperparameters for each benchmark. The base representation combines day-$t$ features with five carbon-price lags; the state representation adds compliance and volatility indicators; and the state-interaction representation adds selected interactions between the base variables and state indicators. Non-naive benchmarks are tuned by average validation RMSE across the five forecast horizons before final-holdout evaluation. The retained grid choices span linear, penalized, smooth nonlinear, tree-based, kernel, nearest-neighbor, and sequence-model alternatives, so the comparison is not driven by a single benchmark family.
\inserttable{\ref{tab:app_benchmark_specifications}}

\subsection{Benchmark Importance Diagnostics}

Table~\ref{tab:app_benchmark_importance} reports the strongest price and non-price signals identified by the interpretable non-naive benchmark families. Scores are interpreted within a model class because coefficient magnitudes, permutation importance, feature importance, and marginal effects use different scales. The diagnostics show that price history remains a dominant signal across conventional models, but non-price variables also enter materially: Google Trends attention is selected by penalized regressions, deadline days enters the spline model, fuel and utility information matter for kernel and nearest-neighbor specifications, and power prices are prominent in tree-based models. This pattern supports the maintained view that EUA predictability is multi-source and horizon-dependent.
\inserttable{\ref{tab:app_benchmark_importance}}

\clearpage
\begin{landscapetablepage}
\captionof{table}[Benchmark Specifications and Tuning]{\textbf{Benchmark Specifications and Tuning.} Representation indicates whether a benchmark uses price history only, the state block, the state-plus-interaction design, or the full rolling sequence tensor. Hyperparameters are selected on the validation window and then frozen for the final holdout. Estimator labels are omitted because they are already implied by the benchmark names.}
\label{tab:app_benchmark_specifications}
\begin{tabularx}{\linewidth}{p{0.21\linewidth}p{0.11\linewidth}p{0.35\linewidth}X}
\toprule
Benchmark & Repr. & Validation grid & Selected hyperparameters \\
\midrule
Random Walk & Price hist. & None & Deterministic \\
Drift & Price hist. & None & Deterministic \\
Historical Mean & Price hist. & None & Deterministic \\
ARX Dynamic Reg. & State + inter. & $\alpha \in \{0.1, 1.0, 5.0\}$ & $\alpha = 0.1$ \\
LASSO Dynamic & State + inter. & $\alpha \in \{0.0005, 0.001, 0.002, 0.005\}$ & $\alpha = 0.001$ \\
Elastic Net Dynamic & State + inter. & $\alpha \in \{0.001, 0.002, 0.005\}$; $l_1 \in \{0.3, 0.6, 0.8\}$ & $\alpha = 0.001$, $l_1 = 0.8$ \\
Spline-Ridge GAM & State & Knots $\in \{4, 5\}$; ridge $\alpha \in \{0.5, 1.0, 2.0\}$ & Knots $= 4$, $\alpha = 2.0$ \\
Random Forest & State & $(n, depth, leaf) \in \{(240, 6, 4), (360, 8, 4)\}$ & $(360, 8, 4)$ \\
Gradient Boosting & State & $(n,\eta,\mathrm{depth})$: (200, 0.05, 2); (320, 0.05, 2); (220, 0.03, 3) & $(200, 0.05, 2)$ \\
SVR (RBF) & State & $C \in \{5, 10, 25\}$; $\epsilon \in \{0.01, 0.02\}$ & $C = 5$, $\epsilon = 0.02$ \\
KNN & State & Neighbors: 8, 12, 18; weights: uniform, distance & Neighbors $= 18$, weights $= \mathrm{distance}$ \\
LSTM Sequence & Seq. tensor & Hidden $\in \{64, 96\}$; layers $\in \{1, 2\}$ & Hidden $= 96$, layers $= 2$, dropout $= 0.1$, lr $= 0.0008$ \\
Temp. Fusion Transf. & Seq. tensor & Hidden $\in \{64, 96\}$; layers $\in \{1, 2\}$ & Hidden $= 64$, heads $= 4$, layers $= 1$, dropout $= 0.1$, lr $= 0.0008$ \\
Informer Sequence & Seq. tensor & $d$-model $\in \{64, 96\}$; layers $= 2$ & $d$-model $= 96$, heads $= 4$, layers $= 2$, dropout $= 0.1$, lr $= 0.0006$ \\
\bottomrule
\end{tabularx}
\end{landscapetablepage}
 
\clearpage
\begin{landscapetablepage}
\captionof{table}[Benchmark-Specific Importance Measures]{\textbf{Benchmark-Specific Importance Measures.} Scores are comparable within rows, not across rows. The last column reports the strongest non-price signal retained by each interpretable non-naive benchmark. Deterministic benchmarks and sequence models are omitted from this diagnostic table.}
\label{tab:app_benchmark_importance}
\begin{tabularx}{\linewidth}{lllclcXc}
\toprule
Model & Metric & Top price signal 1 & Score & Top price signal 2 & Score & Top non-price signal & Score \\
\midrule
ARX Dynamic Reg. & Mean abs. coefficient & Carbon $(t)$ & 0.216 & Carbon $(t\!-\!1)$ & 0.216 & Carbon $(t\!-\!1)\times$ transition & 0.060 \\
LASSO Dynamic & Mean abs. coefficient & Carbon $(t)$ & 0.955 & Carbon $(t\!-\!2)$ & 0.008 & GT attention & 0.005 \\
Elastic Net Dynamic & Mean abs. coefficient & Carbon $(t)$ & 0.418 & Carbon $(t\!-\!1)$ & 0.411 & GT attention & 0.008 \\
Spline-Ridge GAM & Permutation importance & Carbon $(t)$ & 0.219 & Carbon $(t\!-\!1)$ & 0.219 & Deadline days & 0.016 \\
Random Forest & Feature importance & Carbon $(t\!-\!5)$ & 0.231 & Carbon $(t)$ & 0.195 & EU power price & 0.096 \\
Gradient Boosting & Feature importance & Carbon $(t\!-\!1)$ & 0.286 & Carbon $(t)$ & 0.237 & EU power price & 0.130 \\
SVR (RBF) & Permutation importance & Carbon $(t\!-\!1)$ & 0.177 & Carbon $(t)$ & 0.177 & Fuel idx. & 0.076 \\
KNN & Marginal effect & Carbon $(t\!-\!1)$ & 0.007 & Carbon $(t)$ & 0.007 & EU utilities index & 0.038 \\
\bottomrule
\end{tabularx}
\end{landscapetablepage}
  
\clearpage
\section{Supplementary Forecast Tests}
\label{app:forecast-tests}

This appendix reports supplementary inference and metric panels for the final holdout. The evidence covers candidate-versus-released comparisons, pairwise benchmark tests, block-bootstrap sensitivity, and the complete cross-horizon benchmark results. The additional tables are designed to separate three questions that are compressed in the main text: whether the validation-fixed release improves the raw candidate path, whether the released forecast beats individual alternatives rather than only the best benchmark summary, and whether the statistical evidence is concentrated at particular horizons.

\subsection{Candidate versus Released Forecast Tests}

Table~\ref{tab:app_holdout_reporting_comparison} reports RMSE, MAE, percentage errors, bias, directional accuracy, $R^2$, and Diebold--Mariano inference for the candidate and released paths on the same 169 forecast origins. Negative Diebold--Mariano statistics indicate lower squared-error loss for the released forecast. The statistics favor the released path at all five horizons, with rejection at the 5\% level at $h=1$ and at the 10\% level at $h=3$.

The released layer reduces RMSE from 1.064 to 0.976 at $h=1$ and from 1.750 to 1.700 at $h=5$, with the largest absolute RMSE reduction at $h=3$ (1.586 to 1.424). MAE and percentage-error measures move in the same direction. Bias is not uniformly improved, especially at $h=2$ and $h=4$, and directional accuracy is unchanged from $h=3$ onward. The released rule therefore improves the price-level forecast path without claiming a general solution to signed daily timing.
\inserttable{\ref{tab:app_holdout_reporting_comparison}}

\subsection{Released Forecast versus Benchmark Pairwise Tests}

Table~\ref{tab:app_pairwise_dm_tests} reports two-sided Diebold--Mariano tests between the released forecast and all fourteen benchmarks at horizons one through five. Table~\ref{tab:app_pairwise_cw_tests} reports one-sided Clark--West tests for the maintained nested-model pairs in the candidate and released layers. The reported signs follow the conventions stated in each table.

The pairwise tests show that the evidence is broadest at the middle horizons. Against the random-walk and drift benchmarks, the released forecast is favored at every horizon, with $p$-values below 5\% for $h=2$ through $h=5$ and a weaker 10\% result at $h=1$. Comparisons with machine-learning benchmarks are also mostly favorable, but not uniformly decisive at one day. This pattern is consistent with the main result that the economic value comes from ranking the multi-day path rather than from dominating every short-run directional alternative.
\inserttable{\ref{tab:app_pairwise_dm_tests}}
\inserttable{\ref{tab:app_pairwise_cw_tests}}

\subsection{Block-Bootstrap Sensitivity}

Table~\ref{tab:app_block_bootstrap_sensitivity} reports moving-block-bootstrap inference against the horizon-specific best benchmark under block sizes 5, 10, and 20. The table presents 95\% intervals and two-sided $p$-values for the squared-loss gain. The intervals remain above zero at $h=3$ and $h=4$ under all three block sizes.

The bootstrap exercise gives a sharper view of sampling uncertainty. At $h=3$ and $h=4$, every confidence interval is positive, including the widest block-size-20 intervals. At $h=1$, $h=2$, and $h=5$, the point estimates remain positive but at least one interval includes zero. The appendix therefore supports a bounded inference statement: forecast gains are strongest and most stable in the middle of the one- to five-day path.
\inserttable{\ref{tab:app_block_bootstrap_sensitivity}}

\subsection{Complete Benchmark Comparison}

Tables~\ref{tab:app_holdout_benchmark_summary}--\ref{tab:app_holdout_benchmark_da} report the complete benchmark comparison on the common final holdout. The tables present RMSE and random-walk-relative $R^2_{OOS}$, MAE and bias, percentage-error measures, and directional accuracy, respectively. This organization separates price-level loss, signed error, scale-adjusted loss, and directional performance.

The full panel confirms that the released forecast has the lowest RMSE at every horizon, but it also reveals why the paper avoids overstating the result. Some benchmarks are close at one day, and directional accuracy does not rank models the same way as price-level loss. The strongest evidence is instead the cross-horizon price-path pattern: random-walk-relative $R^2_{OOS}$ rises from 1.2\% at $h=1$ to 15.5\% at $h=5$, while several conventional and flexible benchmarks deteriorate sharply at longer horizons.
\inserttable{\ref{tab:app_holdout_benchmark_summary}}
\inserttable{\ref{tab:app_holdout_benchmark_mae_bias}}
\inserttable{\ref{tab:app_holdout_benchmark_percentage_errors}}
\inserttable{\ref{tab:app_holdout_benchmark_da}}

\clearpage
\begin{table}[p]
\centering
\standardtablelayout
\caption[Released and Candidate Forecast Metrics]{\textbf{Released and Candidate Forecast Metrics.} The validation-fixed released forecast and uncalibrated candidate path are evaluated on the same 169 final-holdout origins. Bias is the mean signed forecast error. Diebold--Mariano tests use squared-error loss and Newey--West lag $h-1$; negative statistics favor the released forecast.}
\label{tab:app_holdout_reporting_comparison}
\begin{tabular*}{0.92\textwidth}{@{\extracolsep{\fill}}YYZZZZZ}
\toprule
Metric & Series & $h=1$ & $h=2$ & $h=3$ & $h=4$ & $h=5$ \\
\midrule
RMSE & Released & 0.976 & 1.316 & 1.424 & 1.615 & 1.700 \\
RMSE & Candidate & 1.064 & 1.344 & 1.586 & 1.640 & 1.750 \\
MAE & Released & 0.781 & 1.065 & 1.179 & 1.313 & 1.360 \\
MAE & Candidate & 0.873 & 1.079 & 1.301 & 1.338 & 1.404 \\
MAPE & Released & 1.044 & 1.422 & 1.571 & 1.748 & 1.809 \\
MAPE & Candidate & 1.170 & 1.439 & 1.732 & 1.779 & 1.864 \\
sMAPE & Released & 1.047 & 1.425 & 1.574 & 1.756 & 1.817 \\
sMAPE & Candidate & 1.165 & 1.439 & 1.721 & 1.784 & 1.866 \\
Bias & Released & -0.107 & -0.153 & -0.173 & -0.379 & -0.275 \\
Bias & Candidate & 0.353 & 0.077 & 0.541 & -0.192 & 0.049 \\
DA & Released & 0.491 & 0.580 & 0.627 & 0.604 & 0.633 \\
DA & Candidate & 0.479 & 0.544 & 0.627 & 0.604 & 0.633 \\
$R^2$ & Released & 0.958 & 0.924 & 0.912 & 0.887 & 0.877 \\
$R^2$ & Candidate & 0.950 & 0.921 & 0.891 & 0.884 & 0.869 \\
DM stat. & Candidate vs. released & -2.312 & -0.593 & -1.767 & -0.425 & -0.793 \\
$p$-value & Candidate vs. released & 0.0220 & 0.5537 & 0.0791 & 0.6714 & 0.4289 \\
\bottomrule
\end{tabular*}
\end{table}
 
\clearpage
\begin{landscapetablepage}
\captionof{table}[Pairwise Diebold--Mariano Tests]{\textbf{Pairwise Diebold--Mariano Tests.} Each cell reports the Diebold--Mariano statistic followed by its two-sided $p$-value in parentheses. Tests use squared-error loss and Newey--West lag $h-1$. Negative statistics favor the released forecast.}
\label{tab:app_pairwise_dm_tests}
\begin{adjustbox}{width=\linewidth}
\begin{tabular}{lrrrrr}
\toprule
Benchmark & $h=1$ & $h=2$ & $h=3$ & $h=4$ & $h=5$ \\
\midrule
Random Walk & -1.821 (0.070) & -2.381 (0.018) & -2.582 (0.011) & -2.577 (0.011) & -2.325 (0.021) \\
Drift & -1.821 (0.070) & -2.381 (0.018) & -2.582 (0.011) & -2.577 (0.011) & -2.325 (0.021) \\
Historical Mean & -2.991 (0.003) & -3.487 (0.001) & -3.948 ($<0.001$) & -4.137 ($<0.001$) & -4.102 ($<0.001$) \\
ARX Dynamic Reg. & -3.510 (0.001) & -0.659 (0.511) & -8.192 ($<0.001$) & -7.567 ($<0.001$) & -6.855 ($<0.001$) \\
LASSO Dynamic & -1.796 (0.074) & -2.359 (0.019) & -2.207 (0.029) & -3.237 (0.001) & -6.073 ($<0.001$) \\
Elastic Net Dynamic & -1.813 (0.072) & -2.113 (0.036) & -2.013 (0.046) & -5.867 ($<0.001$) & -6.380 ($<0.001$) \\
Spline-Ridge GAM & -0.668 (0.505) & -1.221 (0.224) & -1.838 (0.068) & -2.240 (0.026) & -2.438 (0.016) \\
Random Forest & -2.031 (0.044) & -3.750 ($<0.001$) & -2.720 (0.007) & -2.363 (0.019) & -2.265 (0.025) \\
Gradient Boosting & -1.625 (0.106) & -2.401 (0.017) & -2.161 (0.032) & -3.121 (0.002) & -3.474 (0.001) \\
SVR (RBF) & -5.769 ($<0.001$) & -2.926 (0.004) & -3.613 ($<0.001$) & -3.917 ($<0.001$) & -3.831 ($<0.001$) \\
KNN & -1.266 (0.207) & -2.491 (0.014) & -2.902 (0.004) & -3.098 (0.002) & -2.648 (0.009) \\
LSTM Sequence & -2.565 (0.011) & -1.818 (0.071) & -1.421 (0.157) & -1.763 (0.080) & -3.023 (0.003) \\
Temp. Fusion Transf. & -1.938 (0.054) & -2.743 (0.007) & -2.574 (0.011) & -2.670 (0.008) & -2.421 (0.017) \\
Informer Sequence & -1.733 (0.085) & -2.409 (0.017) & -2.570 (0.011) & -2.529 (0.012) & -1.065 (0.289) \\
\bottomrule
\end{tabular}
\end{adjustbox}
\end{landscapetablepage}

\clearpage
\begin{landscapetablepage}
\captionof{table}[Pairwise Clark--West Tests]{\textbf{Pairwise Clark--West Tests.} Each cell reports the Clark--West statistic followed by its one-sided $p$-value in parentheses. Positive statistics favor the second model in each comparison.}
\label{tab:app_pairwise_cw_tests}
\begin{adjustbox}{width=\linewidth}
\begin{tabular}{llrrrrr}
\toprule
Forecast layer & Comparison & $h=1$ & $h=2$ & $h=3$ & $h=4$ & $h=5$ \\
\midrule
Candidate & Random Walk vs. ARX Dynamic Reg. & 1.772 (0.039) & 2.579 (0.005) & 3.331 (0.001) & 4.106 ($<0.001$) & 4.833 ($<0.001$) \\
Candidate & ARX Dynamic Reg. vs. LASSO Dynamic & 15.001 ($<0.001$) & 15.475 ($<0.001$) & 15.613 ($<0.001$) & 15.659 ($<0.001$) & 15.690 ($<0.001$) \\
Candidate & ARX Dynamic Reg. vs. Elastic Net Dynamic & 15.011 ($<0.001$) & 15.481 ($<0.001$) & 15.614 ($<0.001$) & 15.661 ($<0.001$) & 15.686 ($<0.001$) \\
Candidate & Random Forest vs. Gradient Boosting & 2.542 (0.006) & -8.662 (1.000) & -6.912 (1.000) & -10.252 (1.000) & -10.937 (1.000) \\
Released & Random Walk vs. ARX Dynamic Reg. & -2.093 (0.981) & 3.059 (0.001) & 3.860 ($<0.001$) & 4.739 ($<0.001$) & 5.483 ($<0.001$) \\
Released & ARX Dynamic Reg. vs. LASSO Dynamic & 5.190 ($<0.001$) & 2.414 (0.008) & 15.006 ($<0.001$) & 15.500 ($<0.001$) & 15.600 ($<0.001$) \\
Released & ARX Dynamic Reg. vs. Elastic Net Dynamic & 5.183 ($<0.001$) & 2.464 (0.007) & 15.009 ($<0.001$) & 15.505 ($<0.001$) & 15.574 ($<0.001$) \\
Released & Random Forest vs. Gradient Boosting & 0.340 (0.367) & -0.377 (0.647) & 2.844 (0.003) & -3.259 (0.999) & -4.454 (1.000) \\
\bottomrule
\end{tabular}
\end{adjustbox}
\end{landscapetablepage}
 
\clearpage
\begin{landscapetablepage}
\captionof{table}[Moving-Block-Bootstrap Sensitivity]{\textbf{Moving-Block-Bootstrap Sensitivity.} Each horizon compares the released forecast with its lowest-RMSE competing model. Entries report 95\% confidence intervals for the squared-loss gain, defined as benchmark loss minus released-forecast loss, and two-sided bootstrap $p$-values from 2,000 draws.}
\label{tab:app_block_bootstrap_sensitivity}
\begin{adjustbox}{width=\linewidth}
\begin{tabular}{llcccccc}
\toprule
& & \multicolumn{2}{c}{Block size 5} & \multicolumn{2}{c}{Block size 10} & \multicolumn{2}{c}{Block size 20} \\
\cmidrule(lr){3-4}\cmidrule(lr){5-6}\cmidrule(lr){7-8}
Horizon & Best benchmark & 95\% interval & $p$-value & 95\% interval & $p$-value & 95\% interval & $p$-value \\
\midrule
$h=1$ & KNN & $[-0.002,\ 0.016]$ & 0.124 & $[-0.001,\ 0.015]$ & 0.095 & $[-0.001,\ 0.014]$ & 0.071 \\
$h=2$ & Spline-Ridge GAM & $[-0.017,\ 0.119]$ & 0.198 & $[-0.014,\ 0.117]$ & 0.162 & $[-0.001,\ 0.101]$ & 0.072 \\
$h=3$ & Gradient Boosting & $[0.011,\ 0.237]$ & 0.026 & $[0.024,\ 0.241]$ & 0.027 & $[0.037,\ 0.235]$ & 0.016 \\
$h=4$ & Random Forest & $[0.042,\ 0.375]$ & 0.018 & $[0.052,\ 0.377]$ & 0.012 & $[0.051,\ 0.359]$ & 0.009 \\
$h=5$ & Informer Sequence & $[-0.141,\ 0.482]$ & 0.281 & $[-0.135,\ 0.478]$ & 0.272 & $[-0.178,\ 0.499]$ & 0.325 \\
\bottomrule
\end{tabular}
\end{adjustbox}
\end{landscapetablepage}

\clearpage
\begin{landscapetablepage}
\captionof{table}[Benchmark Comparison in the Final Holdout]{\textbf{Benchmark Comparison in the Final Holdout.} All models are evaluated on the same 169 final-holdout origins. $R^2_{OOS,h}$ is relative to the Random Walk, and negative values indicate underperformance.}
\label{tab:app_holdout_benchmark_summary}
\begin{tabular*}{\linewidth}{@{\extracolsep{\fill}}YZZZZZZZZZZ}
\toprule
Model & RMSE$_1$ & RMSE$_2$ & RMSE$_3$ & RMSE$_4$ & RMSE$_5$ & $R^2_{1}$ & $R^2_{2}$ & $R^2_{3}$ & $R^2_{4}$ & $R^2_{5}$ \\
\midrule
Released forecast & 0.976 & 1.316 & 1.424 & 1.615 & 1.700 & 0.012 & 0.046 & 0.097 & 0.093 & 0.155 \\
Random Walk & 0.982 & 1.347 & 1.498 & 1.696 & 1.849 & 0.000 & 0.000 & 0.000 & 0.000 & 0.000 \\
Drift & 0.982 & 1.347 & 1.498 & 1.696 & 1.849 & 0.000 & 0.000 & 0.000 & 0.000 & 0.000 \\
Historical Mean & 1.036 & 1.515 & 1.858 & 2.232 & 2.592 & -0.113 & -0.265 & -0.537 & -0.733 & -0.964 \\
ARX Dynamic Reg. & 1.034 & 1.336 & 5.844 & 11.573 & 16.355 & -0.108 & 0.017 & -14.219 & -45.570 & -77.217 \\
LASSO Dynamic & 0.982 & 1.348 & 1.468 & 2.130 & 5.437 & 0.000 & -0.002 & 0.039 & -0.578 & -7.644 \\
Elastic Net Dynamic & 0.982 & 1.342 & 1.568 & 3.662 & 7.373 & 0.000 & 0.007 & -0.095 & -3.664 & -14.896 \\
Spline-Ridge GAM & 0.980 & 1.333 & 1.476 & 1.683 & 1.855 & 0.003 & 0.020 & 0.029 & 0.015 & -0.006 \\
Random Forest & 0.982 & 1.341 & 1.485 & 1.677 & 1.835 & 0.000 & 0.010 & 0.017 & 0.022 & 0.015 \\
Gradient Boosting & 0.989 & 1.351 & 1.467 & 1.735 & 2.037 & -0.014 & -0.005 & 0.041 & -0.047 & -0.213 \\
SVR (RBF) & 1.232 & 1.442 & 1.744 & 2.096 & 2.305 & -0.575 & -0.146 & -0.356 & -0.527 & -0.554 \\
KNN & 0.979 & 1.352 & 1.536 & 1.751 & 1.890 & 0.005 & -0.007 & -0.051 & -0.066 & -0.044 \\
LSTM Sequence & 0.997 & 1.408 & 1.502 & 1.809 & 2.193 & -0.031 & -0.093 & -0.006 & -0.139 & -0.407 \\
Temp. Fusion Transf. & 0.983 & 1.372 & 1.500 & 1.705 & 1.866 & -0.003 & -0.038 & -0.002 & -0.011 & -0.018 \\
Informer Sequence & 0.981 & 1.348 & 1.497 & 1.692 & 1.750 & 0.001 & -0.002 & 0.001 & 0.004 & 0.104 \\
Released-forecast $R^2_{OOS}$ (\%) & --- & --- & --- & --- & --- & 1.2 & 4.6 & 9.7 & 9.3 & 15.5 \\
\bottomrule
\end{tabular*}
\end{landscapetablepage}

\clearpage
\begin{landscapetablepage}
\captionof{table}[Benchmark MAE and Bias in the Final Holdout]{\textbf{Benchmark MAE and Bias in the Final Holdout.} All models are evaluated on the same 169 final-holdout origins. Bias is the mean signed forecast error.}
\label{tab:app_holdout_benchmark_mae_bias}
\begin{adjustbox}{width=\linewidth}
\begin{tabular}{lrrrrrrrrrr}
\toprule
& \multicolumn{5}{c}{MAE} & \multicolumn{5}{c}{Bias} \\
\cmidrule(lr){2-6}\cmidrule(lr){7-11}
Model & $h=1$ & $h=2$ & $h=3$ & $h=4$ & $h=5$ & $h=1$ & $h=2$ & $h=3$ & $h=4$ & $h=5$ \\
\midrule
Released forecast & 0.781 & 1.065 & 1.179 & 1.313 & 1.360 & -0.107 & -0.153 & -0.173 & -0.379 & -0.275 \\
Random Walk & 0.780 & 1.088 & 1.241 & 1.381 & 1.476 & -0.152 & -0.301 & -0.422 & -0.559 & -0.671 \\
Drift & 0.780 & 1.088 & 1.241 & 1.381 & 1.476 & -0.152 & -0.301 & -0.422 & -0.559 & -0.671 \\
Historical Mean & 0.812 & 1.236 & 1.541 & 1.863 & 2.157 & -0.362 & -0.744 & -1.125 & -1.485 & -1.832 \\
ARX Dynamic Reg. & 0.810 & 1.070 & 4.937 & 9.487 & 13.267 & -0.318 & 0.086 & 4.273 & 8.697 & 12.349 \\
LASSO Dynamic & 0.780 & 1.089 & 1.216 & 1.775 & 4.646 & -0.154 & -0.308 & -0.326 & 0.939 & 3.850 \\
Elastic Net Dynamic & 0.780 & 1.084 & 1.301 & 3.158 & 6.202 & -0.152 & -0.292 & 0.365 & 2.415 & 5.419 \\
Spline-Ridge GAM & 0.777 & 1.074 & 1.221 & 1.370 & 1.486 & -0.146 & -0.285 & -0.414 & -0.601 & -0.795 \\
Random Forest & 0.789 & 1.081 & 1.229 & 1.367 & 1.464 & -0.090 & -0.195 & -0.346 & -0.471 & -0.614 \\
Gradient Boosting & 0.781 & 1.091 & 1.213 & 1.418 & 1.649 & -0.208 & -0.317 & -0.297 & -0.664 & -1.059 \\
SVR (RBF) & 0.995 & 1.191 & 1.447 & 1.748 & 1.892 & -0.647 & -0.504 & -0.770 & -1.082 & -1.212 \\
KNN & 0.779 & 1.093 & 1.276 & 1.433 & 1.521 & -0.138 & -0.317 & -0.513 & -0.682 & -0.764 \\
LSTM Sequence & 0.787 & 1.134 & 1.254 & 1.482 & 1.827 & -0.216 & 0.290 & 0.213 & 0.490 & 0.900 \\
Temp. Fusion Transf. & 0.781 & 1.110 & 1.242 & 1.388 & 1.493 & -0.159 & -0.380 & -0.427 & -0.584 & -0.716 \\
Informer Sequence & 0.780 & 1.089 & 1.241 & 1.378 & 1.391 & -0.148 & -0.305 & -0.421 & -0.550 & -0.315 \\
\bottomrule
\end{tabular}
\end{adjustbox}
\end{landscapetablepage}

\clearpage
\begin{landscapetablepage}
\captionof{table}[Benchmark Percentage Errors in the Final Holdout]{\textbf{Benchmark Percentage Errors in the Final Holdout.} MAPE and sMAPE are reported in percent for the same 169 final-holdout origins.}
\label{tab:app_holdout_benchmark_percentage_errors}
\begin{adjustbox}{width=\linewidth}
\begin{tabular}{lrrrrrrrrrr}
\toprule
& \multicolumn{5}{c}{MAPE (\%)} & \multicolumn{5}{c}{sMAPE (\%)} \\
\cmidrule(lr){2-6}\cmidrule(lr){7-11}
Model & $h=1$ & $h=2$ & $h=3$ & $h=4$ & $h=5$ & $h=1$ & $h=2$ & $h=3$ & $h=4$ & $h=5$ \\
\midrule
Released forecast & 1.044 & 1.422 & 1.571 & 1.748 & 1.809 & 1.047 & 1.425 & 1.574 & 1.756 & 1.817 \\
Random Walk & 1.044 & 1.450 & 1.651 & 1.837 & 1.959 & 1.046 & 1.457 & 1.659 & 1.849 & 1.976 \\
Drift & 1.044 & 1.450 & 1.651 & 1.837 & 1.959 & 1.046 & 1.457 & 1.659 & 1.849 & 1.976 \\
Historical Mean & 1.080 & 1.626 & 2.012 & 2.422 & 2.787 & 1.085 & 1.640 & 2.036 & 2.457 & 2.837 \\
ARX Dynamic Reg. & 1.081 & 1.431 & 6.396 & 12.197 & 16.989 & 1.086 & 1.431 & 6.138 & 11.212 & 15.093 \\
LASSO Dynamic & 1.044 & 1.452 & 1.619 & 2.355 & 6.024 & 1.046 & 1.458 & 1.625 & 2.334 & 5.810 \\
Elastic Net Dynamic & 1.044 & 1.445 & 1.736 & 4.126 & 8.008 & 1.046 & 1.450 & 1.730 & 4.033 & 7.606 \\
Spline-Ridge GAM & 1.039 & 1.433 & 1.626 & 1.825 & 1.971 & 1.042 & 1.439 & 1.634 & 1.838 & 1.992 \\
Random Forest & 1.055 & 1.443 & 1.636 & 1.820 & 1.945 & 1.057 & 1.447 & 1.642 & 1.830 & 1.961 \\
Gradient Boosting & 1.043 & 1.454 & 1.615 & 1.883 & 2.171 & 1.047 & 1.461 & 1.621 & 1.897 & 2.198 \\
SVR (RBF) & 1.309 & 1.577 & 1.903 & 2.290 & 2.473 & 1.318 & 1.586 & 1.918 & 2.315 & 2.503 \\
KNN & 1.042 & 1.457 & 1.694 & 1.900 & 2.014 & 1.045 & 1.463 & 1.703 & 1.915 & 2.033 \\
LSTM Sequence & 1.052 & 1.515 & 1.675 & 1.977 & 2.426 & 1.055 & 1.512 & 1.672 & 1.968 & 2.407 \\
Temp. Fusion Transf. & 1.044 & 1.478 & 1.652 & 1.846 & 1.980 & 1.047 & 1.485 & 1.660 & 1.859 & 1.998 \\
Informer Sequence & 1.043 & 1.452 & 1.650 & 1.834 & 1.857 & 1.046 & 1.458 & 1.658 & 1.846 & 1.868 \\
\bottomrule
\end{tabular}
\end{adjustbox}
\end{landscapetablepage}

\clearpage
\begin{landscapetablepage}
\captionof{table}[Benchmark Directional Accuracy in the Final Holdout]{\textbf{Benchmark Directional Accuracy in the Final Holdout.} Entries report directional accuracy for the same 169 final-holdout origins.}
\label{tab:app_holdout_benchmark_da}
\begin{tabular*}{\linewidth}{@{\extracolsep{\fill}}YZZZZZ}
\toprule
Model & $h=1$ & $h=2$ & $h=3$ & $h=4$ & $h=5$ \\
\midrule
Released forecast & 0.491 & 0.580 & 0.627 & 0.604 & 0.633 \\
Random Walk & 0.503 & 0.432 & 0.402 & 0.391 & 0.373 \\
Drift & 0.503 & 0.432 & 0.402 & 0.391 & 0.373 \\
Historical Mean & 0.521 & 0.450 & 0.426 & 0.414 & 0.396 \\
ARX Dynamic Reg. & 0.503 & 0.586 & 0.574 & 0.592 & 0.609 \\
LASSO Dynamic & 0.503 & 0.432 & 0.544 & 0.533 & 0.598 \\
Elastic Net Dynamic & 0.503 & 0.432 & 0.544 & 0.562 & 0.609 \\
Spline-Ridge GAM & 0.533 & 0.521 & 0.538 & 0.521 & 0.503 \\
Random Forest & 0.432 & 0.527 & 0.385 & 0.396 & 0.373 \\
Gradient Boosting & 0.533 & 0.432 & 0.556 & 0.391 & 0.373 \\
SVR (RBF) & 0.527 & 0.438 & 0.462 & 0.432 & 0.432 \\
KNN & 0.485 & 0.432 & 0.373 & 0.373 & 0.373 \\
LSTM Sequence & 0.503 & 0.580 & 0.544 & 0.586 & 0.556 \\
Temp. Fusion Transf. & 0.503 & 0.432 & 0.402 & 0.391 & 0.373 \\
Informer Sequence & 0.503 & 0.432 & 0.402 & 0.391 & 0.586 \\
\bottomrule
\end{tabular*}
\end{landscapetablepage}
  
\clearpage
\section{Information and State Diagnostics}
\label{app:information-state}

This appendix reports supplementary evidence on the predictive-information inputs and rolling-state representation. The analysis covers raw-output ablation and cross-horizon comparisons of alternative causal state constructions. These tests are diagnostics rather than a post-holdout re-selection exercise: they ask whether the main result depends on one input layer or residual-state choice.

\subsection{Raw-Output Ablation}

Table~\ref{tab:app_raw_layer_ablation} reports raw-layer RMSE and the horizon-five change relative to the full candidate model. Table~\ref{tab:app_raw_layer_ablation_mae_r2} reports the corresponding MAE and random-walk-relative $R^2_{OOS}$. Removing the VMD components while retaining observed price history raises horizon-five raw RMSE from 1.750 to 1.811.

The ablation results indicate that no single non-price block mechanically drives the short-horizon forecast. Removing information indices leaves horizon-five RMSE almost unchanged, whereas dropping the state IMFs raises horizon-five RMSE and removes most of the candidate layer's positive horizon-five $R^2_{OOS}$. Autoregressive residual lags and a local-level Kalman state are locally competitive at shorter horizons, so the evidence points to horizon-specific state value rather than dominance by one auxiliary construction.
\inserttable{\ref{tab:app_raw_layer_ablation}}
\inserttable{\ref{tab:app_raw_layer_ablation_mae_r2}}

\subsection{Alternative State Representations}

Tables~\ref{tab:app_state_construction_rmse}--\ref{tab:app_state_construction_r2} report RMSE, MAE, and random-walk-relative $R^2_{OOS}$ for the candidate and released layers at every horizon. The alternatives comprise VMD, EMD, autoregressive residual lags, EWMA level and volatility, a local-level Kalman state, and no constructed residual state. Each representation is formed from residual histories available by the forecast origin \citep{dragomiretskiy2014vmd,huang1998emd}.

The released forecasts remain competitive across these state choices, but the alternative-state tables also show meaningful horizon heterogeneity. The maintained VMD representation has the lowest released RMSE at $h=5$, while local-level or no-state variants are close or slightly better at shorter horizons. The main text therefore treats the rolling state as robustness-sensitive: causal state summaries help the forecast path, with the clearest incremental value beyond the next trading day.
\inserttable{\ref{tab:app_state_construction_rmse}}
\inserttable{\ref{tab:app_state_construction_mae}}
\inserttable{\ref{tab:app_state_construction_r2}}

\clearpage
\begin{table}[p]
\centering
\standardtablelayout
\caption[Forecast Performance after Raw-Layer Ablation]{\textbf{Forecast Performance after Raw-Layer Ablation.} Components are removed from the raw forecast layer before reporting-layer calibration and source selection. $\Delta h=5$ is measured against the full raw model.}
\label{tab:app_raw_layer_ablation}
\begin{tabular*}{0.98\textwidth}{@{\extracolsep{\fill}}p{0.23\textwidth}p{0.19\textwidth}rrrrrr}
\toprule
Variant & Change & $h=1$ & $h=2$ & $h=3$ & $h=4$ & $h=5$ & $\Delta h=5$ \\
\midrule
Full model & None & 1.064 & 1.344 & 1.586 & 1.640 & 1.750 & 0.000 \\
Drop information indices & Information idx. & 1.037 & 1.339 & 1.492 & 1.623 & 1.740 & $-0.010$ \\
Key raw only & Other raw predictors & 0.972 & 1.345 & 1.462 & 1.687 & 1.759 & +0.009 \\
Drop state IMFs & VMD IMFs & 0.991 & 1.316 & 1.502 & 1.661 & 1.811 & +0.061 \\
Replace VMD with EMD & VMD $\rightarrow$ EMD & 0.986 & 1.356 & 1.490 & 1.741 & 1.791 & +0.041 \\
AR residual lags & VMD $\rightarrow$ AR lags & 1.012 & 1.329 & 1.517 & 1.607 & 1.718 & $-0.032$ \\
EWMA state & VMD $\rightarrow$ EWMA & 0.980 & 1.345 & 1.500 & 1.648 & 1.789 & +0.039 \\
Local-level Kalman state & VMD $\rightarrow$ Kalman & 0.992 & 1.305 & 1.453 & 1.591 & 1.705 & $-0.045$ \\
No correction & Correction branch & 1.588 & 1.725 & 1.586 & 1.640 & 1.769 & +0.020 \\
\bottomrule
\end{tabular*}
\end{table}
 
\clearpage
\begin{landscapetablepage}
\captionof{table}[Raw-Layer Ablation MAE and Out-of-Sample Fit]{\textbf{Raw-Layer Ablation MAE and Out-of-Sample Fit.} Entries report MAE and random-walk-relative $R^2_{OOS}$ for all nine raw-layer specifications at horizons one through five.}
\label{tab:app_raw_layer_ablation_mae_r2}
\begin{adjustbox}{width=\linewidth}
\begin{tabular}{lrrrrrrrrrr}
\toprule
& \multicolumn{5}{c}{MAE} & \multicolumn{5}{c}{$R^2_{OOS}$} \\
\cmidrule(lr){2-6}\cmidrule(lr){7-11}
Variant & $h=1$ & $h=2$ & $h=3$ & $h=4$ & $h=5$ & $h=1$ & $h=2$ & $h=3$ & $h=4$ & $h=5$ \\
\midrule
Full model & 0.873 & 1.079 & 1.301 & 1.338 & 1.404 & -0.184 & -0.013 & -0.150 & 0.031 & 0.068 \\
Drop information indices & 0.846 & 1.083 & 1.224 & 1.319 & 1.396 & -0.123 & -0.005 & -0.018 & 0.051 & 0.078 \\
Key raw only & 0.782 & 1.091 & 1.215 & 1.374 & 1.411 & 0.013 & -0.015 & 0.023 & -0.025 & 0.058 \\
Drop state IMFs & 0.800 & 1.062 & 1.231 & 1.325 & 1.409 & -0.027 & 0.028 & -0.032 & 0.006 & 0.002 \\
Replace VMD with EMD & 0.789 & 1.099 & 1.233 & 1.416 & 1.432 & -0.017 & -0.031 & -0.015 & -0.092 & 0.024 \\
AR residual lags & 0.806 & 1.069 & 1.245 & 1.289 & 1.373 & -0.070 & 0.009 & -0.052 & 0.069 & 0.102 \\
EWMA state & 0.785 & 1.095 & 1.240 & 1.337 & 1.428 & -0.004 & -0.014 & -0.028 & 0.021 & 0.026 \\
Local-level Kalman state & 0.789 & 1.057 & 1.185 & 1.281 & 1.373 & -0.028 & 0.045 & 0.034 & 0.088 & 0.116 \\
No correction & 1.368 & 1.443 & 1.301 & 1.338 & 1.393 & -1.635 & -0.668 & -0.150 & 0.031 & 0.047 \\
\bottomrule
\end{tabular}
\end{adjustbox}
\end{landscapetablepage}

\clearpage
\begin{table}[p]
\centering
\standardtablelayout
\caption[Alternative State Representation RMSE]{\textbf{Alternative State Representation RMSE.} Candidate and released forecasts are evaluated at every horizon after re-estimation of the forecasting model, correction path, validation calibration, and horizon-specific release rule.}
\label{tab:app_state_construction_rmse}
\begin{tabular*}{0.98\textwidth}{@{\extracolsep{\fill}}p{0.18\textwidth}p{0.30\textwidth}rrrrr}
\toprule
Forecast layer & State representation & $h=1$ & $h=2$ & $h=3$ & $h=4$ & $h=5$ \\
\midrule
Candidate & VMD & 1.064 & 1.344 & 1.586 & 1.640 & 1.750 \\
Candidate & EMD & 0.986 & 1.356 & 1.490 & 1.741 & 1.791 \\
Candidate & AR residual lags & 1.012 & 1.329 & 1.517 & 1.607 & 1.718 \\
Candidate & EWMA level and volatility & 0.980 & 1.345 & 1.500 & 1.648 & 1.789 \\
Candidate & Local-level Kalman state & 0.992 & 1.305 & 1.453 & 1.591 & 1.705 \\
Candidate & No constructed state & 0.991 & 1.316 & 1.502 & 1.661 & 1.811 \\
Released & VMD & 0.976 & 1.316 & 1.424 & 1.615 & 1.700 \\
Released & EMD & 0.986 & 1.335 & 1.490 & 1.741 & 1.791 \\
Released & AR residual lags & 0.970 & 1.314 & 1.517 & 1.595 & 1.718 \\
Released & EWMA level and volatility & 0.980 & 1.345 & 1.500 & 1.648 & 1.789 \\
Released & Local-level Kalman state & 0.968 & 1.311 & 1.645 & 1.591 & 1.705 \\
Released & No constructed state & 0.970 & 1.305 & 1.422 & 1.661 & 1.811 \\
\bottomrule
\end{tabular*}
\end{table}

\clearpage
\begin{table}[p]
\centering
\standardtablelayout
\caption[Alternative State Representation MAE]{\textbf{Alternative State Representation MAE.} Entries report mean absolute error for the candidate and released forecasts at horizons one through five.}
\label{tab:app_state_construction_mae}
\begin{tabular*}{0.98\textwidth}{@{\extracolsep{\fill}}p{0.18\textwidth}p{0.30\textwidth}rrrrr}
\toprule
Forecast layer & State representation & $h=1$ & $h=2$ & $h=3$ & $h=4$ & $h=5$ \\
\midrule
Candidate & VMD & 0.873 & 1.079 & 1.301 & 1.338 & 1.404 \\
Candidate & EMD & 0.789 & 1.099 & 1.233 & 1.416 & 1.432 \\
Candidate & AR residual lags & 0.806 & 1.069 & 1.245 & 1.289 & 1.373 \\
Candidate & EWMA level and volatility & 0.785 & 1.095 & 1.240 & 1.337 & 1.428 \\
Candidate & Local-level Kalman state & 0.789 & 1.057 & 1.185 & 1.281 & 1.373 \\
Candidate & No constructed state & 0.800 & 1.062 & 1.231 & 1.325 & 1.409 \\
Released & VMD & 0.781 & 1.065 & 1.179 & 1.313 & 1.360 \\
Released & EMD & 0.789 & 1.080 & 1.233 & 1.416 & 1.432 \\
Released & AR residual lags & 0.781 & 1.059 & 1.245 & 1.282 & 1.373 \\
Released & EWMA level and volatility & 0.785 & 1.095 & 1.240 & 1.337 & 1.428 \\
Released & Local-level Kalman state & 0.776 & 1.058 & 1.350 & 1.281 & 1.373 \\
Released & No constructed state & 0.781 & 1.054 & 1.172 & 1.325 & 1.409 \\
\bottomrule
\end{tabular*}
\end{table}

\clearpage
\begin{table}[p]
\centering
\standardtablelayout
\caption[Alternative State Representation Out-of-Sample Fit]{\textbf{Alternative State Representation Out-of-Sample Fit.} Entries report random-walk-relative $R^2_{OOS}$ for the candidate and released forecasts at horizons one through five.}
\label{tab:app_state_construction_r2}
\begin{tabular*}{0.98\textwidth}{@{\extracolsep{\fill}}p{0.18\textwidth}p{0.30\textwidth}rrrrr}
\toprule
Forecast layer & State representation & $h=1$ & $h=2$ & $h=3$ & $h=4$ & $h=5$ \\
\midrule
Candidate & VMD & -0.184 & -0.013 & -0.150 & 0.031 & 0.068 \\
Candidate & EMD & -0.017 & -0.031 & -0.015 & -0.092 & 0.024 \\
Candidate & AR residual lags & -0.070 & 0.009 & -0.052 & 0.069 & 0.102 \\
Candidate & EWMA level and volatility & -0.004 & -0.014 & -0.028 & 0.021 & 0.026 \\
Candidate & Local-level Kalman state & -0.028 & 0.045 & 0.034 & 0.088 & 0.116 \\
Candidate & No constructed state & -0.027 & 0.028 & -0.032 & 0.006 & 0.002 \\
Released & VMD & 0.005 & 0.029 & 0.073 & 0.061 & 0.120 \\
Released & EMD & -0.017 & 0.000 & -0.015 & -0.092 & 0.024 \\
Released & AR residual lags & 0.017 & 0.031 & -0.052 & 0.083 & 0.102 \\
Released & EWMA level and volatility & -0.004 & -0.014 & -0.028 & 0.021 & 0.026 \\
Released & Local-level Kalman state & 0.021 & 0.036 & -0.237 & 0.088 & 0.116 \\
Released & No constructed state & 0.017 & 0.045 & 0.075 & 0.006 & 0.002 \\
\bottomrule
\end{tabular*}
\end{table}

\clearpage
\section{Procurement Sensitivity}
\label{app:procurement-sensitivity}

This appendix documents the procurement calibration and reports sensitivity across order size, demand realization, forecast horizon, and sample construction. Every comparison uses the same optimized schedule, TWAP benchmark, and immediate-purchase benchmark defined in Section~5. The appendix expands the economic-value analysis by showing how the result changes when execution size, demand uncertainty, and overlap in procurement windows are varied.

\subsection{Implementation-Cost and Risk-Preference Calibration}

Table~\ref{tab:app_procurement_parameters} reports the fee, spread, market-volume, impact, capacity, demand, imbalance, scenario, CVaR, and risk-aversion parameters. The analysis evaluates four order sizes from 10,000 to one million EUA. Each decision combines the day-0 close with forecasts through day $h$, so horizons two through five correspond to execution windows of three through six trading days. The cost environment is deliberately conservative: all schedules face exchange fees, half-spreads, impact costs, participation limits, order-size constraints, and a tail-risk penalty. A forecast can therefore create value only if it reallocates execution enough to offset these frictions.
\inserttable{\ref{tab:app_procurement_parameters}}

\subsection{Order Size and Market Impact}

Table~\ref{tab:app_procurement_order_size} reports fixed-demand savings by order size and horizon. Savings relative to TWAP are positive in every cell, while the horizon-five gain declines from 39.2 basis points for 10,000 EUA to 29.6 basis points for one million EUA as impact and capacity become more binding. The corresponding euro saving rises with order size, and average gains relative to immediate purchase are positive in 11 of the 16 size--horizon cells.

The order-size gradient clarifies the economic mechanism. Basis-point savings are largest when the order is small enough that capacity constraints are loose, but euro savings increase with scale because the same price-path improvement is applied to a larger compliance purchase. The weaker comparison with immediate purchase is also informative: an immediate buy can be hard to beat when prices subsequently rise, whereas TWAP is the more natural benchmark for a buyer that has already committed to completing the order over the forecast window.
\inserttable{\ref{tab:app_procurement_order_size}}

\subsection{Demand Risk and Non-Overlapping Windows}

Tables~\ref{tab:app_procurement_full_h2}--\ref{tab:app_procurement_full_h5} report full-sample results for fixed and uncertain demand, four order sizes, and three realized-demand levels at horizons two through five. Tables~\ref{tab:app_procurement_nonoverlap_h2}--\ref{tab:app_procurement_nonoverlap_h5} repeat the same design on non-overlapping execution windows. Each table reports savings relative to TWAP and immediate purchase together with win rates, tail outcomes, regret, and inference.

The full-sample tables show that fixed-demand TWAP savings are positive across all reported order sizes and horizons, with stronger average savings at longer horizons. Under uncertain demand, the realized-demand state matters. High realized demand produces large positive savings because the optimized schedule hedges shortfall costs, while low or central realized demand can be weaker when excess purchases are penalized by the surplus discount. The non-overlapping-window panels are based on substantially fewer decisions, so their inference is less precise, but they provide an important check that the fixed-demand gains are not purely an artifact of overlapping execution windows.
\inserttable{\ref{tab:app_procurement_full_h2}}
\inserttable{\ref{tab:app_procurement_full_h3}}
\inserttable{\ref{tab:app_procurement_full_h4}}
\inserttable{\ref{tab:app_procurement_full_h5}}
\inserttable{\ref{tab:app_procurement_nonoverlap_h2}}
\inserttable{\ref{tab:app_procurement_nonoverlap_h3}}
\inserttable{\ref{tab:app_procurement_nonoverlap_h4}}
\inserttable{\ref{tab:app_procurement_nonoverlap_h5}}

\clearpage
\begin{landscapetablepage}
\captionof{table}[Procurement Parameters]{\textbf{Procurement Parameters.} Market volume is the rolling 20-day median contract volume multiplied by the 1,000-EUA contract size. The impact coefficient produces five basis points of impact at 10\% of rolling median daily volume.}
\label{tab:app_procurement_parameters}
\begin{tabular*}{0.82\linewidth}{@{\extracolsep{\fill}}YYZ}
\toprule
Parameter & Value & Unit \\
\midrule
Contract and minimum trade size & 1,000 & EUA \\
Exchange fee & 0.0035 & EUR/EUA \\
Half-spread & 1.0 & basis point \\
Future-day participation cap & 10 & \% of ADV \\
Maximum future-day share & 50 & \% of total purchase \\
Market impact at 10\% ADV & 5.0 & basis points \\
Demand states & 0.75 / 1.00 / 1.25 & multiple of $Q_t$ \\
Demand probabilities & 0.25 / 0.50 / 0.25 & probability \\
Shortfall premium / surplus discount & 10.0 / 0.5 & EUR/EUA \\
Price scenarios & 250 & paths \\
Risk aversion / CVaR level & 0.02 / 0.95 & coefficient / probability \\
\bottomrule
\end{tabular*}
\end{landscapetablepage}
 
\clearpage
\begin{landscapetablepage}
\captionof{table}[Fixed-Quantity Procurement by Order Size]{\textbf{Fixed-Quantity Procurement by Order Size.} Forecast horizon $h$ defines $h+1$ execution days from day 0 through day $h$. Entries report average realized savings for the central demand realization. Positive values indicate that optimized procurement costs less than the benchmark.}
\label{tab:app_procurement_order_size}
\begin{adjustbox}{width=\linewidth}
\begin{tabular}{rrrrrr}
\toprule
Order size & Forecast horizon & TWAP savings (bps) & Immediate savings (bps) & TWAP savings (EUR) & Immediate savings (EUR) \\
\midrule
10,000 & 2 & 8.8 & $-2.1$ & 692 & $-170$ \\
10,000 & 3 & 17.6 & 1.3 & 1,397 & 104 \\
10,000 & 4 & 29.0 & 5.8 & 2,223 & 389 \\
10,000 & 5 & 39.2 & 8.2 & 2,979 & 553 \\
100,000 & 2 & 8.5 & $-2.1$ & 6,708 & $-1,649$ \\
100,000 & 3 & 17.7 & 1.9 & 14,053 & 1,420 \\
100,000 & 4 & 28.7 & 5.8 & 21,972 & 3,932 \\
100,000 & 5 & 38.5 & 7.9 & 29,290 & 5,324 \\
500,000 & 2 & 8.1 & $-0.8$ & 32,136 & $-3,670$ \\
500,000 & 3 & 16.3 & 2.1 & 64,801 & 8,038 \\
500,000 & 4 & 23.0 & 1.9 & 88,750 & 5,035 \\
500,000 & 5 & 33.5 & 4.6 & 127,972 & 14,500 \\
1,000,000 & 2 & 5.5 & $-1.4$ & 45,243 & $-11,427$ \\
1,000,000 & 3 & 12.8 & 0.8 & 103,723 & 6,210 \\
1,000,000 & 4 & 17.7 & $-1.2$ & 138,977 & $-12,246$ \\
1,000,000 & 5 & 29.6 & 2.9 & 228,202 & 17,156 \\
\bottomrule
\end{tabular}
\end{adjustbox}
\end{landscapetablepage}
 
\clearpage
\begin{landscapetablepage}
\captionof{table}[Full-Sample Procurement Results at $h=2$]{\textbf{Full-Sample Procurement Results at $h=2$.} The table reports all fixed-quantity and uncertain-demand combinations for four order sizes and three realized-demand levels. T and I denote TWAP and immediate purchase. Savings, tail measures, maximum loss, and oracle regret are in basis points. HAC $p$-values are two-sided with lag $h$.}
\label{tab:app_procurement_full_h2}
\scriptsize
\setlength{\tabcolsep}{2.2pt}
\renewcommand{\arraystretch}{0.95}
\begin{adjustbox}{width=\linewidth}
\begin{tabular}{lrrrrrrrrrrrrrrrr}
\toprule
Mode & $Q$ & $D/Q$ & $N$ & Planned & Mean T & Median T & Win T & Worst 10\% T & CVaR$_5$ T & Max loss T & Mean I & Win I & CVaR$_5$ I & Regret & HAC $p$ T & HAC $p$ I \\
\midrule
Fixed & 10,000 & 0.75 & 138 & 10,000 & 11.9 & 13.8 & 0.551 & $-124.6$ & $-145.7$ & 243.5 & $-3.0$ & 0.449 & $-115.7$ & 88.3 & 0.143 & 0.477 \\
Fixed & 10,000 & 1.00 & 138 & 10,000 & 8.8 & 10.3 & 0.551 & $-94.0$ & $-110.0$ & 184.1 & $-2.1$ & 0.449 & $-86.2$ & 72.2 & 0.150 & 0.504 \\
Fixed & 10,000 & 1.25 & 138 & 10,000 & 6.8 & 8.1 & 0.551 & $-73.4$ & $-86.0$ & 144.1 & $-1.6$ & 0.449 & $-66.8$ & 340.7 & 0.154 & 0.519 \\
Fixed & 100,000 & 0.75 & 138 & 100,000 & 11.5 & 15.6 & 0.529 & $-124.3$ & $-147.8$ & 243.7 & $-2.9$ & 0.449 & $-112.6$ & 89.0 & 0.153 & 0.483 \\
Fixed & 100,000 & 1.00 & 138 & 100,000 & 8.5 & 11.7 & 0.529 & $-93.8$ & $-111.6$ & 184.2 & $-2.1$ & 0.449 & $-83.8$ & 72.7 & 0.160 & 0.511 \\
Fixed & 100,000 & 1.25 & 138 & 100,000 & 6.5 & 9.1 & 0.529 & $-73.3$ & $-87.2$ & 144.1 & $-1.5$ & 0.449 & $-64.9$ & 341.1 & 0.164 & 0.527 \\
Fixed & 500,000 & 0.75 & 138 & 500,000 & 11.0 & 13.6 & 0.543 & $-111.8$ & $-133.0$ & 215.0 & $-1.3$ & 0.493 & $-121.7$ & 90.5 & 0.144 & 0.797 \\
Fixed & 500,000 & 1.00 & 138 & 500,000 & 8.1 & 10.2 & 0.543 & $-84.3$ & $-100.4$ & 162.5 & $-0.8$ & 0.493 & $-90.7$ & 73.9 & 0.150 & 0.831 \\
Fixed & 500,000 & 1.25 & 138 & 500,000 & 6.3 & 8.0 & 0.543 & $-65.9$ & $-78.4$ & 127.2 & $-0.6$ & 0.493 & $-70.2$ & 342.1 & 0.154 & 0.850 \\
Fixed & 1,000,000 & 0.75 & 138 & 1,000,000 & 7.5 & 10.5 & 0.536 & $-125.1$ & $-150.7$ & 231.7 & $-2.1$ & 0.529 & $-119.8$ & 95.5 & 0.338 & 0.634 \\
Fixed & 1,000,000 & 1.00 & 138 & 1,000,000 & 5.5 & 7.9 & 0.536 & $-94.4$ & $-113.8$ & 175.3 & $-1.4$ & 0.529 & $-89.2$ & 77.6 & 0.351 & 0.668 \\
Fixed & 1,000,000 & 1.25 & 138 & 1,000,000 & 4.2 & 6.1 & 0.536 & $-73.7$ & $-88.9$ & 137.0 & $-1.0$ & 0.529 & $-69.2$ & 345.0 & 0.358 & 0.687 \\
Uncertain & 10,000 & 0.75 & 138 & 12,000 & 4.3 & $-8.8$ & 0.464 & $-169.7$ & $-196.1$ & 354.1 & $-10.4$ & 0.420 & $-149.9$ & 95.9 & 0.696 & 0.059 \\
Uncertain & 10,000 & 1.00 & 138 & 12,000 & 3.0 & $-6.6$ & 0.464 & $-128.0$ & $-148.0$ & 267.7 & $-7.7$ & 0.420 & $-111.8$ & 78.0 & 0.713 & 0.061 \\
Uncertain & 10,000 & 1.25 & 138 & 12,000 & 217.3 & 207.9 & 1.000 & 111.7 & 94.7 & 0.0 & 209.2 & 1.000 & 125.5 & 122.8 & $<0.001$ & $<0.001$ \\
Uncertain & 100,000 & 0.75 & 138 & 125,000 & $-2.5$ & $-10.5$ & 0.449 & $-188.9$ & $-218.5$ & 382.1 & $-16.7$ & 0.370 & $-155.0$ & 103.2 & 0.825 & 0.001 \\
Uncertain & 100,000 & 1.00 & 138 & 125,000 & $-2.1$ & $-7.9$ & 0.449 & $-142.5$ & $-164.9$ & 288.8 & $-12.5$ & 0.370 & $-115.7$ & 83.4 & 0.806 & 0.001 \\
Uncertain & 100,000 & 1.25 & 138 & 125,000 & 267.0 & 257.4 & 1.000 & 154.6 & 135.9 & 0.0 & 259.2 & 1.000 & 176.9 & 71.5 & $<0.001$ & $<0.001$ \\
Uncertain & 500,000 & 0.75 & 138 & 625,000 & $-7.7$ & $-12.9$ & 0.420 & $-189.6$ & $-219.0$ & 342.2 & $-19.8$ & 0.362 & $-146.0$ & 109.6 & 0.493 & $<0.001$ \\
Uncertain & 500,000 & 1.00 & 138 & 625,000 & $-6.0$ & $-9.7$ & 0.420 & $-143.1$ & $-165.3$ & 258.7 & $-14.8$ & 0.362 & $-108.7$ & 88.2 & 0.477 & $<0.001$ \\
Uncertain & 500,000 & 1.25 & 138 & 625,000 & 263.9 & 256.5 & 1.000 & 155.4 & 136.5 & 0.0 & 257.4 & 1.000 & 180.1 & 75.4 & $<0.001$ & $<0.001$ \\
Uncertain & 1,000,000 & 0.75 & 138 & 1,250,000 & $-13.4$ & $-14.2$ & 0.420 & $-219.8$ & $-270.2$ & 424.4 & $-22.7$ & 0.362 & $-149.2$ & 116.7 & 0.260 & $<0.001$ \\
Uncertain & 1,000,000 & 1.00 & 138 & 1,250,000 & $-10.3$ & $-10.7$ & 0.420 & $-165.9$ & $-204.1$ & 321.1 & $-17.0$ & 0.362 & $-111.2$ & 93.5 & 0.250 & $<0.001$ \\
Uncertain & 1,000,000 & 1.25 & 138 & 1,250,000 & 260.6 & 255.5 & 1.000 & 142.2 & 114.6 & 0.0 & 255.6 & 1.000 & 179.2 & 79.6 & $<0.001$ & $<0.001$ \\
\bottomrule
\end{tabular}
\end{adjustbox}
\end{landscapetablepage}

\clearpage
\begin{landscapetablepage}
\captionof{table}[Full-Sample Procurement Results at $h=3$]{\textbf{Full-Sample Procurement Results at $h=3$.} The table reports all fixed-quantity and uncertain-demand combinations for four order sizes and three realized-demand levels. T and I denote TWAP and immediate purchase. Savings, tail measures, maximum loss, and oracle regret are in basis points. HAC $p$-values are two-sided with lag $h$.}
\label{tab:app_procurement_full_h3}
\scriptsize
\setlength{\tabcolsep}{2.2pt}
\renewcommand{\arraystretch}{0.95}
\begin{adjustbox}{width=\linewidth}
\begin{tabular}{lrrrrrrrrrrrrrrrr}
\toprule
Mode & $Q$ & $D/Q$ & $N$ & Planned & Mean T & Median T & Win T & Worst 10\% T & CVaR$_5$ T & Max loss T & Mean I & Win I & CVaR$_5$ I & Regret & HAC $p$ T & HAC $p$ I \\
\midrule
Fixed & 10,000 & 0.75 & 137 & 10,000 & 23.6 & 19.4 & 0.569 & $-143.6$ & $-158.3$ & 210.4 & 1.5 & 0.380 & $-124.4$ & 101.4 & 0.034 & 0.801 \\
Fixed & 10,000 & 1.00 & 137 & 10,000 & 17.6 & 14.5 & 0.569 & $-108.3$ & $-119.3$ & 159.0 & 1.3 & 0.380 & $-92.7$ & 90.3 & 0.035 & 0.774 \\
Fixed & 10,000 & 1.25 & 137 & 10,000 & 13.6 & 11.3 & 0.569 & $-84.5$ & $-93.2$ & 124.6 & 1.1 & 0.380 & $-71.8$ & 362.3 & 0.036 & 0.758 \\
Fixed & 100,000 & 0.75 & 137 & 100,000 & 23.8 & 19.5 & 0.562 & $-140.9$ & $-158.1$ & 205.1 & 2.2 & 0.409 & $-124.7$ & 101.3 & 0.030 & 0.718 \\
Fixed & 100,000 & 1.00 & 137 & 100,000 & 17.7 & 14.7 & 0.562 & $-106.1$ & $-119.2$ & 155.1 & 1.9 & 0.409 & $-92.9$ & 90.2 & 0.032 & 0.692 \\
Fixed & 100,000 & 1.25 & 137 & 100,000 & 13.8 & 11.4 & 0.562 & $-82.8$ & $-93.1$ & 121.5 & 1.5 & 0.409 & $-71.9$ & 362.3 & 0.032 & 0.676 \\
Fixed & 500,000 & 0.75 & 137 & 500,000 & 21.9 & 18.5 & 0.562 & $-133.0$ & $-153.4$ & 198.8 & 2.6 & 0.474 & $-134.4$ & 104.0 & 0.040 & 0.705 \\
Fixed & 500,000 & 1.00 & 137 & 500,000 & 16.3 & 13.9 & 0.562 & $-100.2$ & $-115.7$ & 150.2 & 2.1 & 0.474 & $-100.1$ & 92.3 & 0.042 & 0.678 \\
Fixed & 500,000 & 1.25 & 137 & 500,000 & 12.7 & 10.8 & 0.562 & $-78.3$ & $-90.4$ & 117.7 & 1.7 & 0.474 & $-77.5$ & 363.9 & 0.042 & 0.663 \\
Fixed & 1,000,000 & 0.75 & 137 & 1,000,000 & 17.3 & 19.3 & 0.555 & $-141.0$ & $-162.7$ & 205.4 & 0.8 & 0.518 & $-136.6$ & 109.8 & 0.121 & 0.890 \\
Fixed & 1,000,000 & 1.00 & 137 & 1,000,000 & 12.8 & 14.5 & 0.555 & $-106.4$ & $-122.8$ & 154.8 & 0.8 & 0.518 & $-101.8$ & 96.6 & 0.126 & 0.859 \\
Fixed & 1,000,000 & 1.25 & 137 & 1,000,000 & 9.9 & 11.4 & 0.555 & $-83.0$ & $-96.0$ & 120.7 & 0.7 & 0.518 & $-78.9$ & 367.4 & 0.128 & 0.840 \\
Uncertain & 10,000 & 0.75 & 137 & 12,000 & 12.2 & 5.4 & 0.526 & $-218.6$ & $-240.7$ & 346.9 & $-9.5$ & 0.423 & $-159.6$ & 113.1 & 0.428 & 0.104 \\
Uncertain & 10,000 & 1.00 & 137 & 12,000 & 8.9 & 4.1 & 0.526 & $-164.9$ & $-181.8$ & 262.2 & $-7.2$ & 0.423 & $-119.6$ & 99.0 & 0.440 & 0.105 \\
Uncertain & 10,000 & 1.25 & 137 & 12,000 & 221.7 & 219.5 & 0.993 & 88.5 & 70.3 & 2.0 & 209.5 & 1.000 & 122.2 & 146.4 & $<0.001$ & $<0.001$ \\
Uncertain & 100,000 & 0.75 & 137 & 125,000 & 9.0 & 4.0 & 0.518 & $-226.4$ & $-256.3$ & 342.3 & $-12.2$ & 0.423 & $-189.4$ & 116.5 & 0.578 & 0.070 \\
Uncertain & 100,000 & 1.00 & 137 & 125,000 & 6.5 & 3.0 & 0.518 & $-170.7$ & $-193.5$ & 258.8 & $-9.1$ & 0.423 & $-141.7$ & 101.6 & 0.593 & 0.070 \\
Uncertain & 100,000 & 1.25 & 137 & 125,000 & 273.5 & 271.2 & 1.000 & 136.6 & 114.0 & 0.0 & 261.7 & 1.000 & 161.5 & 92.7 & $<0.001$ & $<0.001$ \\
Uncertain & 500,000 & 0.75 & 137 & 625,000 & 7.1 & 7.0 & 0.504 & $-211.8$ & $-241.9$ & 320.7 & $-11.8$ & 0.431 & $-176.8$ & 119.2 & 0.655 & 0.046 \\
Uncertain & 500,000 & 1.00 & 137 & 625,000 & 5.0 & 5.3 & 0.504 & $-159.7$ & $-182.6$ & 242.4 & $-8.9$ & 0.431 & $-132.4$ & 103.7 & 0.672 & 0.045 \\
Uncertain & 500,000 & 1.25 & 137 & 625,000 & 272.4 & 271.8 & 1.000 & 141.8 & 118.5 & 0.0 & 261.9 & 1.000 & 168.6 & 94.3 & $<0.001$ & $<0.001$ \\
Uncertain & 1,000,000 & 0.75 & 137 & 1,250,000 & 1.3 & 0.0 & 0.504 & $-246.5$ & $-288.8$ & 370.0 & $-14.6$ & 0.401 & $-164.5$ & 126.2 & 0.941 & 0.003 \\
Uncertain & 1,000,000 & 1.00 & 137 & 1,250,000 & 0.6 & 0.0 & 0.504 & $-186.1$ & $-218.3$ & 278.9 & $-11.1$ & 0.401 & $-123.4$ & 108.9 & 0.960 & 0.003 \\
Uncertain & 1,000,000 & 1.25 & 137 & 1,250,000 & 268.9 & 270.3 & 1.000 & 126.9 & 99.2 & 0.0 & 260.1 & 1.000 & 170.4 & 98.5 & $<0.001$ & $<0.001$ \\
\bottomrule
\end{tabular}
\end{adjustbox}
\end{landscapetablepage}

\clearpage
\begin{landscapetablepage}
\captionof{table}[Full-Sample Procurement Results at $h=4$]{\textbf{Full-Sample Procurement Results at $h=4$.} The table reports all fixed-quantity and uncertain-demand combinations for four order sizes and three realized-demand levels. T and I denote TWAP and immediate purchase. Savings, tail measures, maximum loss, and oracle regret are in basis points. HAC $p$-values are two-sided with lag $h$.}
\label{tab:app_procurement_full_h4}
\scriptsize
\setlength{\tabcolsep}{2.2pt}
\renewcommand{\arraystretch}{0.95}
\begin{adjustbox}{width=\linewidth}
\begin{tabular}{lrrrrrrrrrrrrrrrr}
\toprule
Mode & $Q$ & $D/Q$ & $N$ & Planned & Mean T & Median T & Win T & Worst 10\% T & CVaR$_5$ T & Max loss T & Mean I & Win I & CVaR$_5$ I & Regret & HAC $p$ T & HAC $p$ I \\
\midrule
Fixed & 10,000 & 0.75 & 136 & 10,000 & 38.8 & 30.2 & 0.588 & $-159.3$ & $-196.2$ & 268.0 & 7.3 & 0.301 & $-174.6$ & 102.6 & 0.005 & 0.538 \\
Fixed & 10,000 & 1.00 & 136 & 10,000 & 29.0 & 22.7 & 0.588 & $-119.9$ & $-147.7$ & 202.0 & 5.8 & 0.301 & $-129.8$ & 98.0 & 0.005 & 0.520 \\
Fixed & 10,000 & 1.25 & 136 & 10,000 & 22.6 & 17.7 & 0.588 & $-93.6$ & $-115.3$ & 158.0 & 4.6 & 0.301 & $-100.6$ & 374.2 & 0.005 & 0.511 \\
Fixed & 100,000 & 0.75 & 136 & 100,000 & 38.3 & 28.7 & 0.588 & $-153.2$ & $-185.5$ & 247.9 & 7.3 & 0.324 & $-181.6$ & 103.3 & 0.005 & 0.540 \\
Fixed & 100,000 & 1.00 & 136 & 100,000 & 28.7 & 21.6 & 0.588 & $-115.3$ & $-139.7$ & 186.9 & 5.8 & 0.324 & $-135.0$ & 98.5 & 0.005 & 0.521 \\
Fixed & 100,000 & 1.25 & 136 & 100,000 & 22.3 & 16.8 & 0.588 & $-89.9$ & $-109.1$ & 146.2 & 4.6 & 0.324 & $-104.7$ & 374.6 & 0.005 & 0.511 \\
Fixed & 500,000 & 0.75 & 136 & 500,000 & 30.8 & 20.8 & 0.581 & $-149.2$ & $-183.9$ & 233.8 & 2.1 & 0.397 & $-179.4$ & 111.6 & 0.017 & 0.832 \\
Fixed & 500,000 & 1.00 & 136 & 500,000 & 23.0 & 15.6 & 0.581 & $-112.3$ & $-138.5$ & 176.2 & 1.9 & 0.397 & $-133.4$ & 104.7 & 0.017 & 0.803 \\
Fixed & 500,000 & 1.25 & 136 & 500,000 & 17.8 & 12.2 & 0.581 & $-87.6$ & $-108.1$ & 137.8 & 1.6 & 0.397 & $-103.5$ & 379.6 & 0.018 & 0.788 \\
Fixed & 1,000,000 & 0.75 & 136 & 1,000,000 & 23.8 & 14.9 & 0.551 & $-152.2$ & $-183.5$ & 224.5 & $-1.9$ & 0.449 & $-188.5$ & 119.5 & 0.074 & 0.821 \\
Fixed & 1,000,000 & 1.00 & 136 & 1,000,000 & 17.7 & 11.2 & 0.551 & $-114.7$ & $-138.3$ & 169.2 & $-1.2$ & 0.449 & $-140.2$ & 110.7 & 0.076 & 0.854 \\
Fixed & 1,000,000 & 1.25 & 136 & 1,000,000 & 13.7 & 8.7 & 0.551 & $-89.5$ & $-108.0$ & 132.3 & $-0.8$ & 0.449 & $-108.7$ & 384.4 & 0.077 & 0.872 \\
Uncertain & 10,000 & 0.75 & 136 & 12,000 & 31.0 & 19.8 & 0.559 & $-211.6$ & $-261.4$ & 362.1 & $-0.1$ & 0.456 & $-194.4$ & 110.6 & 0.081 & 0.990 \\
Uncertain & 10,000 & 1.00 & 136 & 12,000 & 23.1 & 14.8 & 0.559 & $-159.4$ & $-196.8$ & 273.0 & 0.0 & 0.456 & $-145.0$ & 104.0 & 0.083 & 0.996 \\
Uncertain & 10,000 & 1.25 & 136 & 12,000 & 232.6 & 219.8 & 0.993 & 88.3 & 57.0 & 10.0 & 215.1 & 1.000 & 98.0 & 155.8 & $<0.001$ & $<0.001$ \\
Uncertain & 100,000 & 0.75 & 136 & 125,000 & 29.4 & 13.3 & 0.529 & $-227.4$ & $-276.1$ & 376.5 & $-1.1$ & 0.449 & $-191.7$ & 112.4 & 0.117 & 0.921 \\
Uncertain & 100,000 & 1.00 & 136 & 125,000 & 21.9 & 10.0 & 0.529 & $-171.4$ & $-208.0$ & 283.8 & $-0.7$ & 0.449 & $-143.0$ & 105.3 & 0.120 & 0.931 \\
Uncertain & 100,000 & 1.25 & 136 & 125,000 & 285.2 & 270.6 & 1.000 & 131.5 & 103.3 & 0.0 & 268.2 & 1.000 & 152.7 & 101.1 & $<0.001$ & $<0.001$ \\
Uncertain & 500,000 & 0.75 & 136 & 625,000 & 19.5 & $-0.1$ & 0.500 & $-223.5$ & $-270.7$ & 332.9 & $-8.6$ & 0.434 & $-198.7$ & 123.1 & 0.300 & 0.319 \\
Uncertain & 500,000 & 1.00 & 136 & 625,000 & 14.4 & $-0.1$ & 0.500 & $-168.6$ & $-203.9$ & 250.9 & $-6.4$ & 0.434 & $-148.3$ & 113.4 & 0.308 & 0.324 \\
Uncertain & 500,000 & 1.25 & 136 & 625,000 & 279.4 & 265.7 & 1.000 & 134.7 & 106.1 & 0.0 & 263.7 & 1.000 & 148.1 & 107.5 & $<0.001$ & $<0.001$ \\
Uncertain & 1,000,000 & 0.75 & 136 & 1,250,000 & 9.9 & $-2.1$ & 0.493 & $-247.4$ & $-303.3$ & 446.4 & $-15.2$ & 0.463 & $-192.0$ & 133.7 & 0.615 & 0.044 \\
Uncertain & 1,000,000 & 1.00 & 136 & 1,250,000 & 7.1 & $-1.6$ & 0.493 & $-186.7$ & $-229.1$ & 338.9 & $-11.4$ & 0.463 & $-143.3$ & 121.3 & 0.629 & 0.044 \\
Uncertain & 1,000,000 & 1.25 & 136 & 1,250,000 & 273.7 & 264.0 & 1.000 & 123.6 & 92.9 & 0.0 & 259.8 & 1.000 & 150.3 & 113.9 & $<0.001$ & $<0.001$ \\
\bottomrule
\end{tabular}
\end{adjustbox}
\end{landscapetablepage}

\clearpage
\begin{landscapetablepage}
\captionof{table}[Full-Sample Procurement Results at $h=5$]{\textbf{Full-Sample Procurement Results at $h=5$.} The table reports all fixed-quantity and uncertain-demand combinations for four order sizes and three realized-demand levels. T and I denote TWAP and immediate purchase. Savings, tail measures, maximum loss, and oracle regret are in basis points. HAC $p$-values are two-sided with lag $h$.}
\label{tab:app_procurement_full_h5}
\scriptsize
\setlength{\tabcolsep}{2.2pt}
\renewcommand{\arraystretch}{0.95}
\begin{adjustbox}{width=\linewidth}
\begin{tabular}{lrrrrrrrrrrrrrrrr}
\toprule
Mode & $Q$ & $D/Q$ & $N$ & Planned & Mean T & Median T & Win T & Worst 10\% T & CVaR$_5$ T & Max loss T & Mean I & Win I & CVaR$_5$ I & Regret & HAC $p$ T & HAC $p$ I \\
\midrule
Fixed & 10,000 & 0.75 & 135 & 10,000 & 52.3 & 48.0 & 0.652 & $-170.5$ & $-209.1$ & 270.6 & 10.4 & 0.319 & $-254.4$ & 101.5 & 0.001 & 0.521 \\
Fixed & 10,000 & 1.00 & 135 & 10,000 & 39.2 & 36.0 & 0.652 & $-128.1$ & $-157.4$ & 204.4 & 8.2 & 0.319 & $-189.4$ & 102.7 & 0.001 & 0.504 \\
Fixed & 10,000 & 1.25 & 135 & 10,000 & 30.5 & 28.0 & 0.652 & $-99.9$ & $-122.8$ & 159.6 & 6.6 & 0.319 & $-146.9$ & 382.5 & 0.001 & 0.495 \\
Fixed & 100,000 & 0.75 & 135 & 100,000 & 51.4 & 47.5 & 0.659 & $-170.3$ & $-209.5$ & 270.9 & 10.0 & 0.333 & $-259.1$ & 102.5 & 0.001 & 0.539 \\
Fixed & 100,000 & 1.00 & 135 & 100,000 & 38.5 & 35.6 & 0.659 & $-128.0$ & $-157.6$ & 204.6 & 7.9 & 0.333 & $-192.9$ & 103.4 & 0.001 & 0.521 \\
Fixed & 100,000 & 1.25 & 135 & 100,000 & 30.0 & 27.6 & 0.659 & $-99.8$ & $-123.0$ & 159.7 & 6.3 & 0.333 & $-149.6$ & 383.1 & 0.001 & 0.512 \\
Fixed & 500,000 & 0.75 & 135 & 500,000 & 44.7 & 39.7 & 0.667 & $-165.2$ & $-200.4$ & 271.9 & 5.6 & 0.407 & $-257.6$ & 109.9 & 0.004 & 0.696 \\
Fixed & 500,000 & 1.00 & 135 & 500,000 & 33.5 & 29.9 & 0.667 & $-124.2$ & $-150.8$ & 205.4 & 4.6 & 0.407 & $-191.9$ & 109.0 & 0.004 & 0.672 \\
Fixed & 500,000 & 1.25 & 135 & 500,000 & 26.0 & 23.4 & 0.667 & $-96.8$ & $-117.7$ & 160.4 & 3.7 & 0.407 & $-148.8$ & 387.6 & 0.004 & 0.660 \\
Fixed & 1,000,000 & 0.75 & 135 & 1,000,000 & 39.6 & 30.1 & 0.637 & $-155.6$ & $-185.7$ & 256.9 & 3.4 & 0.481 & $-262.4$ & 115.8 & 0.011 & 0.789 \\
Fixed & 1,000,000 & 1.00 & 135 & 1,000,000 & 29.6 & 22.5 & 0.637 & $-117.0$ & $-139.7$ & 194.1 & 2.9 & 0.481 & $-195.5$ & 113.4 & 0.011 & 0.762 \\
Fixed & 1,000,000 & 1.25 & 135 & 1,000,000 & 23.0 & 17.6 & 0.637 & $-91.3$ & $-109.0$ & 151.5 & 2.4 & 0.481 & $-151.7$ & 391.1 & 0.011 & 0.748 \\
Uncertain & 10,000 & 0.75 & 135 & 12,000 & 54.8 & 33.9 & 0.622 & $-215.7$ & $-279.1$ & 382.8 & 13.4 & 0.533 & $-263.1$ & 99.0 & 0.009 & 0.405 \\
Uncertain & 10,000 & 1.00 & 135 & 12,000 & 41.0 & 25.5 & 0.622 & $-162.3$ & $-210.2$ & 289.2 & 10.3 & 0.533 & $-196.0$ & 100.8 & 0.009 & 0.400 \\
Uncertain & 10,000 & 1.25 & 135 & 12,000 & 246.3 & 232.4 & 1.000 & 83.7 & 50.9 & 0.0 & 223.0 & 0.985 & 56.1 & 157.7 & $<0.001$ & $<0.001$ \\
Uncertain & 100,000 & 0.75 & 135 & 125,000 & 52.0 & 27.3 & 0.593 & $-227.7$ & $-296.0$ & 411.3 & 11.1 & 0.533 & $-276.9$ & 102.0 & 0.017 & 0.496 \\
Uncertain & 100,000 & 1.00 & 135 & 125,000 & 38.8 & 20.5 & 0.593 & $-171.4$ & $-223.0$ & 310.7 & 8.5 & 0.533 & $-206.3$ & 103.1 & 0.018 & 0.490 \\
Uncertain & 100,000 & 1.25 & 135 & 125,000 & 298.2 & 281.2 & 1.000 & 129.1 & 94.7 & 0.0 & 275.3 & 1.000 & 100.5 & 103.7 & $<0.001$ & $<0.001$ \\
Uncertain & 500,000 & 0.75 & 135 & 625,000 & 41.3 & 19.6 & 0.563 & $-223.6$ & $-282.1$ & 413.1 & 2.8 & 0.533 & $-279.7$ & 113.6 & 0.055 & 0.834 \\
Uncertain & 500,000 & 1.00 & 135 & 625,000 & 30.7 & 14.7 & 0.563 & $-168.2$ & $-212.5$ & 312.1 & 2.2 & 0.533 & $-208.4$ & 111.7 & 0.057 & 0.827 \\
Uncertain & 500,000 & 1.25 & 135 & 625,000 & 291.8 & 278.6 & 1.000 & 131.5 & 101.2 & 0.0 & 270.3 & 1.000 & 98.5 & 110.6 & $<0.001$ & $<0.001$ \\
Uncertain & 1,000,000 & 0.75 & 135 & 1,250,000 & 33.4 & 19.4 & 0.563 & $-243.9$ & $-302.2$ & 389.2 & $-2.1$ & 0.533 & $-279.0$ & 122.2 & 0.139 & 0.862 \\
Uncertain & 1,000,000 & 1.00 & 135 & 1,250,000 & 24.8 & 14.6 & 0.563 & $-184.0$ & $-228.3$ & 294.1 & $-1.5$ & 0.533 & $-208.0$ & 118.2 & 0.143 & 0.866 \\
Uncertain & 1,000,000 & 1.25 & 135 & 1,250,000 & 287.2 & 278.0 & 1.000 & 123.4 & 99.5 & 0.0 & 267.4 & 1.000 & 100.8 & 115.8 & $<0.001$ & $<0.001$ \\
\bottomrule
\end{tabular}
\end{adjustbox}
\end{landscapetablepage}

\clearpage
\begin{landscapetablepage}
\captionof{table}[Non-Overlapping-Window Procurement Results at $h=2$]{\textbf{Non-Overlapping-Window Procurement Results at $h=2$.} The table reports all fixed-quantity and uncertain-demand combinations for four order sizes and three realized-demand levels. T and I denote TWAP and immediate purchase. Savings, tail measures, maximum loss, and oracle regret are in basis points. HAC $p$-values are two-sided with lag $h$.}
\label{tab:app_procurement_nonoverlap_h2}
\scriptsize
\setlength{\tabcolsep}{2.2pt}
\renewcommand{\arraystretch}{0.95}
\begin{adjustbox}{width=\linewidth}
\begin{tabular}{lrrrrrrrrrrrrrrrr}
\toprule
Mode & $Q$ & $D/Q$ & $N$ & Planned & Mean T & Median T & Win T & Worst 10\% T & CVaR$_5$ T & Max loss T & Mean I & Win I & CVaR$_5$ I & Regret & HAC $p$ T & HAC $p$ I \\
\midrule
Fixed & 10,000 & 0.75 & 46 & 10,000 & 22.3 & 21.1 & 0.609 & $-85.3$ & $-96.2$ & 127.0 & $-5.1$ & 0.478 & $-90.5$ & 61.7 & 0.013 & 0.354 \\
Fixed & 10,000 & 1.00 & 46 & 10,000 & 16.7 & 15.9 & 0.609 & $-64.2$ & $-72.4$ & 95.6 & $-3.8$ & 0.478 & $-67.4$ & 53.3 & 0.013 & 0.362 \\
Fixed & 10,000 & 1.25 & 46 & 10,000 & 12.9 & 12.4 & 0.609 & $-50.1$ & $-56.5$ & 74.6 & $-2.9$ & 0.478 & $-52.3$ & 326.5 & 0.014 & 0.365 \\
Fixed & 100,000 & 0.75 & 46 & 100,000 & 20.9 & 20.7 & 0.587 & $-85.4$ & $-98.3$ & 127.1 & $-6.1$ & 0.478 & $-98.5$ & 63.4 & 0.019 & 0.277 \\
Fixed & 100,000 & 1.00 & 46 & 100,000 & 15.6 & 15.5 & 0.587 & $-64.3$ & $-74.0$ & 95.7 & $-4.5$ & 0.478 & $-73.4$ & 54.6 & 0.020 & 0.284 \\
Fixed & 100,000 & 1.25 & 46 & 100,000 & 12.1 & 12.1 & 0.587 & $-50.2$ & $-57.7$ & 74.7 & $-3.5$ & 0.478 & $-56.8$ & 327.5 & 0.020 & 0.288 \\
Fixed & 500,000 & 0.75 & 46 & 500,000 & 19.8 & 18.9 & 0.630 & $-77.4$ & $-84.0$ & 84.8 & $-5.0$ & 0.500 & $-109.5$ & 65.7 & 0.015 & 0.421 \\
Fixed & 500,000 & 1.00 & 46 & 500,000 & 14.8 & 14.2 & 0.630 & $-58.3$ & $-63.2$ & 64.0 & $-3.6$ & 0.500 & $-81.6$ & 56.4 & 0.016 & 0.432 \\
Fixed & 500,000 & 1.25 & 46 & 500,000 & 11.5 & 11.0 & 0.630 & $-45.5$ & $-49.3$ & 49.9 & $-2.8$ & 0.500 & $-63.2$ & 328.9 & 0.016 & 0.438 \\
Fixed & 1,000,000 & 0.75 & 46 & 1,000,000 & 18.7 & 18.8 & 0.609 & $-80.0$ & $-88.9$ & 99.4 & $-3.2$ & 0.543 & $-95.5$ & 68.3 & 0.019 & 0.612 \\
Fixed & 1,000,000 & 1.00 & 46 & 1,000,000 & 13.9 & 14.1 & 0.609 & $-60.3$ & $-67.1$ & 75.2 & $-2.3$ & 0.543 & $-71.3$ & 58.3 & 0.019 & 0.625 \\
Fixed & 1,000,000 & 1.25 & 46 & 1,000,000 & 10.8 & 11.0 & 0.609 & $-47.1$ & $-52.4$ & 58.8 & $-1.8$ & 0.543 & $-55.3$ & 330.4 & 0.020 & 0.630 \\
Uncertain & 10,000 & 0.75 & 46 & 12,000 & 19.5 & $-0.8$ & 0.500 & $-135.3$ & $-152.8$ & 182.6 & $-7.7$ & 0.565 & $-129.6$ & 64.5 & 0.141 & 0.279 \\
Uncertain & 10,000 & 1.00 & 46 & 12,000 & 14.5 & $-0.6$ & 0.500 & $-101.9$ & $-115.2$ & 137.5 & $-5.8$ & 0.565 & $-97.0$ & 55.5 & 0.144 & 0.276 \\
Uncertain & 10,000 & 1.25 & 46 & 12,000 & 226.4 & 215.2 & 1.000 & 135.1 & 122.9 & 0.0 & 210.9 & 1.000 & 134.7 & 105.7 & $<0.001$ & $<0.001$ \\
Uncertain & 100,000 & 0.75 & 46 & 125,000 & 10.9 & 1.9 & 0.500 & $-147.1$ & $-167.7$ & 194.9 & $-15.9$ & 0.478 & $-154.2$ & 73.5 & 0.390 & 0.066 \\
Uncertain & 100,000 & 1.00 & 46 & 125,000 & 8.1 & 1.4 & 0.500 & $-110.8$ & $-126.5$ & 146.7 & $-11.9$ & 0.478 & $-114.9$ & 62.2 & 0.398 & 0.064 \\
Uncertain & 100,000 & 1.25 & 46 & 125,000 & 275.2 & 265.0 & 1.000 & 181.9 & 168.3 & 0.0 & 260.1 & 1.000 & 177.2 & 55.4 & $<0.001$ & $<0.001$ \\
Uncertain & 500,000 & 0.75 & 46 & 625,000 & 9.3 & $-8.0$ & 0.457 & $-148.5$ & $-162.6$ & 170.1 & $-15.3$ & 0.413 & $-133.7$ & 76.4 & 0.452 & 0.065 \\
Uncertain & 500,000 & 1.00 & 46 & 625,000 & 6.8 & $-6.0$ & 0.457 & $-112.0$ & $-122.7$ & 128.4 & $-11.5$ & 0.413 & $-99.8$ & 64.3 & 0.462 & 0.064 \\
Uncertain & 500,000 & 1.25 & 46 & 625,000 & 274.2 & 263.2 & 1.000 & 182.3 & 173.5 & 0.0 & 260.4 & 1.000 & 185.5 & 57.1 & $<0.001$ & $<0.001$ \\
Uncertain & 1,000,000 & 0.75 & 46 & 1,250,000 & 8.4 & $-8.3$ & 0.478 & $-156.6$ & $-182.3$ & 236.6 & $-13.2$ & 0.435 & $-117.5$ & 78.7 & 0.519 & 0.098 \\
Uncertain & 1,000,000 & 1.00 & 46 & 1,250,000 & 6.1 & $-6.3$ & 0.478 & $-118.2$ & $-137.7$ & 178.9 & $-10.0$ & 0.435 & $-88.3$ & 66.1 & 0.531 & 0.097 \\
Uncertain & 1,000,000 & 1.25 & 46 & 1,250,000 & 273.6 & 261.7 & 1.000 & 179.3 & 162.1 & 0.0 & 261.4 & 1.000 & 184.2 & 58.5 & $<0.001$ & $<0.001$ \\
\bottomrule
\end{tabular}
\end{adjustbox}
\end{landscapetablepage}

\clearpage
\begin{landscapetablepage}
\captionof{table}[Non-Overlapping-Window Procurement Results at $h=3$]{\textbf{Non-Overlapping-Window Procurement Results at $h=3$.} The table reports all fixed-quantity and uncertain-demand combinations for four order sizes and three realized-demand levels. T and I denote TWAP and immediate purchase. Savings, tail measures, maximum loss, and oracle regret are in basis points. HAC $p$-values are two-sided with lag $h$.}
\label{tab:app_procurement_nonoverlap_h3}
\scriptsize
\setlength{\tabcolsep}{2.2pt}
\renewcommand{\arraystretch}{0.95}
\begin{adjustbox}{width=\linewidth}
\begin{tabular}{lrrrrrrrrrrrrrrrr}
\toprule
Mode & $Q$ & $D/Q$ & $N$ & Planned & Mean T & Median T & Win T & Worst 10\% T & CVaR$_5$ T & Max loss T & Mean I & Win I & CVaR$_5$ I & Regret & HAC $p$ T & HAC $p$ I \\
\midrule
Fixed & 10,000 & 0.75 & 35 & 10,000 & 13.2 & 13.6 & 0.514 & $-142.0$ & $-150.0$ & 155.3 & $-6.3$ & 0.371 & $-144.2$ & 117.9 & 0.345 & 0.324 \\
Fixed & 10,000 & 1.00 & 35 & 10,000 & 9.7 & 10.2 & 0.514 & $-107.2$ & $-113.4$ & 117.5 & $-4.6$ & 0.371 & $-107.3$ & 99.3 & 0.353 & 0.339 \\
Fixed & 10,000 & 1.25 & 35 & 10,000 & 7.5 & 8.0 & 0.514 & $-83.8$ & $-88.8$ & 92.1 & $-3.5$ & 0.371 & $-83.2$ & 366.5 & 0.357 & 0.349 \\
Fixed & 100,000 & 0.75 & 35 & 100,000 & 13.5 & 1.6 & 0.514 & $-140.2$ & $-146.3$ & 147.8 & $-5.5$ & 0.429 & $-142.5$ & 117.7 & 0.327 & 0.386 \\
Fixed & 100,000 & 1.00 & 35 & 100,000 & 10.0 & 1.2 & 0.514 & $-105.8$ & $-110.6$ & 111.9 & $-3.9$ & 0.429 & $-106.1$ & 99.2 & 0.334 & 0.405 \\
Fixed & 100,000 & 1.25 & 35 & 100,000 & 7.7 & 1.0 & 0.514 & $-82.7$ & $-86.6$ & 87.6 & $-3.0$ & 0.429 & $-82.2$ & 366.4 & 0.339 & 0.417 \\
Fixed & 500,000 & 0.75 & 35 & 500,000 & 15.5 & 17.1 & 0.514 & $-136.2$ & $-150.1$ & 154.6 & $-1.2$ & 0.514 & $-128.3$ & 116.5 & 0.207 & 0.872 \\
Fixed & 500,000 & 1.00 & 35 & 500,000 & 11.5 & 12.8 & 0.514 & $-102.8$ & $-113.5$ & 117.0 & $-0.7$ & 0.514 & $-95.5$ & 98.3 & 0.213 & 0.902 \\
Fixed & 500,000 & 1.25 & 35 & 500,000 & 8.9 & 10.0 & 0.514 & $-80.3$ & $-88.8$ & 91.6 & $-0.4$ & 0.514 & $-74.0$ & 365.7 & 0.217 & 0.920 \\
Fixed & 1,000,000 & 0.75 & 35 & 1,000,000 & 14.2 & 16.8 & 0.514 & $-131.5$ & $-145.9$ & 159.2 & 0.6 & 0.543 & $-128.9$ & 118.9 & 0.243 & 0.939 \\
Fixed & 1,000,000 & 1.00 & 35 & 1,000,000 & 10.5 & 12.6 & 0.514 & $-99.2$ & $-110.2$ & 120.5 & 0.6 & 0.543 & $-96.0$ & 100.0 & 0.250 & 0.910 \\
Fixed & 1,000,000 & 1.25 & 35 & 1,000,000 & 8.1 & 9.8 & 0.514 & $-77.6$ & $-86.1$ & 94.4 & 0.6 & 0.543 & $-74.4$ & 367.1 & 0.254 & 0.893 \\
Uncertain & 10,000 & 0.75 & 35 & 12,000 & $-0.2$ & 4.9 & 0.514 & $-228.9$ & $-246.0$ & 265.5 & $-19.4$ & 0.371 & $-144.5$ & 131.6 & 0.990 & 0.071 \\
Uncertain & 10,000 & 1.00 & 35 & 12,000 & $-0.4$ & 3.6 & 0.514 & $-172.9$ & $-185.9$ & 200.9 & $-14.5$ & 0.371 & $-107.6$ & 109.6 & 0.976 & 0.070 \\
Uncertain & 10,000 & 1.25 & 35 & 12,000 & 214.2 & 217.0 & 1.000 & 76.6 & 60.6 & 0.0 & 203.6 & 1.000 & 124.7 & 152.0 & $<0.001$ & $<0.001$ \\
Uncertain & 100,000 & 0.75 & 35 & 125,000 & 1.9 & 4.0 & 0.514 & $-238.4$ & $-267.9$ & 288.6 & $-16.6$ & 0.400 & $-156.1$ & 129.6 & 0.920 & 0.104 \\
Uncertain & 100,000 & 1.00 & 35 & 125,000 & 1.2 & 3.0 & 0.514 & $-179.9$ & $-202.5$ & 218.4 & $-12.5$ & 0.400 & $-117.2$ & 108.1 & 0.935 & 0.103 \\
Uncertain & 100,000 & 1.25 & 35 & 125,000 & 269.2 & 262.2 & 1.000 & 124.2 & 99.2 & 0.0 & 258.9 & 1.000 & 168.0 & 95.1 & $<0.001$ & $<0.001$ \\
Uncertain & 500,000 & 0.75 & 35 & 625,000 & $-1.7$ & 7.0 & 0.514 & $-233.5$ & $-269.0$ & 289.5 & $-17.9$ & 0.400 & $-155.6$ & 134.1 & 0.926 & 0.095 \\
Uncertain & 500,000 & 1.00 & 35 & 625,000 & $-1.6$ & 5.3 & 0.514 & $-176.2$ & $-203.3$ & 219.0 & $-13.4$ & 0.400 & $-116.8$ & 111.4 & 0.911 & 0.095 \\
Uncertain & 500,000 & 1.25 & 35 & 625,000 & 267.0 & 248.8 & 1.000 & 127.0 & 98.5 & 0.0 & 258.1 & 1.000 & 168.2 & 97.8 & $<0.001$ & $<0.001$ \\
Uncertain & 1,000,000 & 0.75 & 35 & 1,250,000 & $-9.2$ & $-13.1$ & 0.486 & $-233.9$ & $-261.1$ & 290.2 & $-22.4$ & 0.343 & $-157.9$ & 142.8 & 0.610 & 0.023 \\
Uncertain & 1,000,000 & 1.00 & 35 & 1,250,000 & $-7.2$ & $-9.8$ & 0.486 & $-176.8$ & $-197.4$ & 219.6 & $-16.8$ & 0.343 & $-118.5$ & 118.0 & 0.596 & 0.023 \\
Uncertain & 1,000,000 & 1.25 & 35 & 1,250,000 & 262.6 & 247.3 & 1.000 & 131.3 & 103.2 & 0.0 & 255.4 & 1.000 & 166.9 & 103.0 & $<0.001$ & $<0.001$ \\
\bottomrule
\end{tabular}
\end{adjustbox}
\end{landscapetablepage}

\clearpage
\begin{landscapetablepage}
\captionof{table}[Non-Overlapping-Window Procurement Results at $h=4$]{\textbf{Non-Overlapping-Window Procurement Results at $h=4$.} The table reports all fixed-quantity and uncertain-demand combinations for four order sizes and three realized-demand levels. T and I denote TWAP and immediate purchase. Savings, tail measures, maximum loss, and oracle regret are in basis points. HAC $p$-values are two-sided with lag $h$.}
\label{tab:app_procurement_nonoverlap_h4}
\scriptsize
\setlength{\tabcolsep}{2.2pt}
\renewcommand{\arraystretch}{0.95}
\begin{adjustbox}{width=\linewidth}
\begin{tabular}{lrrrrrrrrrrrrrrrr}
\toprule
Mode & $Q$ & $D/Q$ & $N$ & Planned & Mean T & Median T & Win T & Worst 10\% T & CVaR$_5$ T & Max loss T & Mean I & Win I & CVaR$_5$ I & Regret & HAC $p$ T & HAC $p$ I \\
\midrule
Fixed & 10,000 & 0.75 & 28 & 10,000 & 38.3 & 43.6 & 0.571 & $-117.9$ & $-136.5$ & 150.7 & 1.7 & 0.250 & $-242.3$ & 102.5 & 0.002 & 0.933 \\
Fixed & 10,000 & 1.00 & 28 & 10,000 & 28.7 & 32.7 & 0.571 & $-88.9$ & $-102.9$ & 113.5 & 1.6 & 0.250 & $-180.0$ & 95.4 & 0.002 & 0.914 \\
Fixed & 10,000 & 1.25 & 28 & 10,000 & 22.3 & 25.5 & 0.571 & $-69.5$ & $-80.3$ & 88.7 & 1.4 & 0.250 & $-139.2$ & 369.9 & 0.003 & 0.903 \\
Fixed & 100,000 & 0.75 & 28 & 100,000 & 38.5 & 43.4 & 0.571 & $-112.3$ & $-132.6$ & 150.8 & 2.5 & 0.250 & $-241.3$ & 102.4 & 0.002 & 0.902 \\
Fixed & 100,000 & 1.00 & 28 & 100,000 & 28.9 & 32.6 & 0.571 & $-84.7$ & $-99.9$ & 113.7 & 2.3 & 0.250 & $-179.2$ & 95.4 & 0.002 & 0.883 \\
Fixed & 100,000 & 1.25 & 28 & 100,000 & 22.5 & 25.4 & 0.571 & $-66.2$ & $-78.0$ & 88.8 & 1.9 & 0.250 & $-138.6$ & 369.8 & 0.002 & 0.872 \\
Fixed & 500,000 & 0.75 & 28 & 500,000 & 24.5 & 14.4 & 0.536 & $-123.9$ & $-133.3$ & 151.6 & $-8.8$ & 0.286 & $-228.8$ & 117.4 & 0.010 & 0.600 \\
Fixed & 500,000 & 1.00 & 28 & 500,000 & 18.3 & 10.8 & 0.536 & $-93.5$ & $-100.5$ & 114.3 & $-6.3$ & 0.286 & $-170.0$ & 106.6 & 0.011 & 0.622 \\
Fixed & 500,000 & 1.25 & 28 & 500,000 & 14.2 & 8.4 & 0.536 & $-73.1$ & $-78.5$ & 89.3 & $-4.7$ & 0.286 & $-131.5$ & 378.8 & 0.011 & 0.635 \\
Fixed & 1,000,000 & 0.75 & 28 & 1,000,000 & 9.7 & 6.6 & 0.536 & $-126.7$ & $-133.9$ & 145.5 & $-20.3$ & 0.321 & $-252.0$ & 133.3 & 0.395 & 0.143 \\
Fixed & 1,000,000 & 1.00 & 28 & 1,000,000 & 7.2 & 4.9 & 0.536 & $-95.7$ & $-101.0$ & 109.6 & $-14.9$ & 0.321 & $-187.3$ & 118.6 & 0.405 & 0.153 \\
Fixed & 1,000,000 & 1.25 & 28 & 1,000,000 & 5.5 & 3.8 & 0.536 & $-74.7$ & $-78.9$ & 85.7 & $-11.5$ & 0.321 & $-144.9$ & 388.4 & 0.410 & 0.159 \\
Uncertain & 10,000 & 0.75 & 28 & 12,000 & 33.4 & 32.7 & 0.607 & $-163.8$ & $-186.9$ & 218.3 & $-3.0$ & 0.500 & $-212.4$ & 107.5 & 0.007 & 0.877 \\
Uncertain & 10,000 & 1.00 & 28 & 12,000 & 24.9 & 24.5 & 0.607 & $-123.6$ & $-140.9$ & 164.5 & $-2.0$ & 0.500 & $-157.7$ & 99.2 & 0.008 & 0.891 \\
Uncertain & 10,000 & 1.25 & 28 & 12,000 & 233.7 & 236.7 & 1.000 & 116.9 & 106.4 & 0.0 & 213.4 & 1.000 & 96.5 & 150.0 & $<0.001$ & $<0.001$ \\
Uncertain & 100,000 & 0.75 & 28 & 125,000 & 27.0 & 26.5 & 0.607 & $-196.0$ & $-206.8$ & 235.5 & $-8.8$ & 0.429 & $-207.4$ & 114.2 & 0.036 & 0.668 \\
Uncertain & 100,000 & 1.00 & 28 & 125,000 & 20.1 & 19.8 & 0.607 & $-147.9$ & $-156.0$ & 177.5 & $-6.4$ & 0.429 & $-154.0$ & 104.2 & 0.038 & 0.681 \\
Uncertain & 100,000 & 1.25 & 28 & 125,000 & 283.6 & 288.8 & 1.000 & 151.2 & 134.9 & 0.0 & 263.7 & 1.000 & 153.9 & 98.2 & $<0.001$ & $<0.001$ \\
Uncertain & 500,000 & 0.75 & 28 & 625,000 & 6.2 & 18.0 & 0.536 & $-207.4$ & $-223.9$ & 236.8 & $-26.7$ & 0.429 & $-225.5$ & 136.0 & 0.672 & 0.084 \\
Uncertain & 500,000 & 1.00 & 28 & 625,000 & 4.4 & 13.5 & 0.536 & $-156.5$ & $-168.9$ & 178.4 & $-19.9$ & 0.429 & $-167.5$ & 120.6 & 0.688 & 0.086 \\
Uncertain & 500,000 & 1.25 & 28 & 625,000 & 271.4 & 270.5 & 1.000 & 144.4 & 124.8 & 0.0 & 253.1 & 1.000 & 143.3 & 111.4 & $<0.001$ & $<0.001$ \\
Uncertain & 1,000,000 & 0.75 & 28 & 1,250,000 & $-12.0$ & 14.7 & 0.536 & $-254.3$ & $-267.5$ & 274.4 & $-41.5$ & 0.393 & $-196.7$ & 155.5 & 0.544 & 0.013 \\
Uncertain & 1,000,000 & 1.00 & 28 & 1,250,000 & $-9.3$ & 11.0 & 0.536 & $-192.0$ & $-202.1$ & 207.3 & $-31.1$ & 0.393 & $-146.2$ & 135.2 & 0.533 & 0.012 \\
Uncertain & 1,000,000 & 1.25 & 28 & 1,250,000 & 260.6 & 268.0 & 1.000 & 121.3 & 119.1 & 0.0 & 244.2 & 1.000 & 140.8 & 123.1 & $<0.001$ & $<0.001$ \\
\bottomrule
\end{tabular}
\end{adjustbox}
\end{landscapetablepage}

\clearpage
\begin{landscapetablepage}
\captionof{table}[Non-Overlapping-Window Procurement Results at $h=5$]{\textbf{Non-Overlapping-Window Procurement Results at $h=5$.} The table reports all fixed-quantity and uncertain-demand combinations for four order sizes and three realized-demand levels. T and I denote TWAP and immediate purchase. Savings, tail measures, maximum loss, and oracle regret are in basis points. HAC $p$-values are two-sided with lag $h$.}
\label{tab:app_procurement_nonoverlap_h5}
\scriptsize
\setlength{\tabcolsep}{2.2pt}
\renewcommand{\arraystretch}{0.95}
\begin{adjustbox}{width=\linewidth}
\begin{tabular}{lrrrrrrrrrrrrrrrr}
\toprule
Mode & $Q$ & $D/Q$ & $N$ & Planned & Mean T & Median T & Win T & Worst 10\% T & CVaR$_5$ T & Max loss T & Mean I & Win I & CVaR$_5$ I & Regret & HAC $p$ T & HAC $p$ I \\
\midrule
Fixed & 10,000 & 0.75 & 23 & 10,000 & 49.0 & 68.4 & 0.609 & $-148.4$ & $-166.6$ & 186.1 & 9.9 & 0.435 & $-274.3$ & 94.1 & 0.027 & 0.758 \\
Fixed & 10,000 & 1.00 & 23 & 10,000 & 36.7 & 51.3 & 0.609 & $-111.3$ & $-125.1$ & 140.0 & 7.7 & 0.435 & $-204.3$ & 104.4 & 0.027 & 0.749 \\
Fixed & 10,000 & 1.25 & 23 & 10,000 & 28.6 & 40.0 & 0.609 & $-86.7$ & $-97.6$ & 109.2 & 6.1 & 0.435 & $-158.6$ & 389.9 & 0.027 & 0.745 \\
Fixed & 100,000 & 0.75 & 23 & 100,000 & 47.8 & 68.2 & 0.652 & $-148.9$ & $-166.7$ & 186.3 & 9.2 & 0.435 & $-284.8$ & 95.4 & 0.031 & 0.779 \\
Fixed & 100,000 & 1.00 & 23 & 100,000 & 35.8 & 51.1 & 0.652 & $-111.7$ & $-125.2$ & 140.1 & 7.2 & 0.435 & $-212.1$ & 105.4 & 0.031 & 0.770 \\
Fixed & 100,000 & 1.25 & 23 & 100,000 & 27.9 & 39.8 & 0.652 & $-87.0$ & $-97.7$ & 109.3 & 5.7 & 0.435 & $-164.7$ & 390.7 & 0.030 & 0.766 \\
Fixed & 500,000 & 0.75 & 23 & 500,000 & 52.4 & 67.4 & 0.696 & $-144.4$ & $-167.3$ & 187.1 & 16.4 & 0.522 & $-286.0$ & 91.2 & 0.005 & 0.642 \\
Fixed & 500,000 & 1.00 & 23 & 500,000 & 39.3 & 50.5 & 0.696 & $-108.5$ & $-125.6$ & 140.7 & 12.6 & 0.522 & $-213.0$ & 102.3 & 0.005 & 0.634 \\
Fixed & 500,000 & 1.25 & 23 & 500,000 & 30.5 & 39.3 & 0.696 & $-84.6$ & $-98.0$ & 109.8 & 9.9 & 0.522 & $-165.4$ & 388.1 & 0.005 & 0.630 \\
Fixed & 1,000,000 & 0.75 & 23 & 1,000,000 & 53.0 & 69.9 & 0.652 & $-123.7$ & $-142.1$ & 147.8 & 20.1 & 0.652 & $-285.3$ & 91.3 & 0.006 & 0.572 \\
Fixed & 1,000,000 & 1.00 & 23 & 1,000,000 & 39.7 & 52.4 & 0.652 & $-92.9$ & $-106.7$ & 110.8 & 15.4 & 0.652 & $-212.5$ & 102.4 & 0.006 & 0.564 \\
Fixed & 1,000,000 & 1.25 & 23 & 1,000,000 & 30.9 & 40.8 & 0.652 & $-72.5$ & $-83.3$ & 86.4 & 12.1 & 0.652 & $-165.0$ & 388.2 & 0.006 & 0.560 \\
Uncertain & 10,000 & 0.75 & 23 & 12,000 & 55.3 & 54.1 & 0.609 & $-175.2$ & $-210.0$ & 246.1 & 16.4 & 0.696 & $-300.8$ & 87.7 & 0.043 & 0.641 \\
Uncertain & 10,000 & 1.00 & 23 & 12,000 & 41.3 & 40.8 & 0.609 & $-131.5$ & $-157.7$ & 185.1 & 12.5 & 0.696 & $-224.0$ & 99.7 & 0.043 & 0.635 \\
Uncertain & 10,000 & 1.25 & 23 & 12,000 & 246.5 & 245.3 & 1.000 & 110.5 & 81.1 & 0.0 & 224.5 & 0.957 & 26.1 & 162.7 & $<0.001$ & $<0.001$ \\
Uncertain & 100,000 & 0.75 & 23 & 125,000 & 57.4 & 46.6 & 0.609 & $-177.1$ & $-214.4$ & 251.9 & 19.2 & 0.696 & $-308.3$ & 85.7 & 0.044 & 0.594 \\
Uncertain & 100,000 & 1.00 & 23 & 125,000 & 42.9 & 35.2 & 0.609 & $-132.9$ & $-161.0$ & 189.5 & 14.5 & 0.696 & $-229.6$ & 98.1 & 0.045 & 0.590 \\
Uncertain & 100,000 & 1.25 & 23 & 125,000 & 301.3 & 294.2 & 1.000 & 160.4 & 129.6 & 0.0 & 279.8 & 1.000 & 71.7 & 105.6 & $<0.001$ & $<0.001$ \\
Uncertain & 500,000 & 0.75 & 23 & 625,000 & 64.7 & 45.7 & 0.652 & $-184.4$ & $-216.8$ & 253.2 & 29.0 & 0.739 & $-305.7$ & 78.7 & 0.010 & 0.446 \\
Uncertain & 500,000 & 1.00 & 23 & 625,000 & 48.4 & 34.5 & 0.652 & $-138.6$ & $-162.8$ & 190.4 & 21.9 & 0.739 & $-227.7$ & 92.9 & 0.010 & 0.442 \\
Uncertain & 500,000 & 1.25 & 23 & 625,000 & 305.6 & 300.3 & 1.000 & 157.9 & 128.2 & 0.0 & 285.5 & 1.000 & 73.2 & 101.4 & $<0.001$ & $<0.001$ \\
Uncertain & 1,000,000 & 0.75 & 23 & 1,250,000 & 58.3 & 44.9 & 0.609 & $-170.0$ & $-195.6$ & 209.5 & 25.8 & 0.652 & $-309.6$ & 86.0 & 0.029 & 0.481 \\
Uncertain & 1,000,000 & 1.00 & 23 & 1,250,000 & 43.6 & 33.9 & 0.609 & $-127.7$ & $-146.9$ & 157.5 & 19.5 & 0.652 & $-230.5$ & 98.3 & 0.029 & 0.477 \\
Uncertain & 1,000,000 & 1.25 & 23 & 1,250,000 & 301.8 & 293.2 & 1.000 & 165.1 & 140.6 & 0.0 & 283.6 & 1.000 & 71.0 & 105.8 & $<0.001$ & $<0.001$ \\
\bottomrule
\end{tabular}
\end{adjustbox}
\end{landscapetablepage}

\clearpage
\section{Directional Diagnostics}
\label{app:directional-diagnostics}

This appendix reports directional performance across horizons one through five for the same 169 final-holdout origins used in the main analysis. Each horizon maintains a separate queue of unsettled outcomes, and expert weights are updated when the corresponding target date is reached. The exercise is intentionally separated from the procurement analysis because a trading rule and a compliance buyer use the same forecast for different objectives.

\subsection{Return, Risk, and Path Diagnostics}

Tables~\ref{tab:app_directional_return_risk}--\ref{tab:app_directional_ce} report return and downside-risk measures, statistical inference and execution outcomes, and certainty-equivalent performance for all six strategies and five horizons. Figure~\ref{fig:directional_equity_panels} plots cumulative-return paths across horizons, while Figure~\ref{fig:directional_drawdown_panels} plots the corresponding drawdowns.

The directional evidence is mixed in exactly the way the main text emphasizes. The released rule earns positive cumulative returns at all five horizons, but buy-and-hold and volatility-scaled passive exposure are competitive at longer horizons. Mean-return inference is strongest at $h=1$, $h=2$, $h=3$, and $h=5$, while the $h=4$ estimate is weaker. Certainty-equivalent gains relative to the volatility-scaled passive benchmark are positive at $h=1$ and $h=2$ but negative from $h=3$ through $h=5$. The figures therefore help bound the interpretation: the forecast is economically useful in the procurement problem, but it is not a uniformly superior stand-alone directional trading signal.
\inserttable{\ref{tab:app_directional_return_risk}}
\inserttable{\ref{tab:app_directional_inference_execution}}
\inserttable{\ref{tab:app_directional_ce}}
\insertfigure{\ref{fig:directional_equity_panels}}
\insertfigure{\ref{fig:directional_drawdown_panels}}

\clearpage
\begin{landscapetablepage}
\captionof{table}[Directional Return and Risk by Horizon]{\textbf{Directional Return and Risk by Horizon.} The table reports maintained trading settings, cumulative return, Sharpe ratio, Sortino ratio, and maximum drawdown for all six strategies at horizons one through five.}
\label{tab:app_directional_return_risk}
\scriptsize
\setlength{\tabcolsep}{3pt}
\renewcommand{\arraystretch}{0.82}
\begin{adjustbox}{width=\linewidth}
\begin{tabular}{rlrrrrrrrr}
\toprule
$h$ & Strategy & One-way fee & Round trip & Short & Max position & Cum. ret. & Sharpe & Sortino & Max DD \\
\midrule
1 & Released rule & 0.001 & 0.002 & 1 & 1.0 & 0.376 & 2.416 & 2.878 & $-0.073$ \\
1 & Buy and hold & 0.001 & 0.002 & 1 & 1.0 & 0.318 & 2.069 & 2.444 & $-0.082$ \\
1 & Vol.-scaled passive & 0.001 & 0.002 & 1 & 1.0 & 0.354 & 2.289 & 2.735 & $-0.082$ \\
1 & Random-walk no-change & 0.001 & 0.002 & 1 & 1.0 & 0.000 & --- & --- & 0.000 \\
1 & Best parsimonious & 0.001 & 0.002 & 1 & 1.0 & $-0.178$ & $-1.277$ & $-1.133$ & $-0.235$ \\
1 & Best flexible & 0.001 & 0.002 & 1 & 1.0 & $-0.023$ & $-0.061$ & $-0.061$ & $-0.216$ \\
2 & Released rule & 0.001 & 0.002 & 1 & 1.0 & 0.382 & 2.708 & 3.351 & $-0.070$ \\
2 & Buy and hold & 0.001 & 0.002 & 1 & 1.0 & 0.310 & 2.151 & 2.580 & $-0.076$ \\
2 & Vol.-scaled passive & 0.001 & 0.002 & 1 & 1.0 & 0.330 & 2.279 & 2.755 & $-0.067$ \\
2 & Random-walk no-change & 0.001 & 0.002 & 1 & 1.0 & 0.000 & --- & --- & 0.000 \\
2 & Best parsimonious & 0.001 & 0.002 & 1 & 1.0 & $-0.209$ & $-1.665$ & $-1.502$ & $-0.234$ \\
2 & Best flexible & 0.001 & 0.002 & 1 & 1.0 & $-0.117$ & $-0.836$ & $-0.815$ & $-0.243$ \\
3 & Released rule & 0.001 & 0.002 & 1 & 1.0 & 0.277 & 2.447 & 2.865 & $-0.051$ \\
3 & Buy and hold & 0.001 & 0.002 & 1 & 1.0 & 0.302 & 2.346 & 2.295 & $-0.069$ \\
3 & Vol.-scaled passive & 0.001 & 0.002 & 1 & 1.0 & 0.324 & 2.514 & 2.466 & $-0.062$ \\
3 & Random-walk no-change & 0.001 & 0.002 & 1 & 1.0 & 0.000 & --- & --- & 0.000 \\
3 & Best parsimonious & 0.001 & 0.002 & 1 & 1.0 & $-0.153$ & $-1.291$ & $-1.380$ & $-0.202$ \\
3 & Best flexible & 0.001 & 0.002 & 1 & 1.0 & $-0.102$ & $-0.808$ & $-0.826$ & $-0.230$ \\
4 & Released rule & 0.001 & 0.002 & 1 & 1.0 & 0.179 & 1.424 & 1.573 & $-0.066$ \\
4 & Buy and hold & 0.001 & 0.002 & 1 & 1.0 & 0.300 & 2.202 & 2.680 & $-0.076$ \\
4 & Vol.-scaled passive & 0.001 & 0.002 & 1 & 1.0 & 0.290 & 2.170 & 2.634 & $-0.076$ \\
4 & Random-walk no-change & 0.001 & 0.002 & 1 & 1.0 & 0.000 & --- & --- & 0.000 \\
4 & Best parsimonious & 0.001 & 0.002 & 1 & 1.0 & $-0.189$ & $-1.544$ & $-1.340$ & $-0.221$ \\
4 & Best flexible & 0.001 & 0.002 & 1 & 1.0 & $-0.161$ & $-1.273$ & $-1.196$ & $-0.265$ \\
5 & Released rule & 0.001 & 0.002 & 1 & 1.0 & 0.281 & 2.472 & 2.410 & $-0.028$ \\
5 & Buy and hold & 0.001 & 0.002 & 1 & 1.0 & 0.311 & 2.625 & 2.982 & $-0.060$ \\
5 & Vol.-scaled passive & 0.001 & 0.002 & 1 & 1.0 & 0.352 & 3.037 & 3.656 & $-0.042$ \\
5 & Random-walk no-change & 0.001 & 0.002 & 1 & 1.0 & 0.000 & --- & --- & 0.000 \\
5 & Best parsimonious & 0.001 & 0.002 & 1 & 1.0 & $-0.146$ & $-1.325$ & $-1.455$ & $-0.205$ \\
5 & Best flexible & 0.001 & 0.002 & 1 & 1.0 & $-0.057$ & $-0.436$ & $-0.444$ & $-0.205$ \\
\bottomrule
\end{tabular}
\end{adjustbox}
\end{landscapetablepage}

\clearpage
\begin{landscapetablepage}
\captionof{table}[Directional Inference and Execution by Horizon]{\textbf{Directional Inference and Execution by Horizon.} Mean return and daily alpha are accompanied by one-sided $p$-values and $t$-statistics. Alpha is measured relative to the volatility-scaled passive benchmark. Position and turnover columns report sample averages, and total turnover is accumulated over the evaluation path.}
\label{tab:app_directional_inference_execution}
\scriptsize
\setlength{\tabcolsep}{3pt}
\renewcommand{\arraystretch}{0.82}
\begin{adjustbox}{width=\linewidth}
\begin{tabular}{rlrrrrrrrrr}
\toprule
$h$ & Strategy & Mean ret. & $p$(ret.) & $t$(ret.) & Daily $\alpha$ & $p(\alpha)$ & $t(\alpha)$ & Avg. position & Avg. turnover & Total turnover \\
\midrule
1 & Released rule & 0.001973 & 0.007 & 2.475 & 0.000093 & 0.084 & 1.382 & 0.939 & 0.065 & 11.056 \\
1 & Buy and hold & 0.001723 & 0.018 & 2.106 & $-0.000157$ & 0.933 & $-1.508$ & 1.000 & 0.006 & 1.000 \\
1 & Vol.-scaled passive & 0.001880 & 0.011 & 2.310 & 0.000000 & --- & --- & 0.949 & 0.040 & 6.700 \\
1 & Random-walk no-change & 0.000000 & --- & --- & $-0.001880$ & 0.989 & $-2.310$ & 0.000 & 0.000 & 0.000 \\
1 & Best parsimonious & $-0.001068$ & 0.900 & $-1.285$ & $-0.002948$ & 0.971 & $-1.912$ & 1.000 & 0.041 & 7.000 \\
1 & Best flexible & $-0.000051$ & 0.523 & $-0.057$ & $-0.001931$ & 0.915 & $-1.378$ & 1.000 & 0.065 & 11.000 \\
2 & Released rule & 0.001974 & 0.002 & 2.986 & 0.000222 & 0.287 & 0.563 & 0.915 & 0.164 & 13.907 \\
2 & Buy and hold & 0.001668 & 0.010 & 2.367 & $-0.000085$ & 0.879 & $-1.180$ & 1.000 & 0.012 & 1.000 \\
2 & Vol.-scaled passive & 0.001753 & 0.007 & 2.518 & 0.000000 & --- & --- & 0.944 & 0.079 & 6.700 \\
2 & Random-walk no-change & 0.000000 & --- & --- & $-0.001753$ & 0.993 & $-2.518$ & 0.000 & 0.000 & 0.000 \\
2 & Best parsimonious & $-0.001301$ & 0.976 & $-2.016$ & $-0.003054$ & 0.990 & $-2.367$ & 1.000 & 0.035 & 3.000 \\
2 & Best flexible & $-0.000656$ & 0.798 & $-0.839$ & $-0.002408$ & 0.988 & $-2.295$ & 1.000 & 0.082 & 7.000 \\
3 & Released rule & 0.001478 & 0.005 & 2.665 & $-0.000225$ & 0.735 & $-0.633$ & 0.849 & 0.219 & 12.468 \\
3 & Buy and hold & 0.001605 & 0.007 & 2.519 & $-0.000098$ & 0.928 & $-1.481$ & 1.000 & 0.018 & 1.000 \\
3 & Vol.-scaled passive & 0.001703 & 0.004 & 2.740 & 0.000000 & --- & --- & 0.950 & 0.084 & 4.800 \\
3 & Random-walk no-change & 0.000000 & --- & --- & $-0.001703$ & 0.996 & $-2.740$ & 0.000 & 0.000 & 0.000 \\
3 & Best parsimonious & $-0.000906$ & 0.927 & $-1.471$ & $-0.002609$ & 0.992 & $-2.485$ & 1.000 & 0.053 & 3.000 \\
3 & Best flexible & $-0.000567$ & 0.760 & $-0.712$ & $-0.002271$ & 0.987 & $-2.285$ & 1.000 & 0.123 & 7.000 \\
4 & Released rule & 0.001021 & 0.084 & 1.400 & $-0.000528$ & 0.694 & $-0.510$ & 0.926 & 0.624 & 26.834 \\
4 & Buy and hold & 0.001592 & 0.008 & 2.514 & 0.000043 & 0.319 & 0.475 & 1.000 & 0.023 & 1.000 \\
4 & Vol.-scaled passive & 0.001549 & 0.009 & 2.465 & 0.000000 & --- & --- & 0.934 & 0.067 & 2.900 \\
4 & Random-walk no-change & 0.000000 & --- & --- & $-0.001549$ & 0.991 & $-2.465$ & 0.000 & 0.000 & 0.000 \\
4 & Best parsimonious & $-0.001144$ & 0.979 & $-2.090$ & $-0.002693$ & 0.991 & $-2.480$ & 1.000 & 0.070 & 3.000 \\
4 & Best flexible & $-0.000948$ & 0.888 & $-1.234$ & $-0.002497$ & 0.997 & $-2.862$ & 1.000 & 0.070 & 3.000 \\
5 & Released rule & 0.001508 & 0.007 & 2.608 & $-0.000319$ & 0.639 & $-0.358$ & 0.935 & 0.728 & 24.758 \\
5 & Buy and hold & 0.001644 & 0.006 & 2.690 & $-0.000183$ & 0.936 & $-1.558$ & 1.000 & 0.029 & 1.000 \\
5 & Vol.-scaled passive & 0.001827 & 0.001 & 3.270 & 0.000000 & --- & --- & 0.944 & 0.141 & 4.800 \\
5 & Random-walk no-change & 0.000000 & --- & --- & $-0.001827$ & 0.999 & $-3.270$ & 0.000 & 0.000 & 0.000 \\
5 & Best parsimonious & $-0.000873$ & 0.918 & $-1.421$ & $-0.002700$ & 0.997 & $-3.006$ & 1.000 & 0.088 & 3.000 \\
5 & Best flexible & $-0.000289$ & 0.636 & $-0.351$ & $-0.002116$ & 0.976 & $-2.045$ & 1.000 & 0.206 & 7.000 \\
\bottomrule
\end{tabular}
\end{adjustbox}
\end{landscapetablepage}

\clearpage
\begin{landscapetablepage}
\captionof{table}[Directional Certainty-Equivalent Performance]{\textbf{Directional Certainty-Equivalent Performance.} CE denotes certainty-equivalent return. Gains are measured relative to the volatility-scaled passive benchmark and are reported at daily and annual frequencies.}
\label{tab:app_directional_ce}
\footnotesize
\setlength{\tabcolsep}{4pt}
\renewcommand{\arraystretch}{0.90}
\begin{tabular*}{\linewidth}{@{\extracolsep{\fill}}YYZZZZ}
\toprule
$h$ & Strategy & CE daily & CE annual & CE gain daily & CE gain annual \\
\midrule
1 & Released rule & 0.001721 & 0.434 & 0.000096 & 0.024 \\
1 & Buy and hold & 0.001461 & 0.368 & $-0.000164$ & $-0.041$ \\
1 & Vol.-scaled passive & 0.001625 & 0.409 & 0.000000 & 0.000 \\
1 & Random-walk no-change & 0.000000 & 0.000 & $-0.001625$ & $-0.409$ \\
1 & Best parsimonious & $-0.001332$ & $-0.336$ & $-0.002957$ & $-0.745$ \\
1 & Best flexible & $-0.000317$ & $-0.080$ & $-0.001942$ & $-0.489$ \\
2 & Released rule & 0.003547 & 0.447 & 0.000489 & 0.062 \\
2 & Buy and hold & 0.002881 & 0.363 & $-0.000177$ & $-0.022$ \\
2 & Vol.-scaled passive & 0.003058 & 0.385 & 0.000000 & 0.000 \\
2 & Random-walk no-change & 0.000000 & 0.000 & $-0.003058$ & $-0.385$ \\
2 & Best parsimonious & $-0.003064$ & $-0.386$ & $-0.006122$ & $-0.771$ \\
2 & Best flexible & $-0.001777$ & $-0.224$ & $-0.004834$ & $-0.609$ \\
3 & Released rule & 0.004020 & 0.338 & $-0.000569$ & $-0.048$ \\
3 & Buy and hold & 0.004285 & 0.360 & $-0.000304$ & $-0.025$ \\
3 & Vol.-scaled passive & 0.004589 & 0.385 & 0.000000 & 0.000 \\
3 & Random-walk no-change & 0.000000 & 0.000 & $-0.004589$ & $-0.385$ \\
3 & Best parsimonious & $-0.003277$ & $-0.275$ & $-0.007865$ & $-0.661$ \\
3 & Best flexible & $-0.002261$ & $-0.190$ & $-0.006850$ & $-0.575$ \\
4 & Released rule & 0.003307 & 0.208 & $-0.002118$ & $-0.133$ \\
4 & Buy and hold & 0.005577 & 0.351 & 0.000151 & 0.010 \\
4 & Vol.-scaled passive & 0.005426 & 0.342 & 0.000000 & 0.000 \\
4 & Random-walk no-change & 0.000000 & 0.000 & $-0.005426$ & $-0.342$ \\
4 & Best parsimonious & $-0.005403$ & $-0.340$ & $-0.010829$ & $-0.682$ \\
4 & Best flexible & $-0.004631$ & $-0.292$ & $-0.010057$ & $-0.634$ \\
5 & Released rule & 0.006838 & 0.345 & $-0.001613$ & $-0.081$ \\
5 & Buy and hold & 0.007481 & 0.377 & $-0.000971$ & $-0.049$ \\
5 & Vol.-scaled passive & 0.008451 & 0.426 & 0.000000 & 0.000 \\
5 & Random-walk no-change & 0.000000 & 0.000 & $-0.008451$ & $-0.426$ \\
5 & Best parsimonious & $-0.005183$ & $-0.261$ & $-0.013635$ & $-0.687$ \\
5 & Best flexible & $-0.002274$ & $-0.115$ & $-0.010726$ & $-0.541$ \\
\bottomrule
\end{tabular*}
\end{landscapetablepage}

\clearpage
\begin{figure}[p]
\centering
\directionalwidepanel{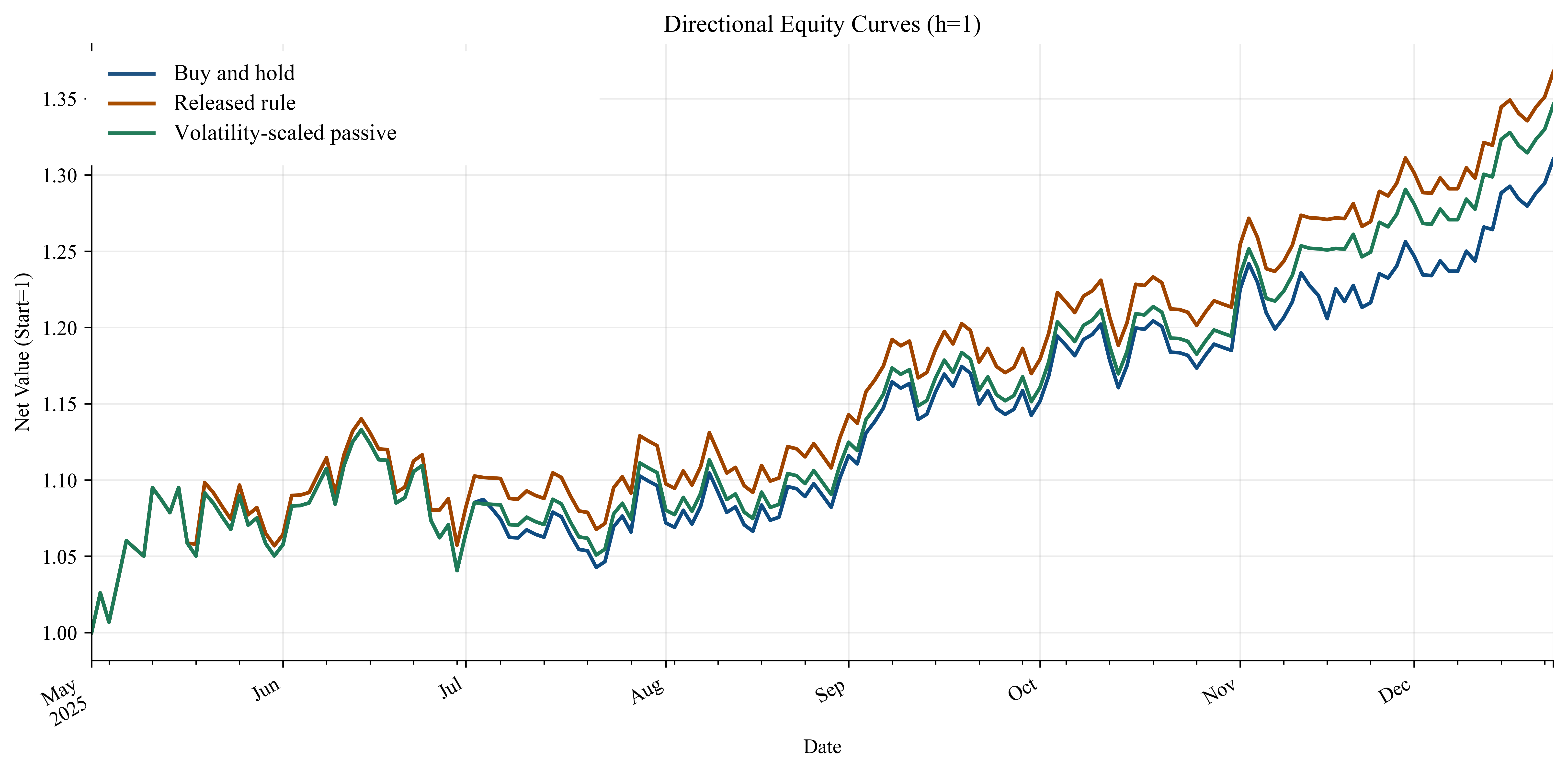}{(a) $h=1$}
\vspace{0.35em}
\directionalwidepanel{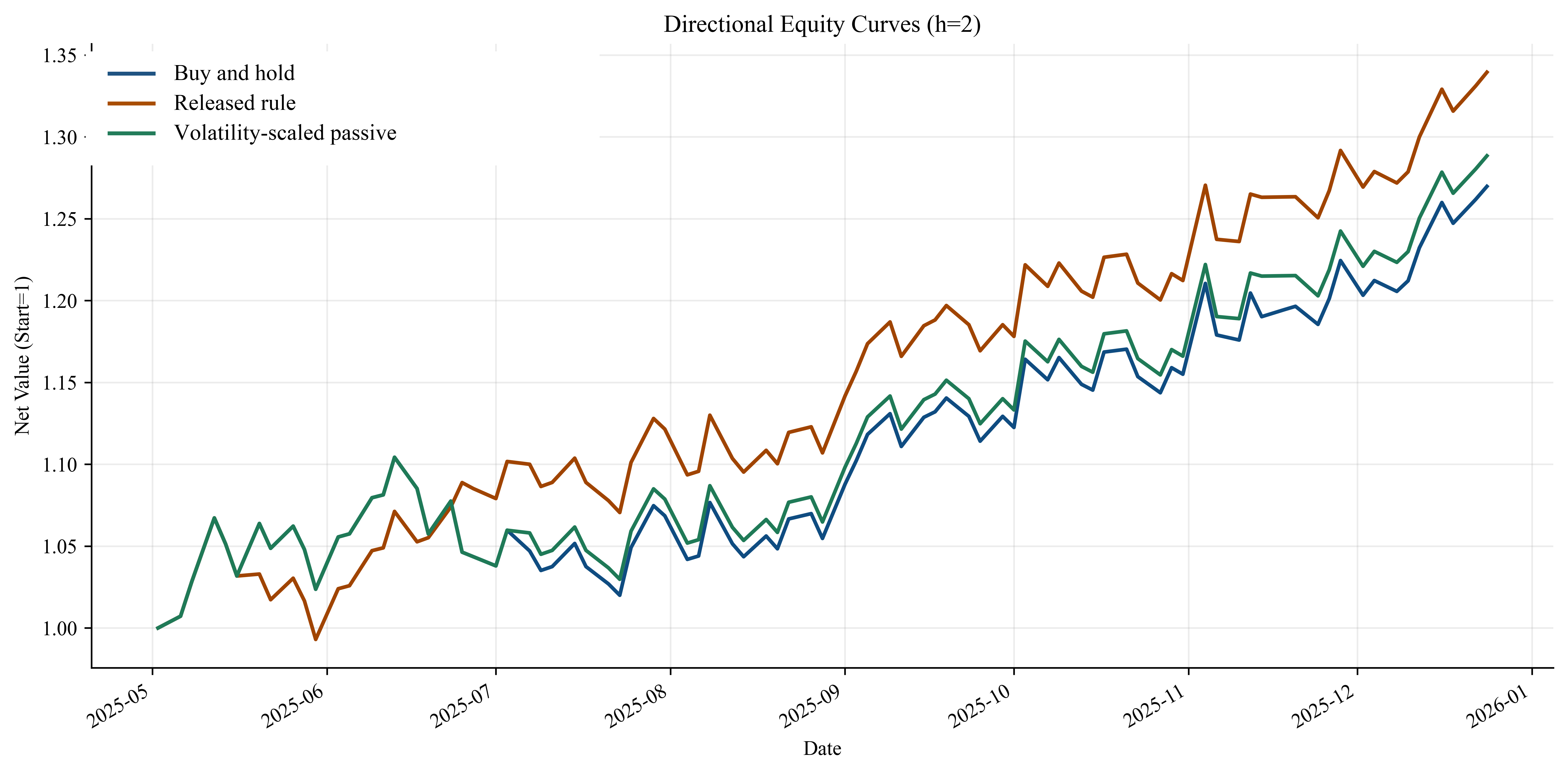}{(b) $h=2$}
\caption[Directional Cumulative Returns by Horizon]{\textbf{Directional Cumulative Returns by Horizon.} Panels plot cumulative returns for the released rule, buy-and-hold, and the volatility-scaled passive benchmark over the final-holdout evaluation window. Returns are accumulated across the same forecast anchors used in Table~\ref{tab:app_directional_return_risk}.}
\label{fig:directional_equity_panels}
\end{figure}

\clearpage
\begin{figure}[p]
\centering
\ContinuedFloat
\directionalwidepanel{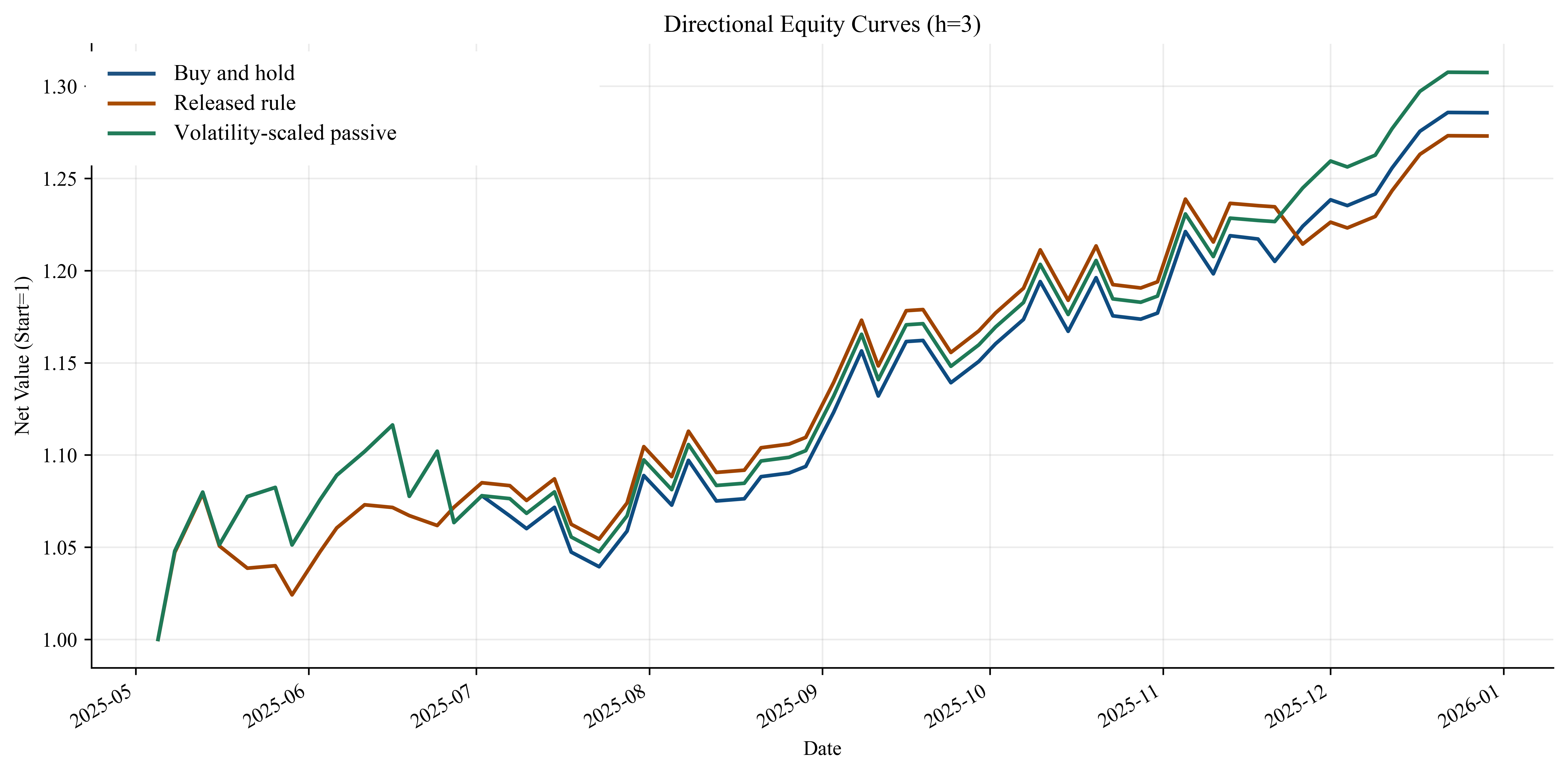}{(c) $h=3$}
\vspace{0.35em}
\directionalwidepanel{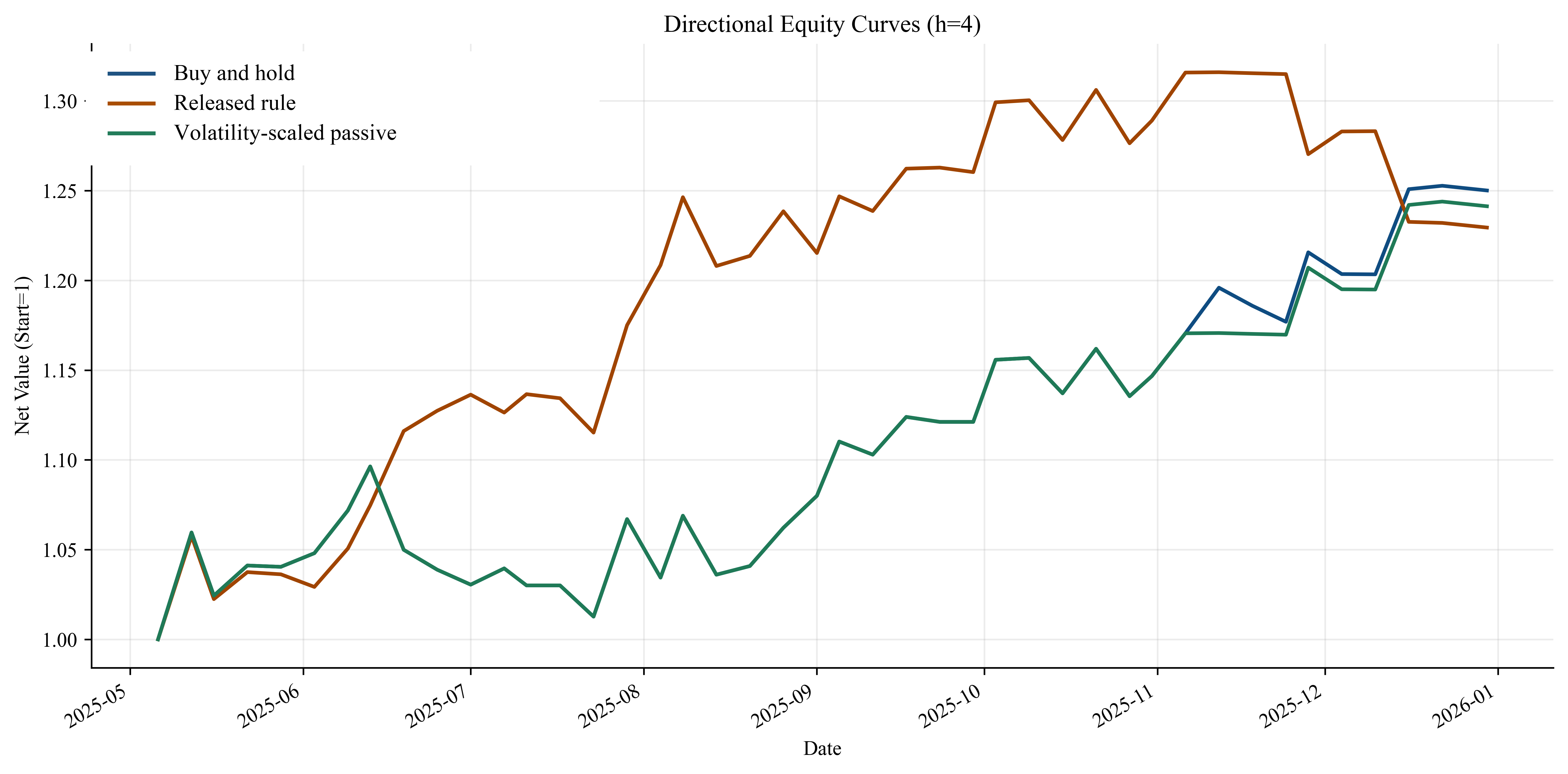}{(d) $h=4$}
\caption[]{\textbf{Directional Cumulative Returns by Horizon (continued).} Panels continue Figure~\ref{fig:directional_equity_panels}.}
\end{figure}

\clearpage
\begin{figure}[p]
\centering
\ContinuedFloat
\directionalwidepanel{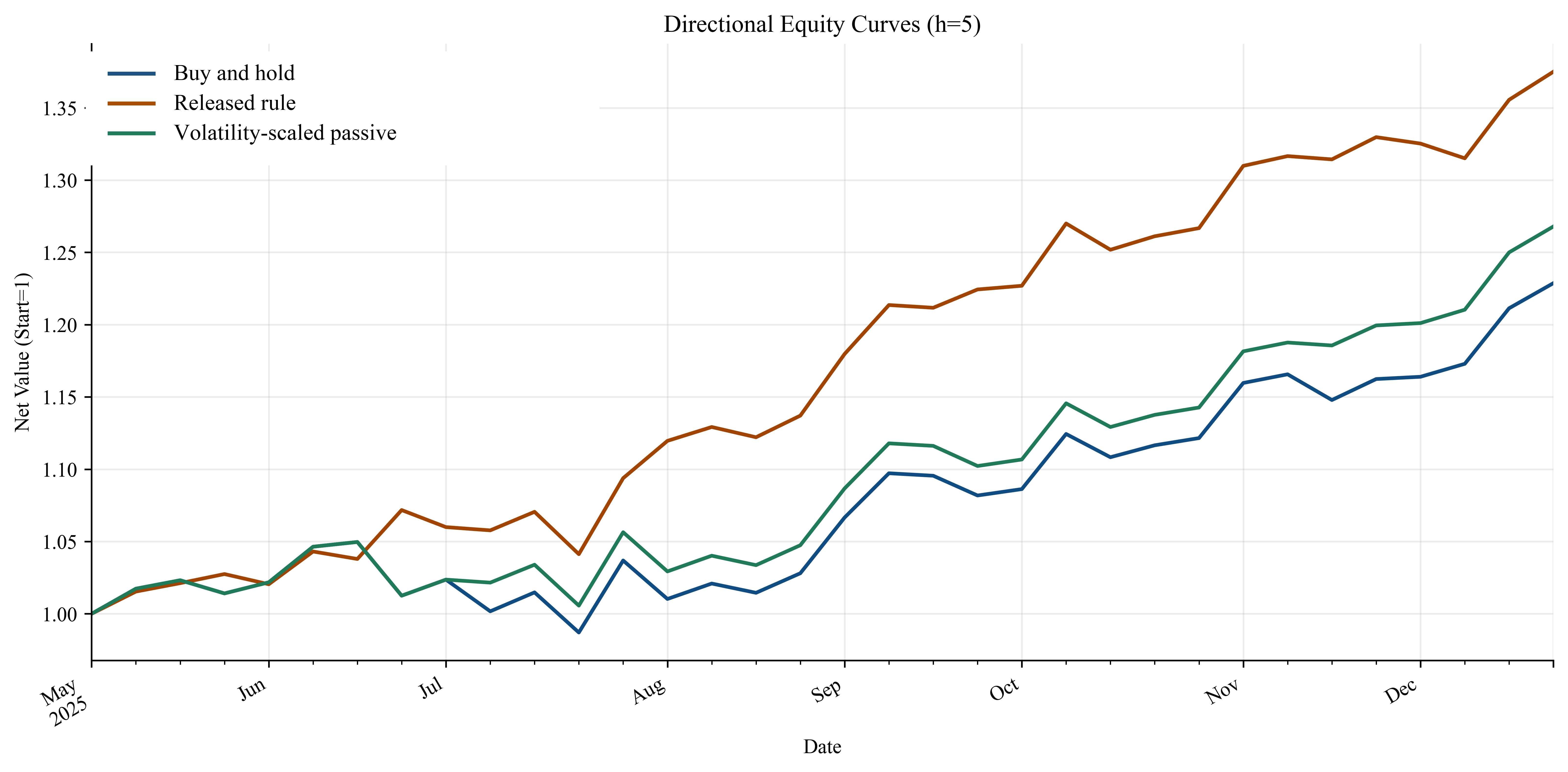}{(e) $h=5$}
\caption[]{\textbf{Directional Cumulative Returns by Horizon (continued).} Panel continues Figure~\ref{fig:directional_equity_panels}.}
\end{figure}

\clearpage
\begin{figure}[p]
\centering
\directionalwidepanel{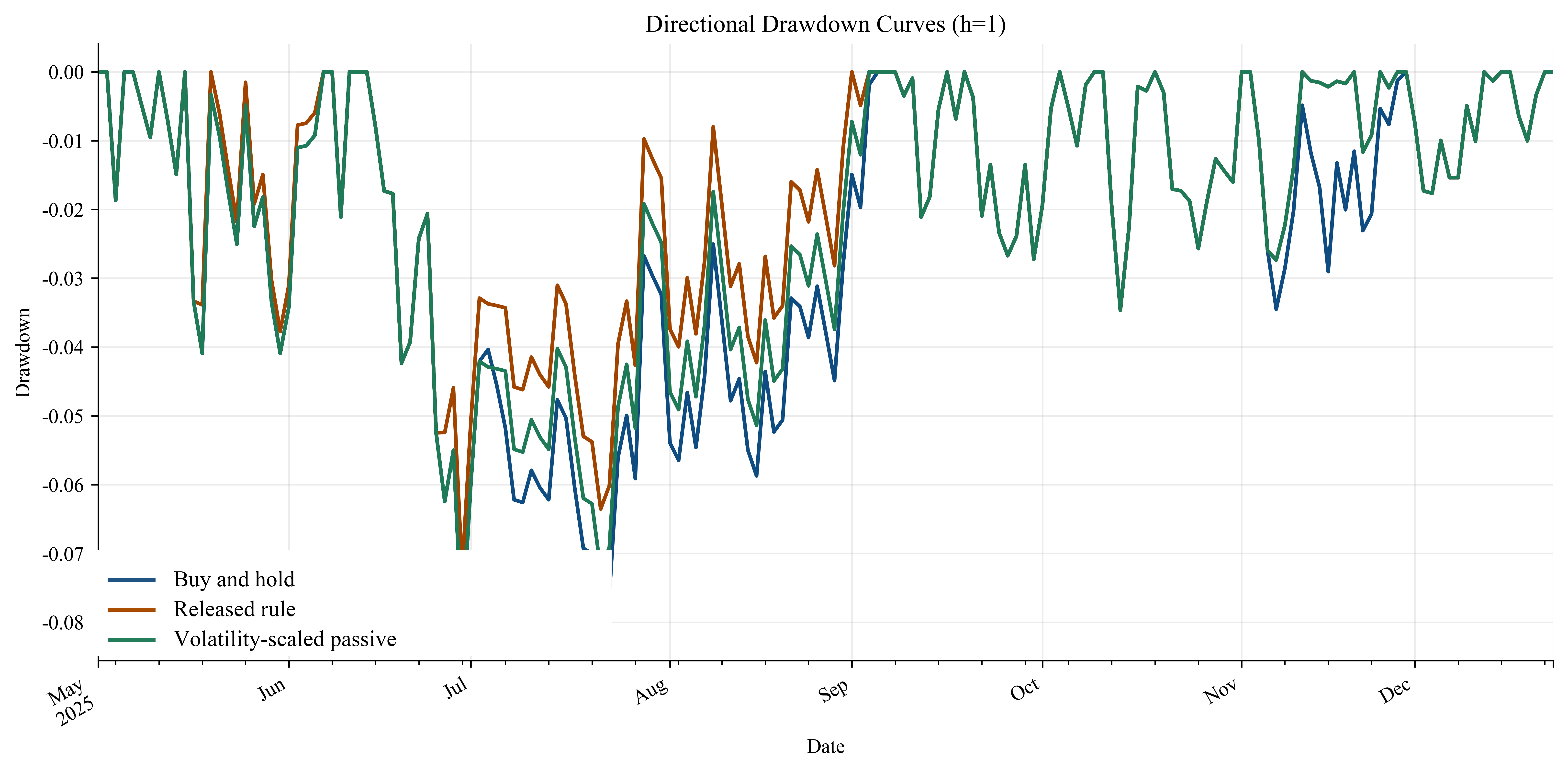}{(a) $h=1$}
\vspace{0.35em}
\directionalwidepanel{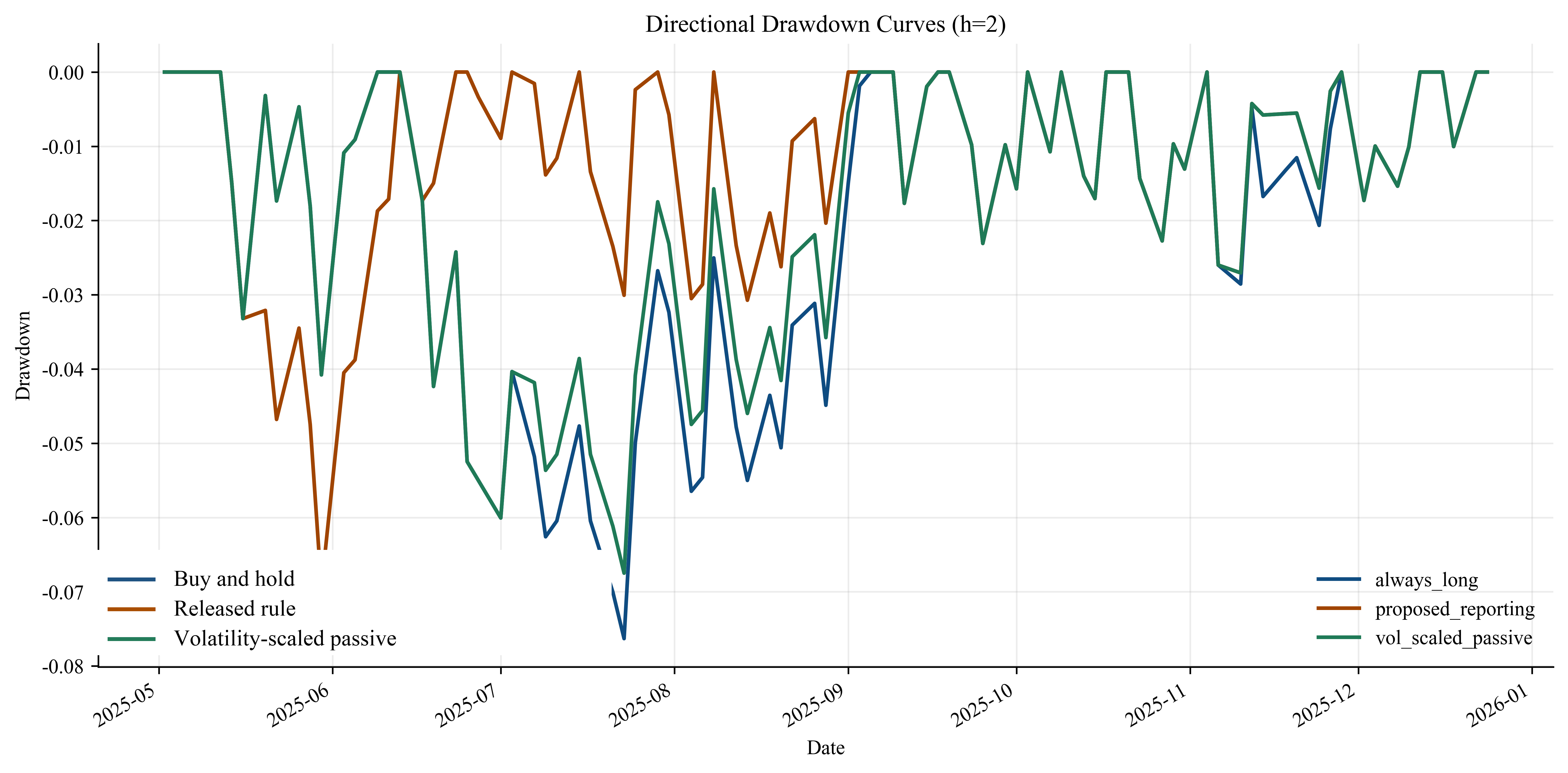}{(b) $h=2$}
\caption[Directional Drawdowns by Horizon]{\textbf{Directional Drawdowns by Horizon.} Panels plot drawdown paths for the released rule, buy-and-hold, and the volatility-scaled passive benchmark over the final-holdout evaluation window. More negative values indicate larger peak-to-trough losses.}
\label{fig:directional_drawdown_panels}
\end{figure}

\clearpage
\begin{figure}[p]
\centering
\ContinuedFloat
\directionalwidepanel{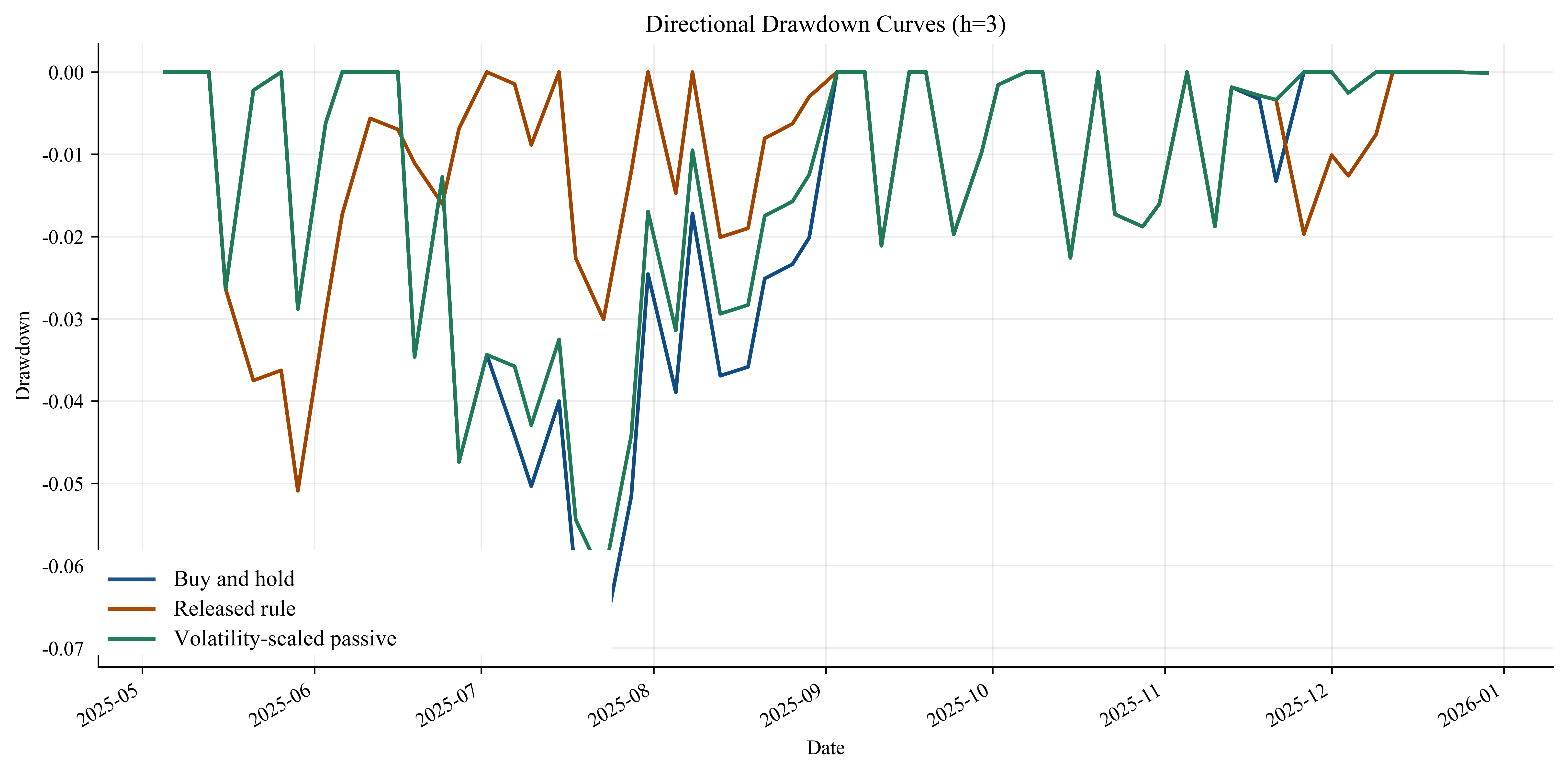}{(c) $h=3$}
\vspace{0.35em}
\directionalwidepanel{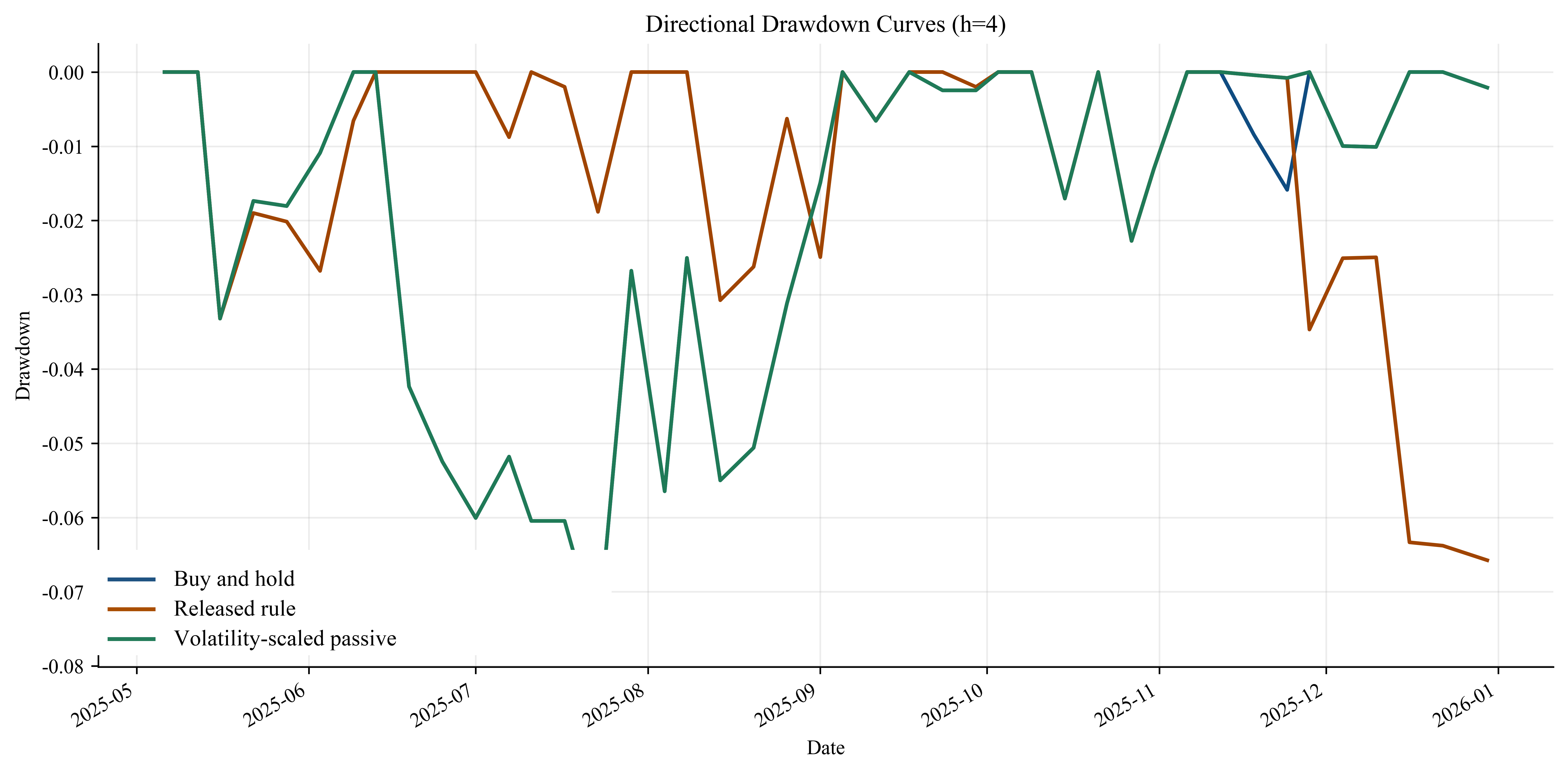}{(d) $h=4$}
\caption[]{\textbf{Directional Drawdowns by Horizon (continued).} Panels continue Figure~\ref{fig:directional_drawdown_panels}.}
\end{figure}

\clearpage
\begin{figure}[p]
\centering
\ContinuedFloat
\directionalwidepanel{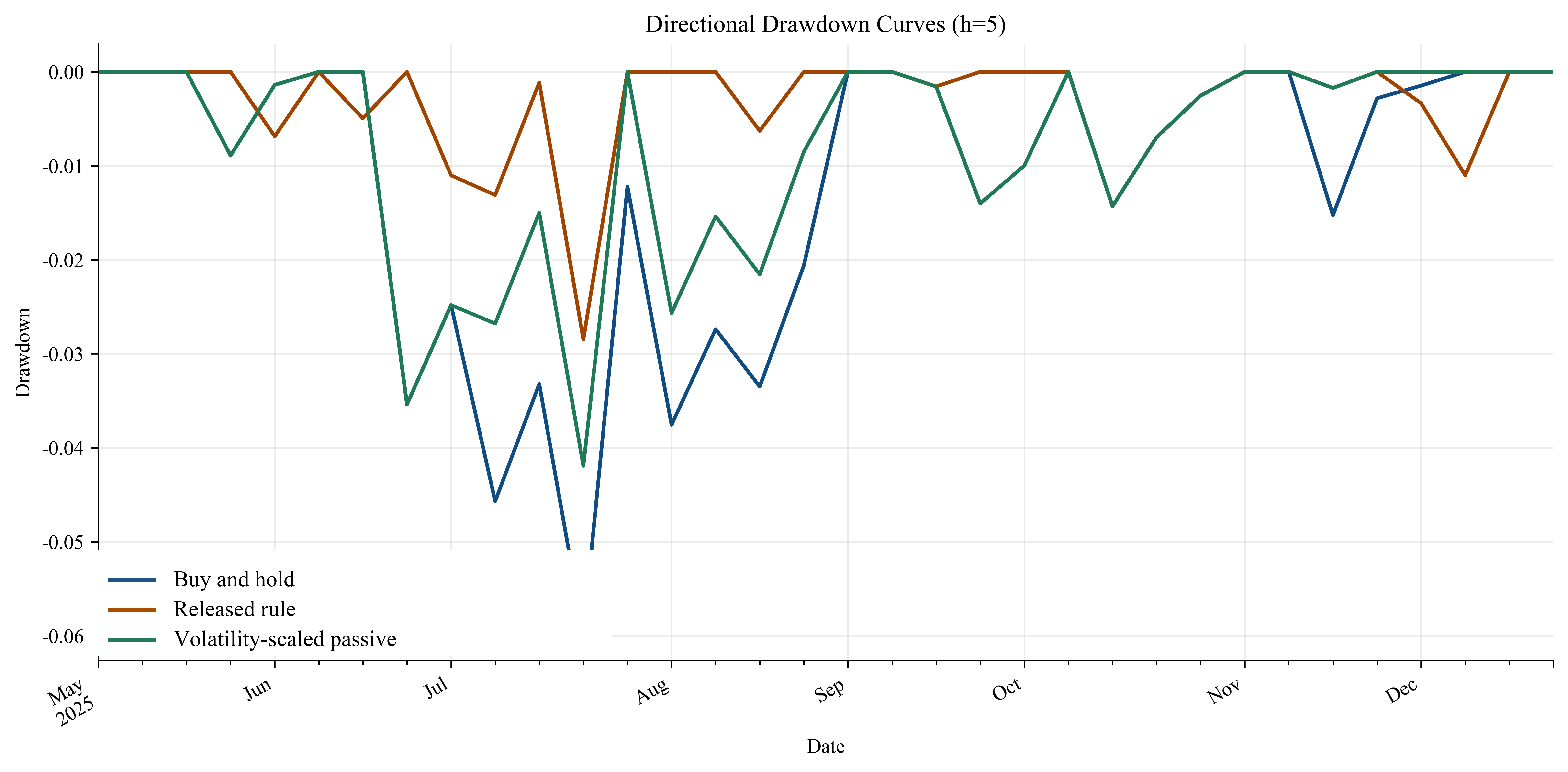}{(e) $h=5$}
\caption[]{\textbf{Directional Drawdowns by Horizon (continued).} Panel continues Figure~\ref{fig:directional_drawdown_panels}.}
\end{figure}
  
\end{document}